\documentclass[aps, prb, twocolumn, superscriptaddress, floatfix]{revtex4-2}
\usepackage[version=4]{mhchem} 
\usepackage[T1]{fontenc} 
\usepackage{colortbl}
\usepackage{hhline}
\usepackage{makecell} 
\usepackage{comment}
\usepackage[normalem]{ulem}

\usepackage{graphicx}
\usepackage{dcolumn}
\usepackage{bm}
\usepackage{tikz} 
\usepackage{soul} 
\usepackage{float}     
\sethlcolor{white}
\usepackage[pdftex,colorlinks=true]{hyperref}
\hypersetup{
  colorlinks=true,
  linkcolor=blue,
  urlcolor=cyan,
}

\usepackage{float}

\begin{document}
\title{Machine Learning-Driven Analytical Models for Threshold Displacement Energy Prediction in Materials} 

\author{Rosty B. Martinez Duque}
  \affiliation{Physics Department, Oklahoma State University, Stillwater, Oklahoma 74075, USA.}
\author{Arman Duha}%
\affiliation{Physics Department, Oklahoma State University, Stillwater, Oklahoma 74075, USA.}
\author{Mario F. Borunda}%
  \affiliation{Physics Department, Oklahoma State University, Stillwater, Oklahoma 74075, USA.}
    \email{mario.borunda@okstate.edu}


\date{\today}

\begin{abstract}
Understanding the behavior of materials under irradiation is crucial for the design and safety of nuclear reactors, spacecraft, and other radiation environments. 
The threshold displacement energy ($E_d$) is a critical parameter for understanding radiation damage in materials, yet its determination often relies on costly experiments or simulations. 
This work leverages the machine learning-based Sure Independence Screening and Sparsifying Operator (SISSO) method to derive accurate, analytical models for predicting $E_d$ using fundamental material properties. 
The models outperform traditional approaches for monoatomic materials, capturing key trends with high accuracy. 
While predictions for polyatomic materials highlight challenges due to dataset complexity, they reveal opportunities for improvement with expanded data. 
This study identifies cohesive energy and melting temperature as key factors influencing $E_d$, offering a robust framework for efficient, data-driven predictions of radiation damage in diverse materials.
\end{abstract}

\maketitle

\section{Introduction}

Grasping the mechanisms of radiation damage is essential across a multitude of fields, including nuclear energy production \cite{dacus2019calculation,chen2021uncertainty,bany2023atomistic}, medical advancements \cite{fang2022integration,lu2021high}, aerospace engineering \cite{de2020applied}, and space exploration \cite{benton2001space, miyazawa2018tolerance,kirmani2022countdown,kirmani2024unraveling,lang2020proton}.
By effectively addressing and minimizing radiation-induced damage, we can drive technological progress, safeguard personnel and the environment, and boldly venture into  space environments where radiation poses significant challenges.

Calculating displacements per atom (dpa) \cite{nordlund2015primary} or vacancies per ion is a common way to quantify the radiation damage induced by different sources \cite{nordlund2018improving,ziegler2010srim}.
Models such as Norgett-Robinson-Torrens (NRT) \cite{norgett1975proposed} and Kinchin-Pease (KP) \cite{kinchin1955displacement} are widely used in binary collision approximation software such as Stopping and Range of Ions in Matter (SRIM) \cite{ziegler2008srim} and Marlowe \cite{robinson1990temporal}. 
These models depend on the threshold displacement energy ($E_d$) of the constituent atoms in the target material. 
In their crudest form, both models predict that the number of displaced atoms resulting from a collision with ions ($\nu$) is inversely proportional to $E_d$ \cite{was2016fundamentals,stoller2013use}.
Therefore, estimations of dpa rely on previously known $E_d$ values, reducing the accuracy of the predicted dpa for novel materials \cite{durant2021radiation}.
The definition of $E_d$ varies slightly according to different measuring methods. 
Still, it can generally be interpreted as the minimum kinetic energy a primary knock-on atom (PKA) needs to produce at least a stable defect on its sublattice \cite{was2016fundamentals,seitz1949disordering}.

The $E_d$ is a very intricate property of a material that can change with temperature \cite{nordlund2015primary,zag1983temperature,saile1985temperature,chen2019atomistic,beeler2016effect}, strain \cite{banisalman2019atomistic,wang2016effect,beeler2016effect,tikhonchev2017threshold}, pressure \cite{zhao2012influence}, and with the direction of the PKA.
There remains some ambiguity in both experimental and simulated values of $E_d$ for most materials \cite{uglov2015physical}.
Most models use an average $E_d$ to predict an approximated dpa \cite{nordlund2015primary}.
In this work, we use the average $E_d$, which will be referred to simply as $E_d$.
Commonly, the $E_d$ is obtained through either experiments or numerical simulations. 
Experimental approaches to measuring $E_d$ are challenging, especially for polyatomic materials, given the equipment requirements and the dependence of $E_d$ on multiple sub-lattices and directions \cite{zinkle1997defect,nordlund2018primary,ghoniem1988binary,gibson1960dynamics}. 
Experimental methods for measuring $E_d$ include changes in resistivity measurements (RM) \cite{corbett1957electron,arnold1960threshold,sosin1962energy,lucasson1962production}, high-voltage electron microscopy (HVEM) \cite{makin1968electron,sharp1973electron,kenik1975orientation,urban1974radiation,yoshiie1979orientation}, time-resolved cathodoluminescence spectroscopy (TRCS) \cite{caulfield1995point,smith2000measured}, and other techniques \cite{smith2005displacement,cooper2001optical,arnold1960threshold}.
These measurement methods are complex and require specialized equipment and resources that are not always readily available.
Similarly, computational methods to calculate the $E_d$ such as molecular dynamics (MD) \cite{torrens1966ionic,devanathan2000displacement,bany2023atomistic,gibson1960dynamics} and ab initio molecular dynamics (AIMD) \cite{lucas2005ab,liu2015ab,yang2021full} tend to be complicated and time consuming since they require a detailed crystal and electronic structure of the material, and sampling several PKAs along several directions to obtain an average $E_d$.

Although the $E_d$ for many materials has been estimated and even compiled for several materials in reports \cite{zinkle1997defect,nordlund2015primary,ASTM_E521,kenik1975orientation,merrill2015threshold}, whenever a prospective material is intended to operate under radiation-hard environments, there will be a need for $E_d$ values for such materials, e.g. variations of halide perovskites \cite{durant2021radiation,kirmani2022countdown,kirmani2024unraveling}. 
There are remarkably successful analytical expressions that can be used to obtain $E_d$.
Still, these require precise experimental measurements, such as finding the threshold electron irradiation energy that would produce defects \cite{zinkle1997defect,hossain1977electron} and measuring displacement probability functions related to resistivity changes \cite{faust1969measurements,iseler1966production}.
There also exists an empirical formula to calculate the effective threshold displacement energy ($E_d^{\text{eff}}$) \cite{ghoniem1988binary}, but it requires individual $E_d$ values of the constituent atoms of the compound.
Additionally, this $E_d^{\text{eff}}$, primarily focuses on stoichiometry and does not explicitly capture the unique atomic interactions specific to each material.
There have been attempts to generalize the estimation of the average $E_d$ by finding its dependency on the fundamental properties of the material's constituent elements in the form of an analytical expression \cite{uglov2015physical,konobeyev2017evaluation}. But ultimately, there is no conclusive analytical expression for the $E_d$ based on general and fundamental properties of materials.

Due to the limited general analytical options to estimate $E_d$, in this work, we employ machine learning, specifically the Sure Independence Screening and Sparsifying Operator (SISSO) method \cite{ouyang2018sisso,purcell2023recent}, to develop analytic expressions to predict $E_d$ in materials based on its fundamental properties.
This method has been proven successful in finding analytical expressions to describe complex quantities, such as a tolerance factor for perovskites \cite{bartel2019new}, classification of metal/nonmetal materials \cite{ouyang2018sisso,ouyang2019simultaneous}, developing predictive models for excited-state properties of molecular crystals \cite{liu2022finding}, and for designing new rare-earth compounds \cite{singh2022machine}.

The outline of this manuscript is as follows. 
In Section II, we present details of our calculations within the SISSO formalism. 
In Section III, we present $E_d$ analytical expressions for monoatomic and polyatomic materials and benchmark our models.
Finally, in Section IV, we present our
conclusions.

\section{Numerical Methods}\label{numericalMethods}
The SISSO model is a machine learning approach that combines regression with compressed sensing to identify high-order analytical models \cite{ouyang2018sisso,ouyang2019simultaneous,purcell2023recent}.
It generates an extended feature space by iteratively combining the primary features using linear and nonlinear algebraic operations ($+$, $\times$, $\div$, $\log$, $\exp$, etc.). 
The sparse regression method filters out the relevant features by analyzing their frequency in descriptor generation, thereby identifying the most predictive models.
SISSO conducts a computer experiment by systematically generating and testing different hypothetical functions against reference data.
SISSO works well with relatively small data sets, and its ability to produce predictive models has been demonstrated using as little as a few hundred \cite{bartel2019new,cao2020artificial}, or even a few tens \cite{foppa2021materials} of training data points.

The dimension (d) of the SISSO model determines the number of relevant features selected from the primary and constructed feature space for use in the final predictive model.
This results in a model that will be composed of different functions ($f_i$) of different parameters ($x_{n_i}$) of the form
\begin{equation}
    \text{model} = \sum_{i=0}^\text{d} c_i f_i(x_0,\dots,x_{n_i}),
\end{equation}
where $c_i$ are coefficients associated with each term that adjusts units between them. 
We perform SISSO calculations with a model dimension of up to two to maintain practicality and simplicity in the models.

We collect $E_d$ values and other material properties to use as primary features for SISSO.
To the best of our knowledge, we compile as many $E_d$ values as are available in the literature.
More details on the datasets are available in Appendix A.
We primarily divide our datasets into two parts: the monoatomic dataset and the polyatomic dataset.

The monoatomic dataset contains 33 different single-atom elements.
This dataset includes the $E_d$ values and fundamental properties of each material such as atomic number ($Z$), atomic mass ($m$), atomic radius ($r$), average bond length ($\ell_{\text{bond}}$), coordination number ($C$), Young's modulus ($E$), density ($\rho$), melting temperature ($T_{\text{melt}}$), thermal coefficient of expansion ($\alpha_T$), cohesive energy ($E_{\text{coh}}$), first ionization energy ($E_{\text{ioniz}}$), and resistivity ($\mathcal{R}$).
We select these particular material properties as features because they correlate to $E_d$
\cite{uglov2015physical,konobeyev2017evaluation}.
The $E_d$ values of each material may come from different sources using various methods, each of which may involve nuanced interpretations of $E_d$ \cite{nordlund2006molecular}.
For example, some methods tend to report the minimum $E_d$, while others provide the average $E_d$ \cite{konobeyev2017evaluation}.
Hence, to mitigate possible effects related to this, we further subdivide the monoatomic dataset by the method used to obtain $E_d$.
Each method is labeled as $M_i$ as follows. $M_1$ encompasses experiments that directly observe the defect generation on the material locally, the data related to it comes mostly from HVEM measurements but also contains transmission electron microscopy \cite{steeds2011orientation}, and Auger electron spectroscopy \cite{steffen1992displacement} data. $M_2$ encompasses experiments that measure $E_d$ indirectly by measuring another property of the material, the data related to it comes mostly from RM, but also deep-level transient spectroscopy \cite{poulin1980threshold}, and annealing studies \cite{vajda1977problem}.
$M_3$ includes MD simulations. Lastly, $M_4$ refers to AIMD simulations.
For $E_d$ values from different sources using the same method, we average the values between the sources.
Additionally, we also implement the multi-task framework \cite{ouyang2019simultaneous} of SISSO by assigning each method as a task. This provides us with a single analytical expression to predict $E_d$ for all the different methods, where the values of the coefficients now adapt to the method.
We label this dataset as $M_0$.

The polyatomic dataset contains materials such as alloys, ceramics, and ionic compounds.
Based on the relevance to $E_d$ and availability in literature, we consider the following material and PKA properties: atomic number ($Z_{\text{PKA}}$), coordination number ($C_{\text{PKA}}$), stoichiometry ($S_{\text{PKA}}$), bond length ($\ell_{\text{PKA}}$), mass ($m_{\text{PKA}}$), atomic radius ($r_{\text{PKA}}$), and the density of the material ($\rho$).
For polyatomic, we subdivide the dataset into two main methods: experimental and simulation. 
We also subdivide the dataset by material type into alloys, ceramics, and non-ceramic semiconductors.
The multi-task framework is used separately for the subdivisions (method and material type).

To benchmark our results, we also calculate $E_d^{\text{eff}}$ for polyatomic materials, which requires the $E_d$ values of the constituent atoms. We do this by using the $E_d$ values of the constituent atoms obtained from our models and comparing them against those obtained from the reported experimental and simulated $E_d$ values.

\section{Results}\label{results}

\subsection{Threshold Displacement Energy in Monoatomic Materials}\label{mono}

One of the most notable models for expressing $E_d$ as a function of fundamental material properties is the one proposed by Konobeyev et al. \cite{konobeyev2017evaluation}, which depends on $E_{\text{coh}}$ and $T_{\text{melt}}$. To test the performance of SISSO, we start by implementing these two properties as features for the samples reported in Konobeyev et al. Our results show that the model obtained from SISSO, labeled as $K_{\text{SR}}$, has a slightly higher $R^2$ compared to Konobeyev et al.'s model, labeled $K$, as shown in Table~\ref{r_squared_table}.
Next, we test the same samples with an expanded feature space, incorporating all features mentioned in the previous section.
This yields the model labeled $K_{\text{S}}$, which achieves an even higher $R^2$ than the previous two models.
We observe that $E_{\text{coh}}$ is present in all three models, while model $K_{\text{S}}$ also includes $C$ and $Z$, resulting in a more accurate model.

Next, we work with the collected dataset, which we divided into subsets $M_i$. In Table~\ref{r_squared_table} we list our results and notice that the $M_4$ subset has the highest $R^2$.
We attribute this to the consistency of this subset, in that, AIMD simulations follow ``ab initio'' equations, which do not rely on empirical force fields.
In contrast, the MD simulation subset, $M_3$, has a significantly lower $R^2$ value.
We attribute this to the nature of MD simulations, which rely on empirical force fields that vary significantly between reports, resulting in less consistent data.
The two experimental subsets $M_1$ and $M_2$ have similar $R^2$ values but the material properties included in the models differ significantly. This possibly signifies the different material properties used in these different experimental methods to measure the $E_d$ values.
Finally, we report our results for the multi-task framework. 
We find a relatively low $R^2$ value, which we attribute to the fact that $M_0$ includes data from various methods, and different methods yield distinct models.
However, we note that the most common material property among all models is $E_{\text{coh}}$, suggesting that it plays an important role in deriving $E_d$ from an analytical expression.
Regarding the increase in SISSO's dimension, we notice that there is not a significant improvement in the $R^2$ value, with a percentage increase ranging from 1\% for $M_3$ to 8\% for $M_1$ and $M_4$.
All the $E_d$ plots comparing the data and models, along with their respective $R^2$, and details about the $K$ models are available in Appendix B.
\begin{table}[t]
    \centering
    \setlength{\tabcolsep}{4pt}
    \setlength\extrarowheight{9pt}
    \begin{tabular}{c c c c}
    \Xhline{1pt}     
       Label & d & $E_d$ & $R^2$  \\ \hline
               $K$ & $\emptyset$ & $c_0 + a_0 E_{\text{coh}} T_{\text{melt}}$ & 0.87 \\
        $K_{\text{SR}}$ & 1 & $c_0 + a_0E_{\text{coh}}T_{\text{melt}}^{\frac{4}{3}}$ & 0.89 \\       
        $K_\text{S}$ & 1 & $c_0 + a_0E_{\text{coh} }\left(\frac{1}{C} -\frac{1}{Z}\right)$ & 0.95 \\
        $M_0$ & 1 & $c_0 + a_0 \frac{E_{\text{coh}}\rho}{\sqrt{m}}$ & 0.74 \\
        $M_0$ & 2 & $c_0 + a_0 \frac{rE_{\text{coh}}\ell_{\text{bond}}}{T_{\text{melt}}} + a_1 \frac{\rho E_{\text{coh}}}{C+Z}$ & 0.78 
        \\
        $M_1$ & 1 & $c_0 + a_0 \alpha_T \ell_{\text{bond}} \sqrt{T_{\text{melt}}}$  & 0.85  \\
        $M_1$ & 2 & $c_0 + a_0 Z^2E_{\text{coh}}^3 + a_1 \alpha_T \ell_{\text{bond}} \sqrt[3]{T_{\text{melt}}}$ & 0.92 \\
        $M_2$ & 1 & $c_0 + a_0 E_{\text{coh}} E_{\text{ioniz}} \sqrt[3]{\rho}$ & 0.82 \\
        $M_2$ & 2 & $c_0 + a_0\frac{E_{\text{coh}}}{E\sqrt[3]{\mathcal{R}}} + a_1 E_{\text{coh}} E_{\text{ioniz}} \sqrt[3]{\rho}$ & 0.86 \\
        $M_3$ & 1 & $c_0 + a_0 \frac{\rho^2}{mC}$ & 0.75 \\
        $M_3$ & 2 & $c_0 + a_0 \frac{T_{\text{melt}}}{E\mathcal{R}C} + a_1 \frac{\rho}{\ell^3_{\text{bond}}}$ & 0.76  \\
        $M_4$ & 1 & $c_0 + a_0 \frac{T_{\text{melt}\sqrt[3]{E}}}{r}$ & 0.91 \\
        $M_4$  & 2 & $c_0 + a_0\frac{\rho E}{E_{\text{coh}}^3} + a_1 \frac{T_{\text{melt}\sqrt[3]{E}}}{r}$ & 0.98 \\
        \vspace{-3ex}  
        \\
     \Xhline{1pt}         
    \end{tabular}
    \caption{Average threshold displacement energy ($E_d$) models with corresponding dataset subdivisions ($M_i$, $K$, $K_{\text{SR}}$, $K_{\text{S}}$), SISSO dimension (d), and coefficient of determination ($R^2$) for each dataset. The coefficients $a_i$ and $c_0$ are unique for each model and are reported in Table~\ref{constants_table}, Appendix B. 
    }
    \label{r_squared_table}
\end{table}
\begin{figure*}[t]
    \includegraphics[width = 17cm]{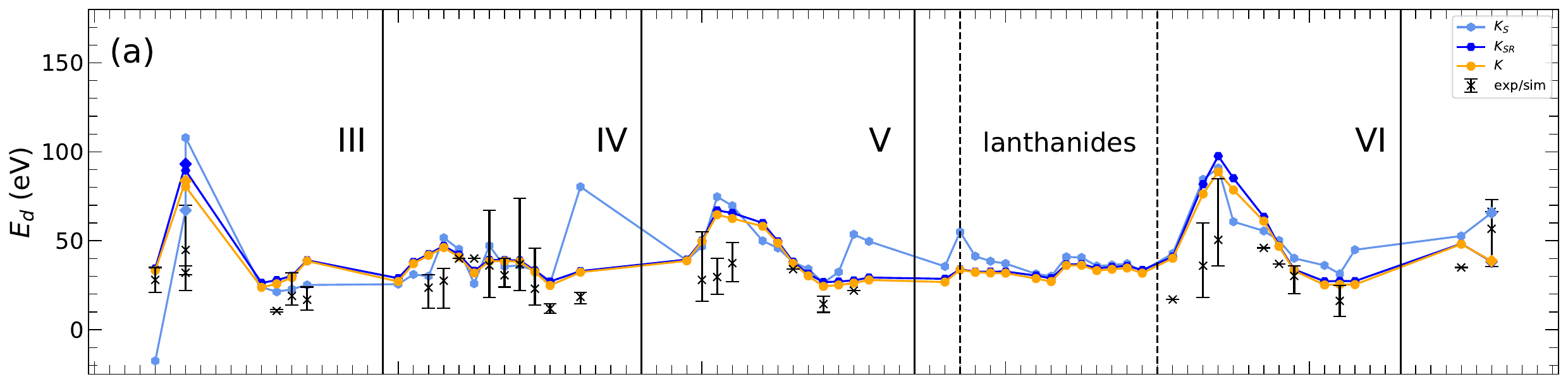}
    \includegraphics[width = 17cm]{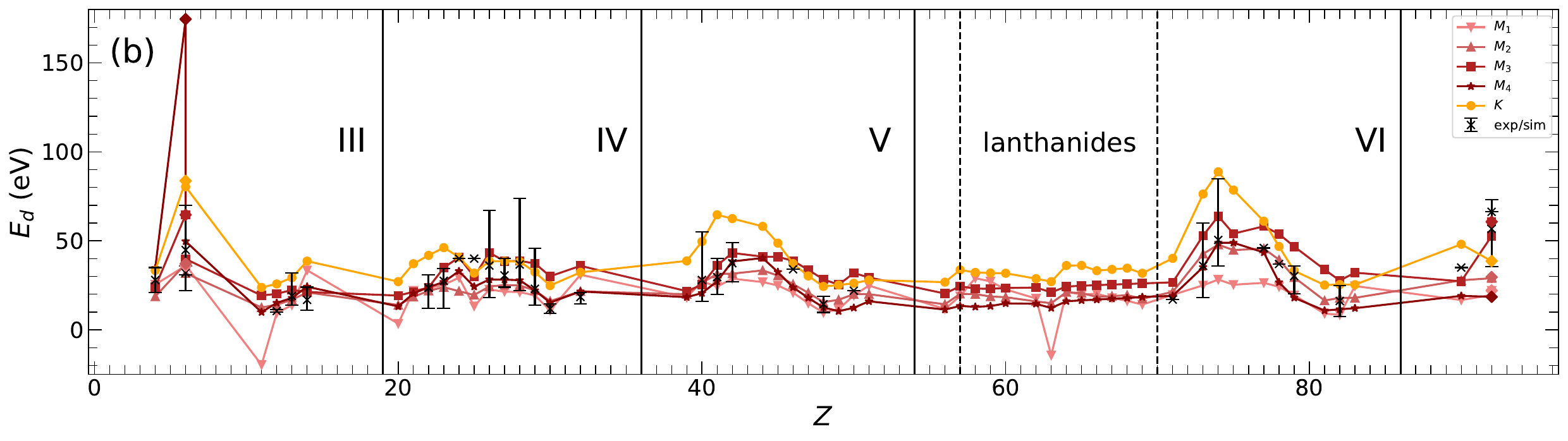}
    \caption{Predicted (solid lines through markers are guides to the eyes) and measured from experiments/simulations (cross markers) average threshold displacement energy ($E_d$) values as a function of the atomic number ($Z$), corresponding to the predictive functions in Table~\ref{r_squared_table}.
    Panel (a) shows the prediction for models $K$, $K_{\text{SR}}$ and $K_{\text{S}}$, while panel (b) displays the prediction of the $M_i$ models, with the $K$ model included as a visual reference. 
    The error bars on the cross markers represent the highest and lowest $E_d$ values from reports in our database, while the cross itself marks the average $E_d$ across different reports.
    The diamond-shaped marker represents different materials that share the same $Z$, specifically, graphite and diamond (diamond marker), as well as gamma uranium and alpha uranium (diamond marker).
    Vertical lines serve as references to denote each row in the periodic table, labeled with Roman numerals.
    The dashed lines serve as references to denote the lanthanides section in the periodic table. 
    }
    \label{Monoatom_Ed_Z_Figure}
\end{figure*}

The $R^2$ value indicates how well the model fits its corresponding dataset, and as such, it is significantly influenced by the consistency of each dataset.
A larger and less consistent dataset does not necessarily result in a less accurate model, although it may yield a lower $R^2$.
Therefore, to better assess their performance, we plot the predicted $E_d$ values of each model alongside the measured experimental or simulated (exp/sim) data against $Z$ in Fig.~\ref{Monoatom_Ed_Z_Figure}. We notice a strong correlation between all models, as well as between the models and the exp/sim data.
Interestingly, for all the models, we observe a pattern along $Z$ that corresponds to the change in rows of the periodic table. 
This is reminiscent of the behavior observed across periodic table rows in material properties, such as first ionization energy or atomic radius \cite{dekock2012atomic}, as also shown in Appendix D, Fig.~\ref{Z_features_figure}.
To help visualize the characteristic ``bump'' in $E_d$ values corresponding to rows III, IV, V, and VI of the periodic table, we include vertical lines in Fig. \ref{Monoatom_Ed_Z_Figure} to denote each row.
These bumps coincide with the patterns identified by Uglov et al.\cite{uglov2015physical}, who suggest that it primarily correlates with the system's binding energy and other related features, such as the Young modulus and the melting temperature.
Furthermore, these parameters are correlated with each other through expressions such as Lindemann's relation \cite{lindemann1910fa}, as demonstrated for $T_{\text{melt}}$ in relation to $E_{\text{coh}}$, $E$, and $\alpha_T$ \cite{grimvall1974correlation,li2002empirical}.
This correlation is also evident in Appendix D, Fig.~\ref{Z_features_figure}, where each feature used in this work is plotted against $Z$.
There are similar bumps when plotting $T_{\text{melt}}$, $E_{\text{coh}}$, $\rho$, and $E$ against $Z$.
Our models thus capture the periodic trend of the material properties, causing the predicted $E_d$ to follow a similar trend.
Another part of this trend is that $E_d$ exhibits a linear (with a small slope) behavior for lanthanides.
To better visualize this, we include vertical dashed lines to indicate the section of lanthanides.
The multi-task model $M_0$ shows a similar behavior (see Appendix B, Fig.~\ref{Monoatom_Ed_Z_Figure_M0}).

It is important to note that models $M_1$ and $K_\text{S}$ yield negative $E_d$ values, which are not allowed since $E_d$ must always be a positive quantity.
Therefore, due to the similar performance of all the models, we recommend against adopting these models, and if they are used, they should be applied with extra caution.

We want to point out that some models are able to predict the differences in $E_d$ values among different materials with the same atomic number $Z$. For example, model $M_4$ predicts the difference between diamond (diamond marker) and graphite. This distinction in $E_d$ is accounted for by the $E$ term present in the model, with $E=1050\text{GPa}$ for diamond \cite{IoffeC-diamond}, compared to $E=15.85\text{GPa}$ for graphite \cite{AZoM}.
On the other hand, the difference in $E_d$ between gamma uranium and alpha uranium (diamond marker) is barely noticeable.



Our results show that there is no significant improvement in the correlation coefficient $R^2$ between $\text{d}=1$ and $\text{d}=2$, as can be observed in Fig.~\ref{Monoatom_Ed_Z_Figure_d2} of Appendix B, which shows similar patterns to Fig.~\ref{Monoatom_Ed_Z_Figure}.
The $\text{d}=1$ models encompass the variables $r$, $m$, $\rho$, $E_{\text{ioniz}}$, $E_{\text{coh}}$, $E$, $C$, and $T_{\text{melt}}$.
All these variables relate directly to material characteristics that relate to how easy it is to disassemble or how easy it would be for an interstitial to move.
The performance of all models is similar, as shown in Fig~\ref{Monoatom_Ed_Z_Figure}, and one model can be used over another based on the availability of known parameters.

We have also looked at the effect of temperature $T$ on $E_d$ and minimum $E_d$ which can be found in Appendix C, and G respectively.

\subsection{Effective Threshold Displacement Energy}\label{section_edeff}

An approach to estimating $E_d$ in complex materials composed of more than one atomic species is to use an effective $E_d$, as described in the empirical formula proposed by Ghoniem et al. \cite{ghoniem1988binary}
\begin{equation}
    E_d^{\text{eff}} = \left[\sum_i^n\frac{S_i}{E_{di}}\right]^{-1},
    \label{ghoniem_eq}
\end{equation}
where $n$  is the number of different constituent atomic species, and $S_i$ and $E_{di}$ represent their corresponding stoichiometry, and threshold displacement energy, respectively.
This formula has been used to calculate the $E_d^{\text{eff}}$ in different polyatomic materials \cite{crocombette2016molecular,agarwal2021use,lin2023predicting}.

We use Eq.~\ref{ghoniem_eq} to benchmark our models ($M_i$). We do this by applying $E_{di}$ values obtained from our models into Eq.~\ref{ghoniem_eq} to calculate the model dependent effective threshold displacement energy, $E_{d_\text{model}}^{\text{eff}}$, and comparing this with literature values, $E_{d_\text{sim/exp}}^{\text{eff}}$, as follows:
\begin{equation}
    \Delta E_d^{\text{eff}} = E_{d_\text{model}}^{\text{eff}} - E_{d_\text{sim/exp}}^{\text{eff}}.
\end{equation}

We have subdivided the polyatomic dataset by material type into the following subsets: alloys (Alloys), such as NiCo and TiAl; ceramics (Cer), such as SiC and ZnO; non-ceramic semiconductors (Semi), such as AlAs and ZnTe; and the entire dataset (All), which encompasses all three categories. Our models are benchmarked separately for these subsets. For each subset, we calculate the mean value ($\overline{\Delta E_d^{\text{eff}}}$) and standard deviation ($\sigma$), as shown in Table~\ref{ghoniem_table}. A good agreement between models and the polyatomic data would yield $\overline{\Delta E_d^{\text{eff}}} \sim0$.
\begin{table}[t]
    \centering
        \setlength{\tabcolsep}{8pt}
    \setlength\extrarowheight{9pt}
    \begin{tabular}{c c c c}
    \Xhline{1pt}     
       Model & subset & $\overline{\Delta E_d^{\text{eff}}} (\text{eV})$ &  $\sigma (\text{eV})$  \\ \hline
       $M_1$ & Alloys & -10.1046 & 9.8237 \\
 & Cer & 0.3133 & 12.8527 \\
 & Semi & 6.0621 & 0.9963 \\
 & All & -1.9303 & 12.4272 \\ \hline
$M_2$ & Alloys & -8.1823 & 9.423 \\
 & Cer & 1.9974 & 12.3339 \\
 & Semi & 11.2540 & 1.3940 \\
 & All & 0.2226 & 12.3005 \\\hline
$M_3$ & Alloys & 6.0022 & 13.1725 \\
 & Cer & 7.3397 & 12.3189 \\
 & Semi & 23.5098 & 4.0426 \\ 
 & All & 8.8226 & 12.9513 \\\hline
$M_4$ & Alloys & -20.4649 & 9.3186 \\
 & Cer & -9.9004 & 12.7403 \\
 & Semi & 5.1960 & 1.7131 \\  
 & All & -11.1122 & 13.3251 \\\hline
$K$ & Alloys & 5.3921 & 9.3498 \\
 & Cer & 20.8951 & 14.7099 \\
 & Semi & 20.5491 & 1.1248 \\   
 & All & 16.5349 & 14.2144 \\
        \vspace{-3ex}  
        \\
     \Xhline{1pt}         
    \end{tabular}
    \caption{The mean difference between $E_d^{\text{eff}}$ obtained using the monoatomic models and the corresponding polyatomic experimental and simulation data ($\overline{\Delta E_d^{\text{eff}}}$), along with their respective standard deviation ($\sigma$), calculated for different subsets (Alloys, Cer, Semi, and All).
    Histograms showing the distributions for $\overline{\Delta E_d^{\text{eff}}}$ are available in Appendix E, Fig.~\ref{ghoniem_hists}.
    }
    \label{ghoniem_table}
\end{table}


The models closest to this condition are $M_1$ and $M_2$, especially for the All and Cer subsets.
However, we note that there is no clear difference in using this formula across different material types.
For example, models $M_1$, $M_2$, and $M_4$ yield a smaller $\overline{\Delta E_d^{\text{eff}}}$ for ceramics, whereas model $K$ yields a smaller $\overline{\Delta E_d^{\text{eff}}}$ for alloys.
Additionally, all models exhibit a standard deviation that ranges from 1.13 eV to 14.71 eV. 
The smaller standard deviations are associated with the Semi category, and we assign this to the fact that the data is significantly smaller than that of the other subsets.
It is important to note that the data from the literature used to compare the model results are averages of the experimental and simulated values, which vary across different reports, and also include their own measurement errors.
In some cases, it can vary within a range similar to the standard deviation range, as shown by the error bars in Fig.~\ref{Monoatom_Ed_Z_Figure} (detailed information about the polyatomic simulations and experimental data is provided in Appendix A).

\subsection{Threshold Displacement Energy in Polyatomic Materials}

Polyatomic materials are more complex, and some of the fundamental properties we use for monoatomic materials are either unknown or difficult to calculate and measure for polyatomic compounds.
Table~\ref{r_squared_table_poly} shows the results of applying SISSO, similar to its use with monoatomic materials, but with a reduced feature set, listed in Section~\ref{numericalMethods}. 
Only the models for alloys and non-ceramic semiconductors yield an $R^2$ larger than 0.5.
This result does not necessarily imply that these models have physical accuracy; rather, it may reflect the fact that they were trained on smaller datasets.
Additional data needs to be gathered and tested over time to confirm the effectiveness of these models or at least the correlation with the implied variables.
The $R^2$ plots and coefficients ($c_0$, $a_i$) for all models (including $\text{d} = 1$ and $\text{d} = 2$) are available in Appendix F.
\begin{table}[t]
    \centering
    \setlength{\tabcolsep}{4pt}
    \setlength\extrarowheight{9pt}
    \begin{tabular}{c c c}
    \Xhline{1pt}     
       Label & $E_d$ & $R^2$  \\ \hline
               All & $c_0 + a_0 \frac{r_{\text{PKA}}^2}{\ell_{\text{PKA}}^3}$ & 0.16 \\  
                Exp & $c_0 + a_0 \frac{Z_{\text{PKA}}^3}{S_{\text{PKA}}^2}$ & 0.18 \\    
                Sim & $c_0 + a_0 \frac{r_{\text{PKA}} \rho}{\ell_{\text{PKA}}^3}$ & 0.19 \\    
                Cer & $c_0 + a_0 \frac{\sqrt{C_{\text{PKA}}}}{\ell_{\text{PKA}}}$ & 0.18 \\    
                Alloys & $c_0 + a_0 \frac{\rho^4}{C_{\text{PKA}}}$ & 0.54 \\    
                Semi & $c_0 + a_0 \frac{\rho}{\ell_{\text{PKA}}^4}$ & 0.87 \\    
        \vspace{-3ex}  
        \\
     \Xhline{1pt}         
    \end{tabular}
    \caption{Threshold displacement energy ($E_d$) models,
dataset subdivision labels, and coefficient of determination ($R^2$) for each dataset, all polyatomic (All), experimental (Exp), simulations (Sim), ceramics (Cer), alloys, and non--ceramic semiconductors (Semi).
All models have $\text{d}=1$, and the labels Exp, Sim, and Semi correspond to the experimental, simulation, and semiconductor subsets.
    }
    \label{r_squared_table_poly}
\end{table}

\section{Conclusions}

This work demonstrates the potential of machine learning, specifically the Sure Independence Screening and Sparsifying Operator (SISSO) method, to develop analytical expressions for predicting threshold displacement energy ($E_d$) values in materials. 
The models constructed using SISSO outperform traditional approaches like that used in Konobeyev et al. \cite{konobeyev2017evaluation} in terms of accuracy and generalizability, leveraging fundamental material properties. 
These results provide a valuable tool for estimating $E_d$ in monoatomic materials, reducing the reliance on complex experiments and simulations.

For polyatomic materials, while the SISSO models showed limited success due to the inherent complexity and diversity of the data, incorporating improved datasets and additional features could enhance their predictive capability. 
The effective $E_d$ formulations for complex materials were previously constrained by the availability of known monoatomic $E_d$ values, considering our monoatomic predictions, this method offers a more robust pragmatic path forward.
This study emphasizes the importance of cohesive energy and melting temperature as key contributors to $E_d$. 
These parameters are strongly related to the atomic bonding strength within the solid's structure \cite{vopson2020generalized,singh2018bond}, reaffirming their role in defect formation dynamics.

Future work should focus on expanding datasets, exploring temperature dependence, and refining models to capture the nuanced behavior of polyatomic materials under varying conditions. 
This approach represents a significant step toward efficient, data-driven prediction of radiation damage parameters, facilitating advancements in materials science applications in nuclear energy, aerospace, and other radiation-intense environments.

\section*{Appendix A: Database}

We compile $E_d$ data from 274 publications covering 129 distinct materials and considering a total of 240 different PKAs. 
Among these, 33 comprise different monoatomic materials, sourcing from 170 articles, while 98 encompass various polyatomic materials, drawn from 104 articles. 
From the reports, we collect the material's stoichiometry, structure (if reported), temperature (if reported), type of measurement or simulation, and the average $E_d$ per PKA. 
If the report only provides directional $E_d$s, we list an average of the reported values. 
In cases where the structure is not explicitly specified in the report, we assume it is the most common configuration.

Other atomic features are taken from the same source to be as consistent as possible.
These are: atomic mass \cite{Mathematica_ElementData}, 
average bond length \cite{MaterialsProject}, 
coordination number \cite{MaterialsProject}, 
first ionization energy \cite{Mathematica_ElementData}, 
thermal coefficient of expansion \cite{Mathematica_ElementData},
Young modulus \cite{Mathematica_ElementData} (except for element Z=32 \cite{IoffeGe}), 
densities \cite{Mathematica_ElementData}, 
melting temperature \cite{Mathematica_ElementData} (except for elements Z =2, and 87 \cite{ColoradoPT}), 
atomic radius\cite{Mathematica_ElementData} (except for elements Z = 57,58, and 87-99 \cite{PubChemAtomicRadius}), 
cohesive energy \cite{Kittel2005},
and resistivity \cite{Mathematica_ElementData}. 
For diamond (C1) and graphite (C2) we use: Young modulus C1 \cite{IoffeC-diamond}, C2 (average) \cite{AZoM}, thermal coefficient of expansion C1 \cite{moelle1997measurement}, C2 \cite{tsang2005graphite}, density C1 and C2 \cite{LibreTextsSolidsDensities}, melting temperature C1 and C2 \cite{IAEA_GuideToGraphite}, cohesive energy C1 \cite{chelikowsky1984first}, C2 \cite{weinert1982total}, and resistivity C1 and C2 \cite{pierson2012handbook}.

The entire data, including the monoatomic and polyatomic sets and their corresponding subsets, is available in the Supplementary Information.

\section*{Appendix B: Monoatomic Models}

\subsection*{Parameter Source Consistency for Model Comparison}

Tables~\ref{constants_table}~and~\ref{constants_table_M0} contain the value of the constants of the models presented in the main manuscript. 

To calculate the values of the constants ($c_0$ and $a_0$) for Konobeyev et al.'s model, we extract 20 arbitrary data points ($E_d$, $Z$) from Fig. 4 in their work \cite{konobeyev2017evaluation}.
We calculate the product $E_{\text{coh}}T_{\text{melt}}$ (using the sources provided in the previous section) for the corresponding extracted $Z$ values.
Then, we use the equation $E_d = c_0 + a_0 E_{\text{coh}}T_{\text{melt}}$ along with pairs of points (extracted $E_d$ and the corresponding $E_{\text{coh}}T_{\text{melt}}$ values) to numerically solve the system of two equations for the constants $a_0$ and $c_0$. 
We apply Cramer's rule recursively to the 20 extracted points to calculate the values for these constants.
Then, we calculate the average and standard error of the different values obtained for $a_0$ and $c_0$, yielding
\begin{eqnarray}
a_0 = 0.0020 \pm 0.0004\\
c_0 = 23 \pm 3.
\end{eqnarray}
It is important to note that the source for the parameter values used in Konobeyev et al.'s work is unavailable.
Therefore, we used the same source cited in the previous section for all features, which may not be identical to the one used in Konobeyev et al.'s study.

\begin{table}[t]
    \centering
    \setlength\extrarowheight{7pt}
    \begin{tabular}{c c c c c}
    \Xhline{1pt}     
       Label & d & $c_0$ & $a_0$ & $a_1$  \\ \hline
              K & $\emptyset$ & 23 & $2\cdot 10^{-3}$ & 0
       \\
                    $K_{\text{SR}}$ & 1 & $2.59404323\cdot10$ & $1.40954371\cdot 10^{-4}$ & 0
       \\
                     $K_{\text{S}}$ & 1 & $2.1292043\cdot10$ & $7.0129589\cdot10$ & 0
       \\
       $M_1$ & 1 & $3.6776365 \cdot 10$ & $-1.1343030\cdot 10^4$ & 0 \\
       $M_1$ & 2 & $3.69216879\cdot 10$ & $-3.14650895\cdot 10^{-6}$ & $-3.4070589\cdot10^4$ \\
       $M_2$ & 1 & $1.1476423\cdot 10$ & $1.9108009\cdot 10^{-1}$ & 0 \\
       $M_2$ & 2 & $8.3050788$ & $3.5187813\cdot 10^{-1}$ & $ 1.9370837\cdot10^{-1}$ \\
       $M_3$ & 1 & $1.8361633 \cdot 10$ & $1.8045909\cdot10^2$ & 0 \\
       $M_3$ & 2 & $1.9316861\cdot 10$ & $-2.9222360 \cdot 10^{-7}$ & $5.0949784\cdot10$ \\
       $M_4$ & 1 & $8.7557539$ & $2.8220473 \cdot 10^{-1}$ & 0 \\
       $M_4$ & 2 & $7.4707180$ & $1.6534782\cdot10^{-1}$ & $2.7430039\cdot 10^{-1}$ \\
     \Xhline{1pt}         
    \end{tabular}
    \caption{Constants $c_0$ and $a_i$ for Konobeyev et al. and SISSO models' with different dimension.
    }
    \label{constants_table}
\end{table}

\begin{table}[t]
    \centering
    \setlength\extrarowheight{7pt}
    \begin{tabular}{c c c c c}
    \Xhline{1pt}     
       Subset & d & $c_0$ & $a_0$ & $a_1$  \\ \hline
       $M_{0M_1}$ & 1 & $1.17287101 \cdot 10$ & $2.16679713$ & 0 \\
       $M_{0M_2}$ & 1 & $1.39086478\cdot 10$ & $2.74425657$ & $0$ \\
       $M_{0M_3}$ & 1 & $1.61839823\cdot 10$ & $4.31459697$ & 0 \\
       $M_{0M_4}$ & 1 & $1.47573984\cdot 10$ & $2.53952087$ & 0 \\       
       $M_{0M_1}$ & 2 & $1.39455351 \cdot 10$ & $-2.61729040$ & $1.02656016\cdot10$ \\
       $M_{0M_2}$ & 2 & $7.08008095$ & $5.96650464$ & $1.34859908 \cdot 10$\\
       $M_{0M_3}$ & 2 & $6.04391458$ & $9.66036430$ & $1.97525642 \cdot 10$ \\
       $M_{0M_4}$ & 2 & $2.03812631\cdot10$ & $-7.13057353$ & $1.63169654\cdot10$ 
       \\
     \Xhline{1pt}         
    \end{tabular}
    \caption{Constants $c_0$ and $a_i$ for each subset (assigned as tasks) of the $M_0$ model. 
    The value of d correspond to the dimension of the $M_0$ model.
    }
    \label{constants_table_M0}
\end{table}

\subsection*{Coefficient of determination}

\begin{figure*}
\includegraphics[width=5.7cm]{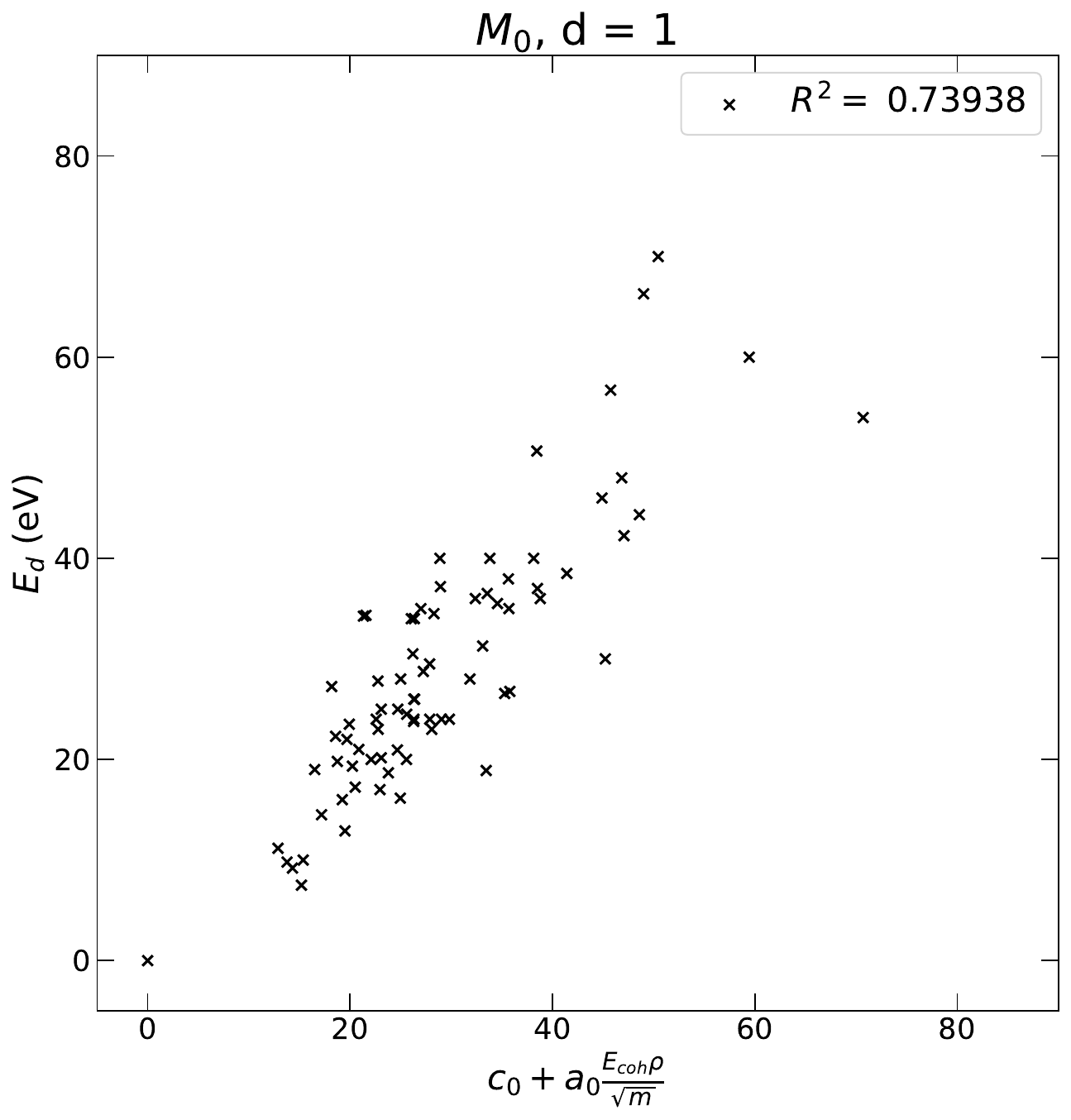}
        \includegraphics[width=5.7cm]{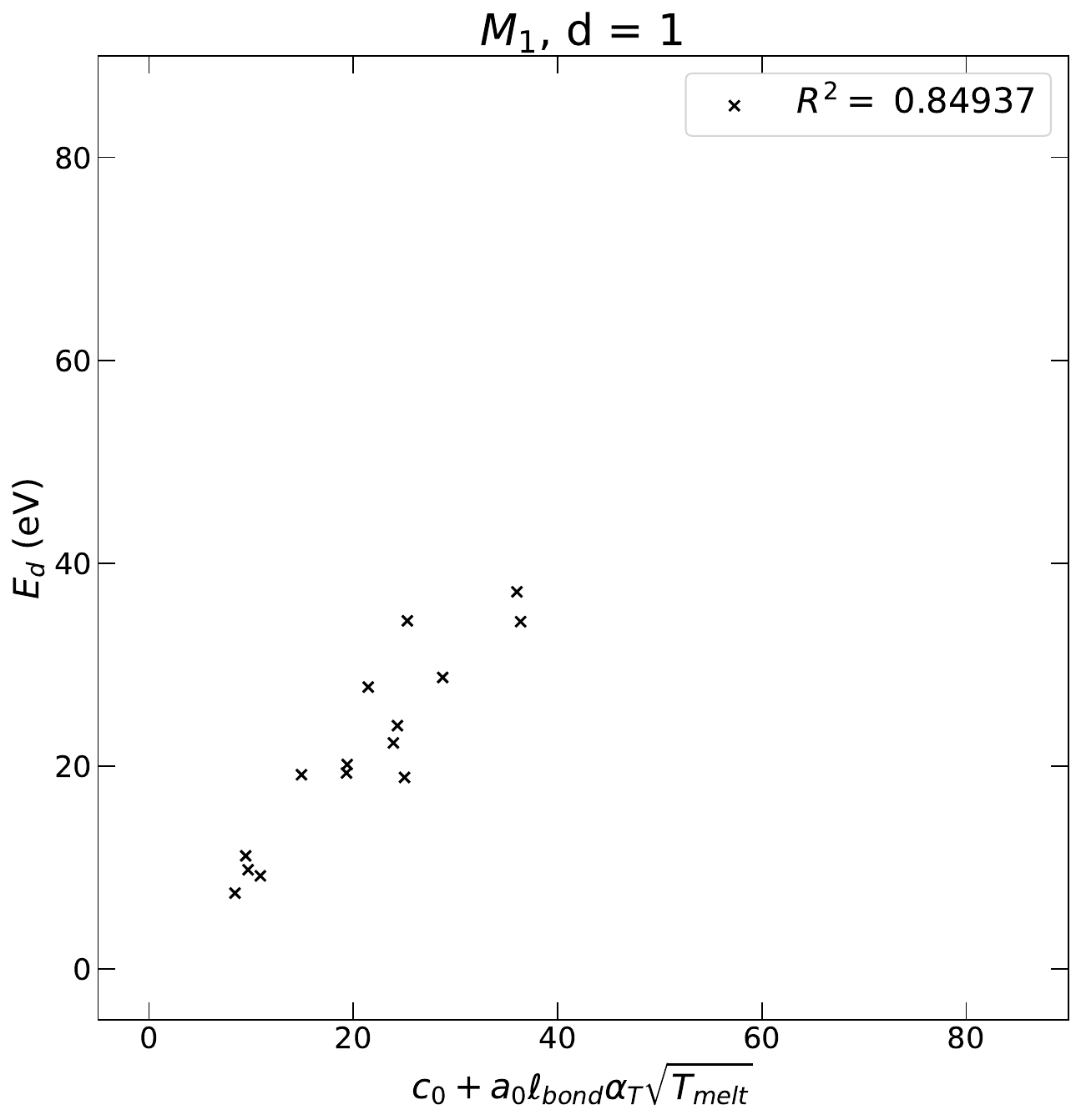}
            \includegraphics[width=5.7cm]{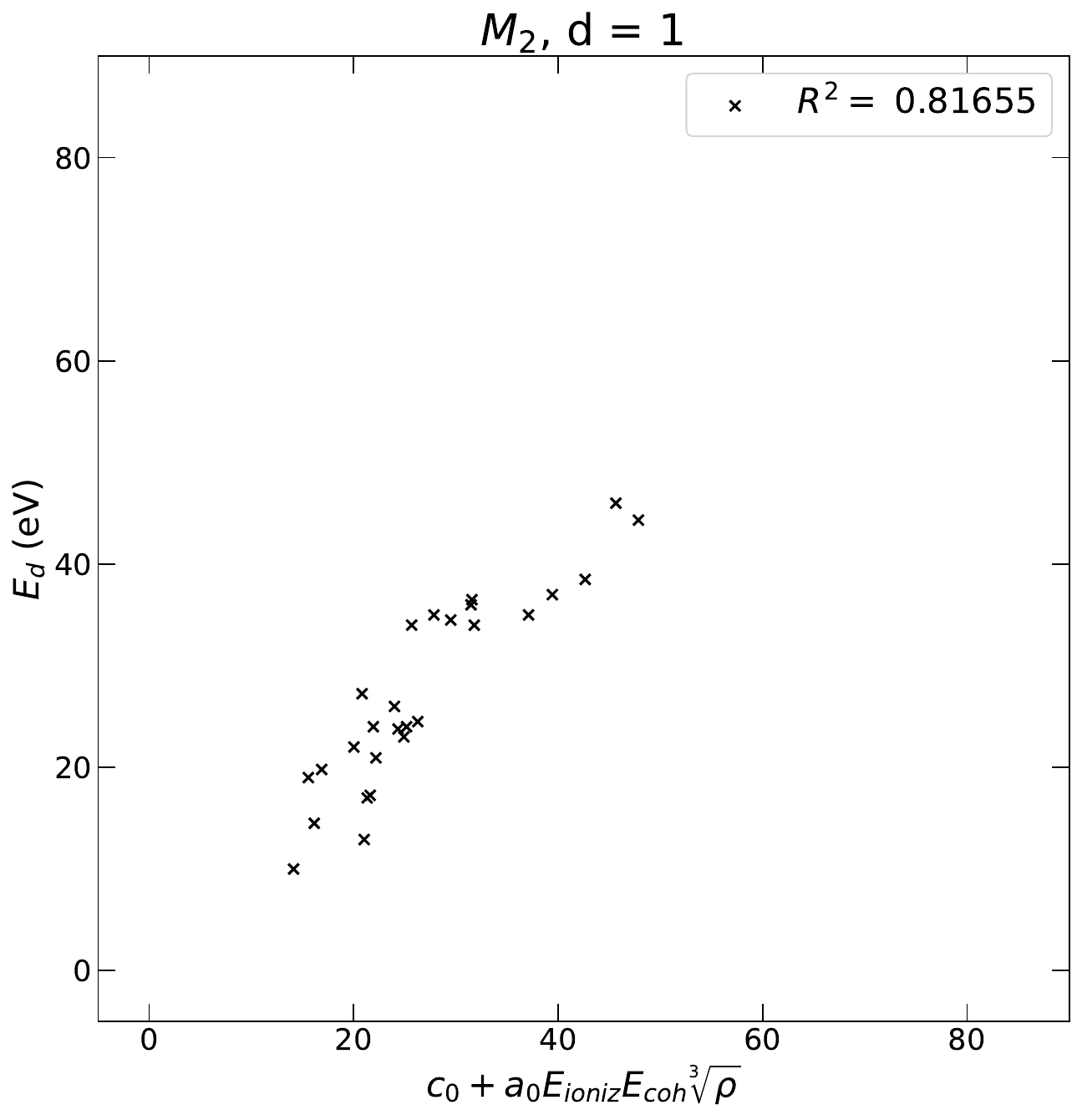}
\includegraphics[width=5.7cm]{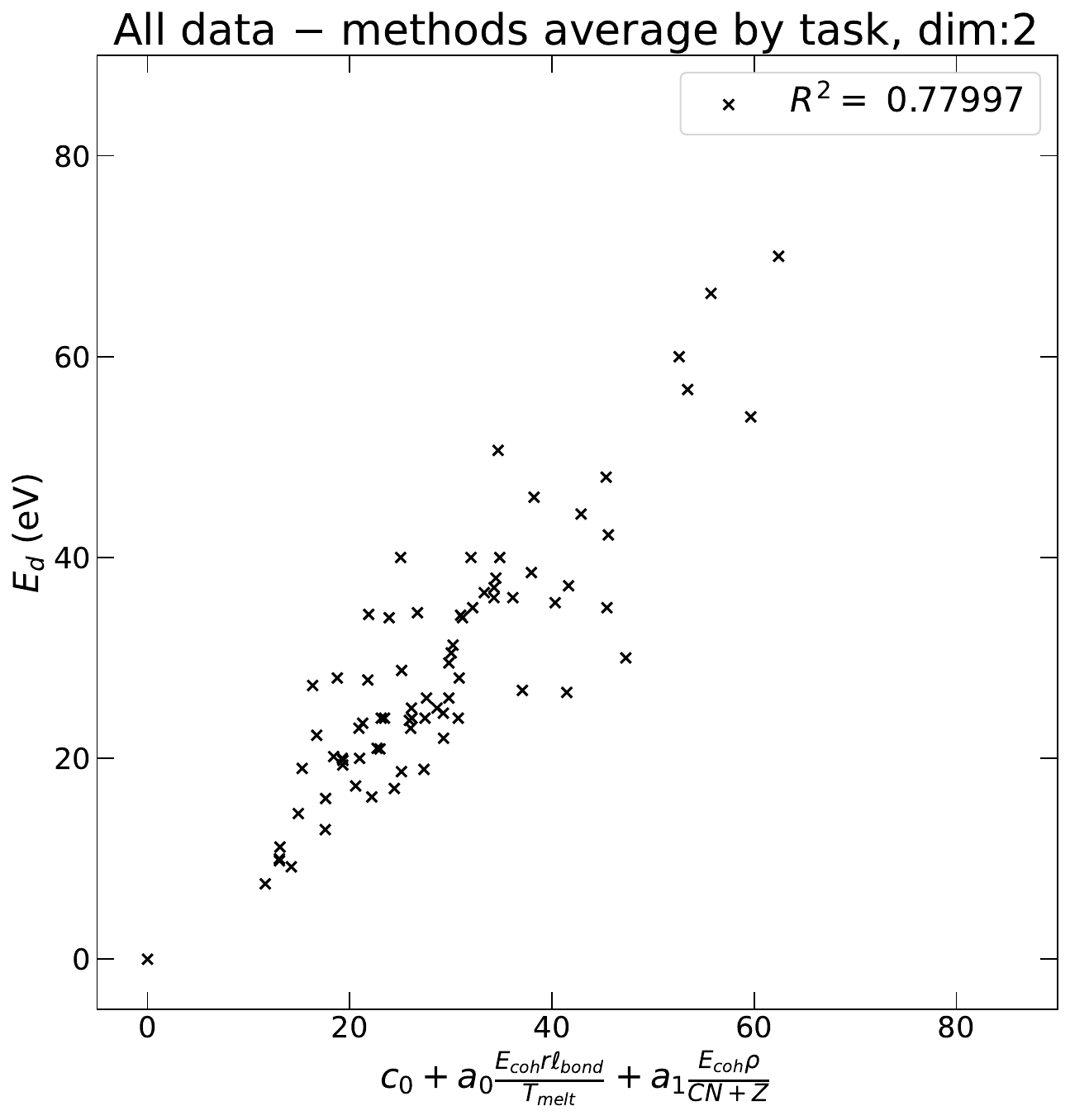}
        \includegraphics[width=5.7cm]{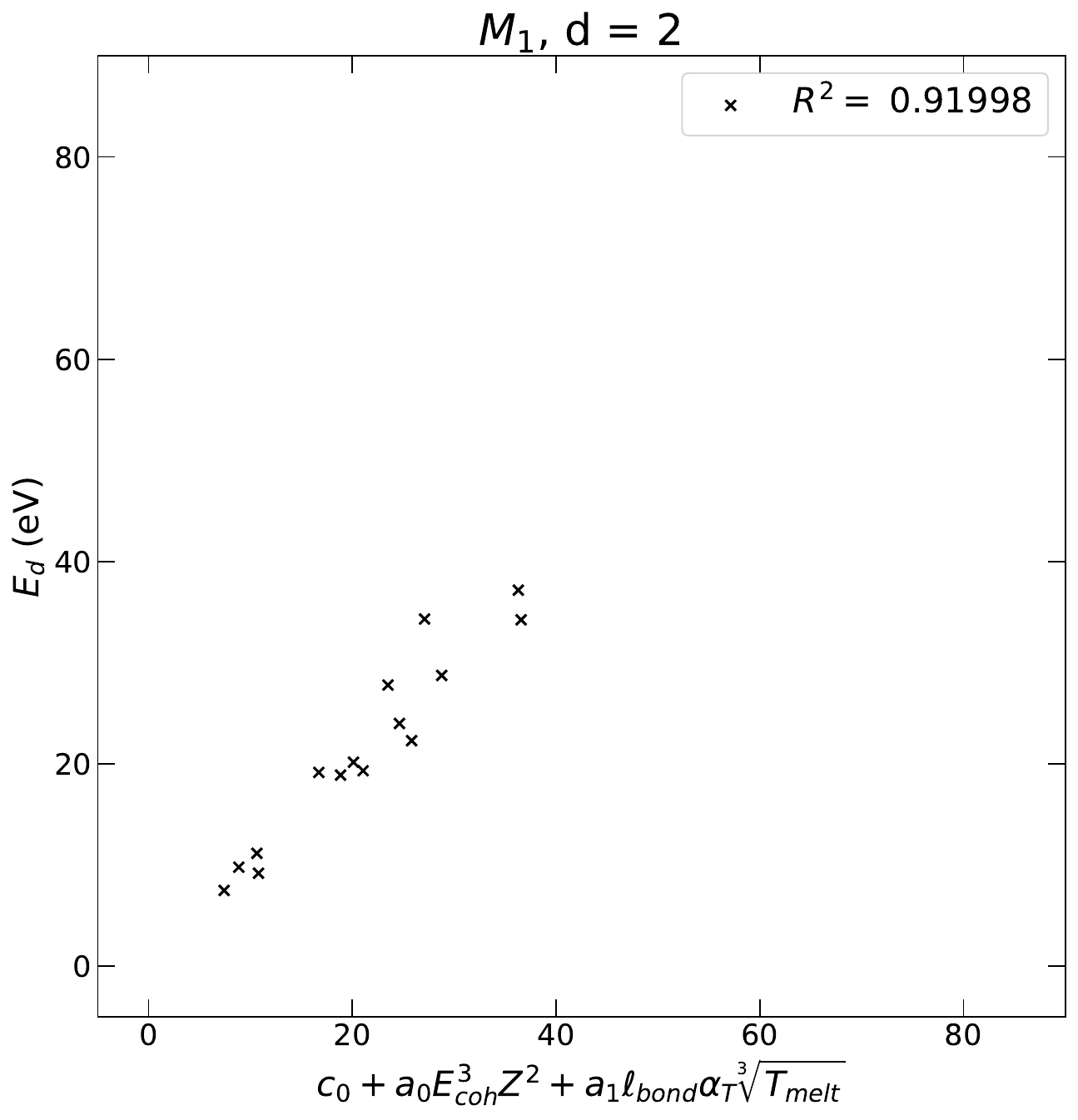}
            \includegraphics[width=5.7cm]{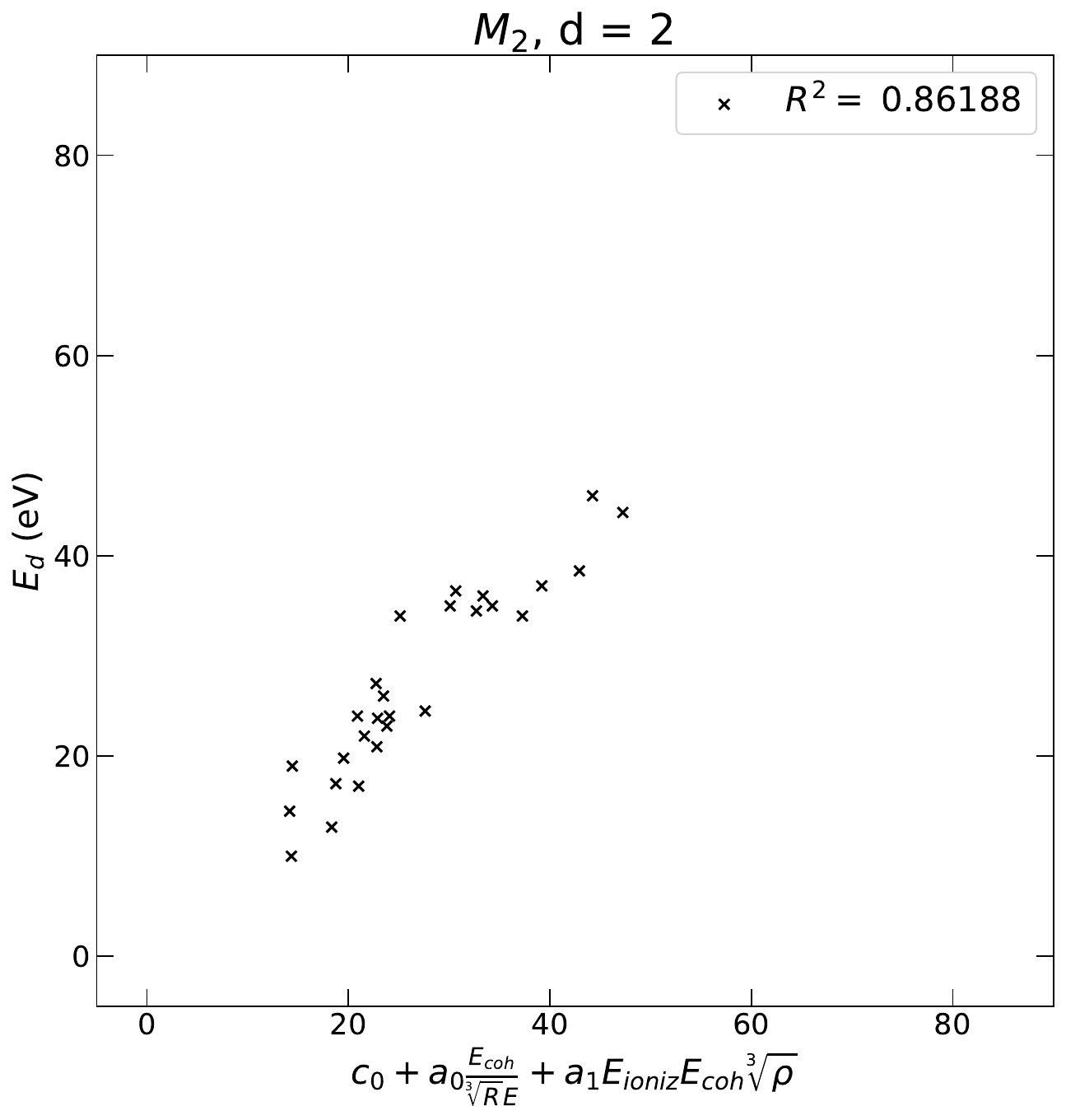}           
\includegraphics[width=5.7cm]{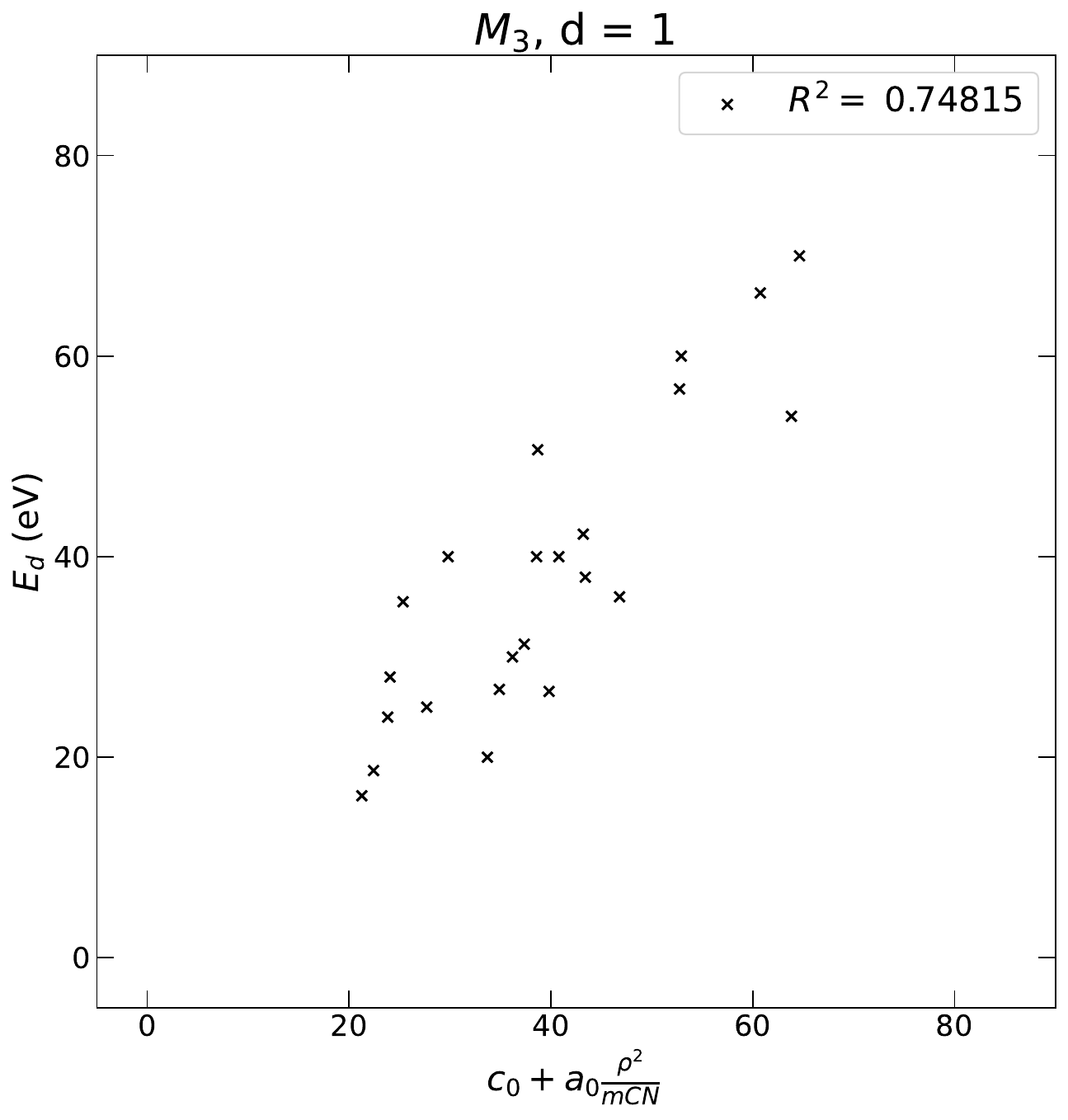}
    \includegraphics[width=5.7cm]{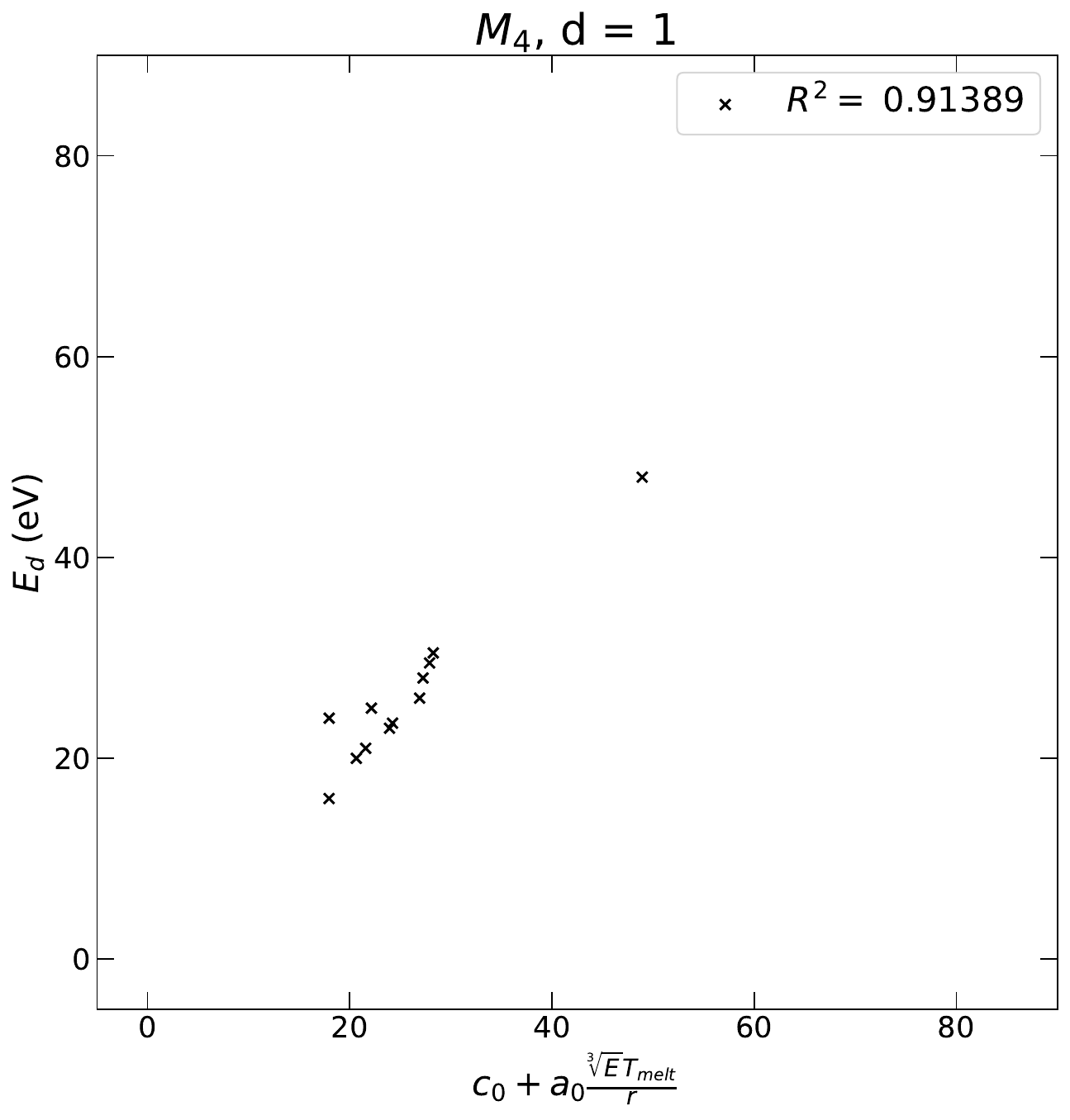}
        \includegraphics[width=5.7cm]{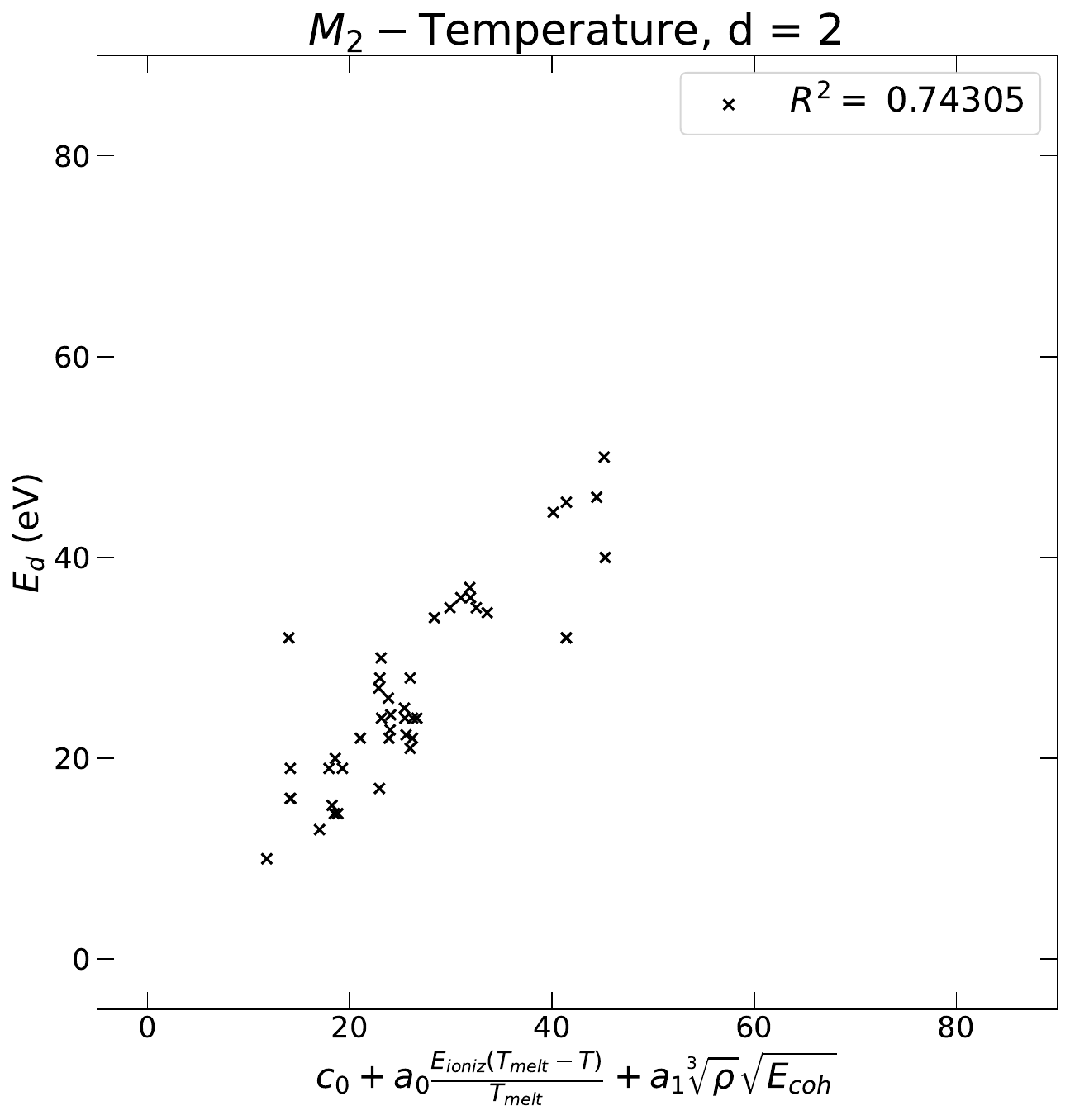}    
\includegraphics[width=5.7cm]{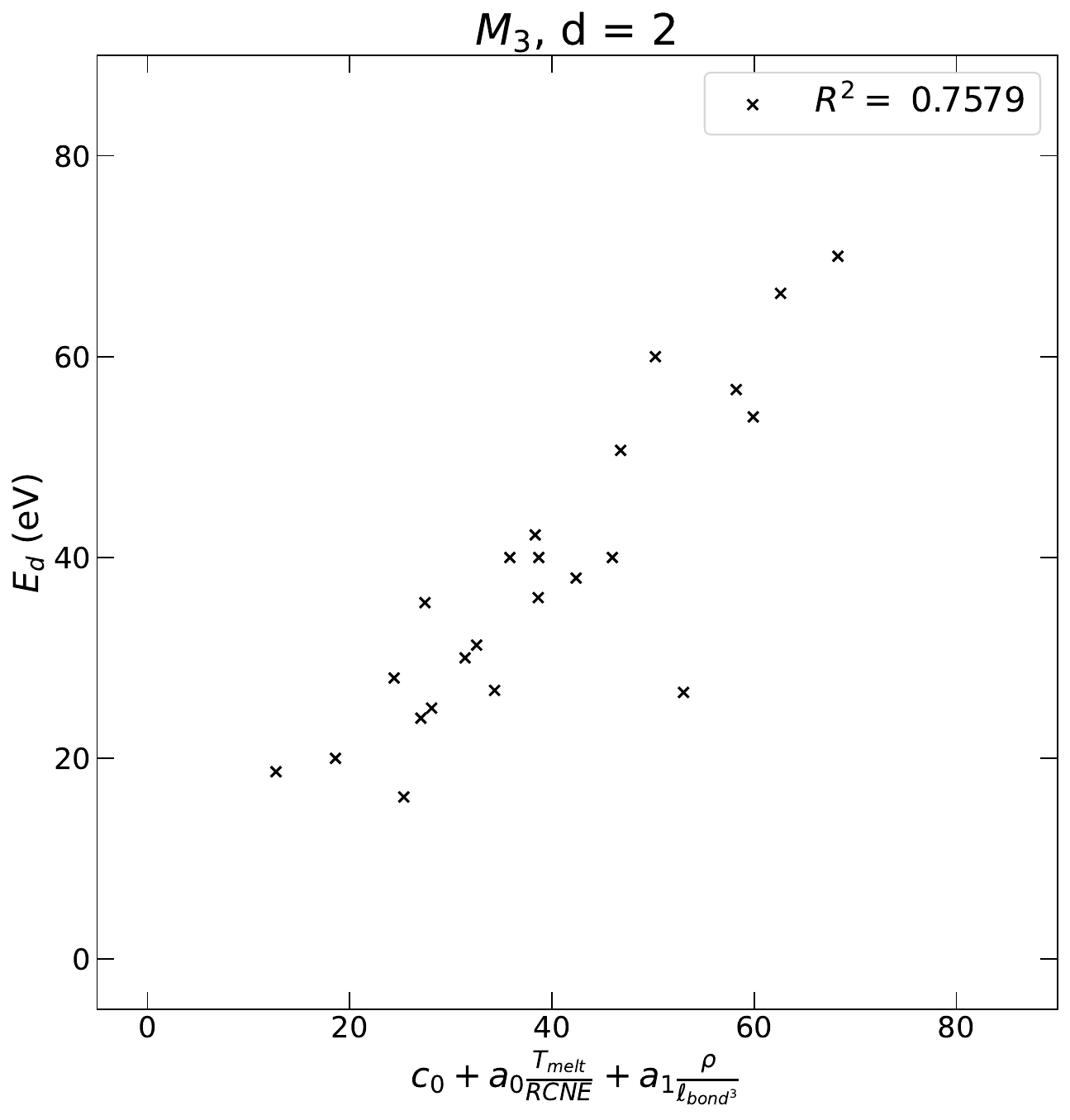}
    \includegraphics[width=5.7cm]{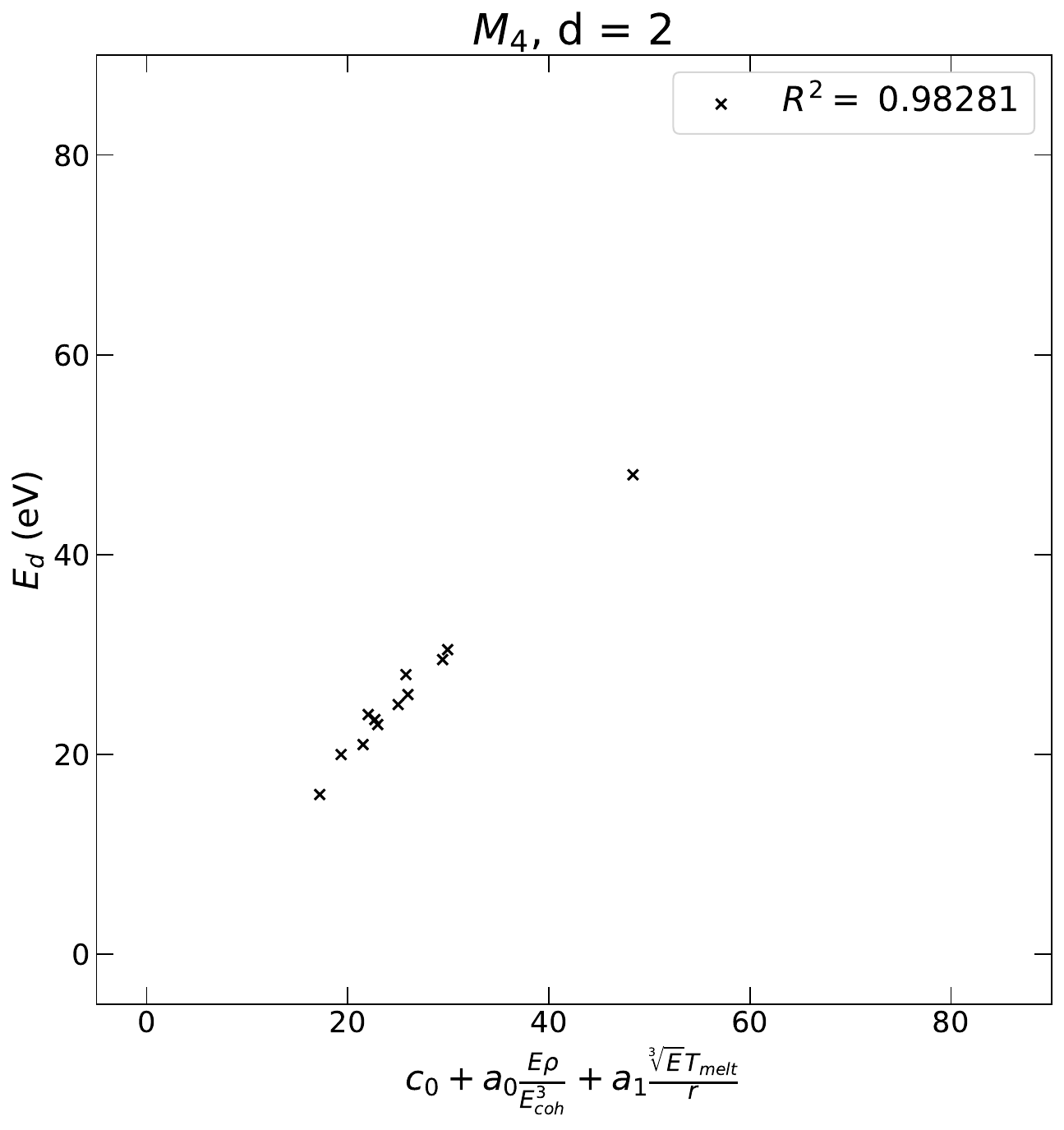} 
        \includegraphics[width=5.7cm]{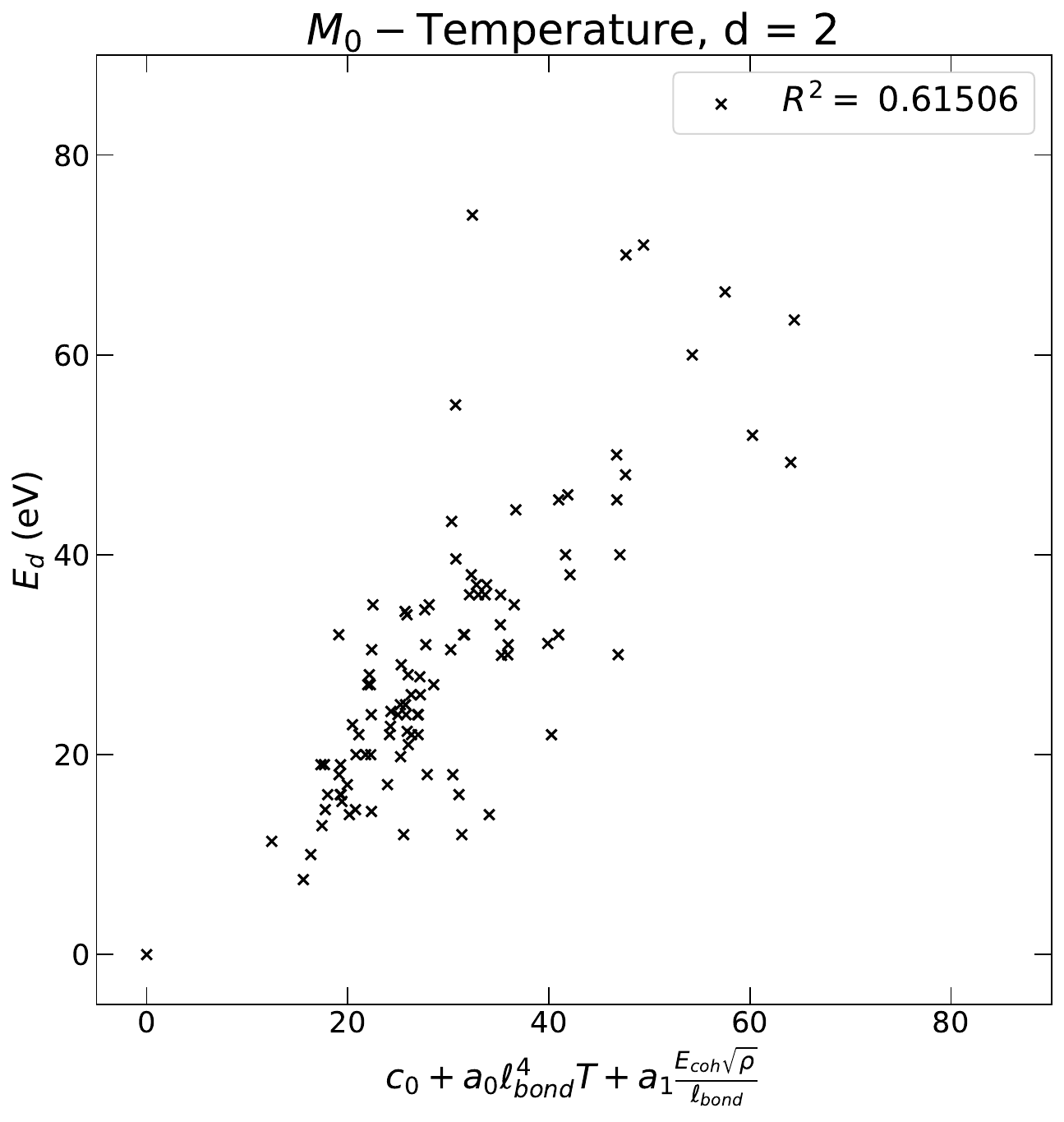}   
    \caption{Plots of $E_d$ data versus $E_d$ SISSO $M_i$ models for each corresponding subset and dimension.}
    \label{R_squared_figure}
\end{figure*}

\begin{figure*}
    \includegraphics[width=5.7cm]{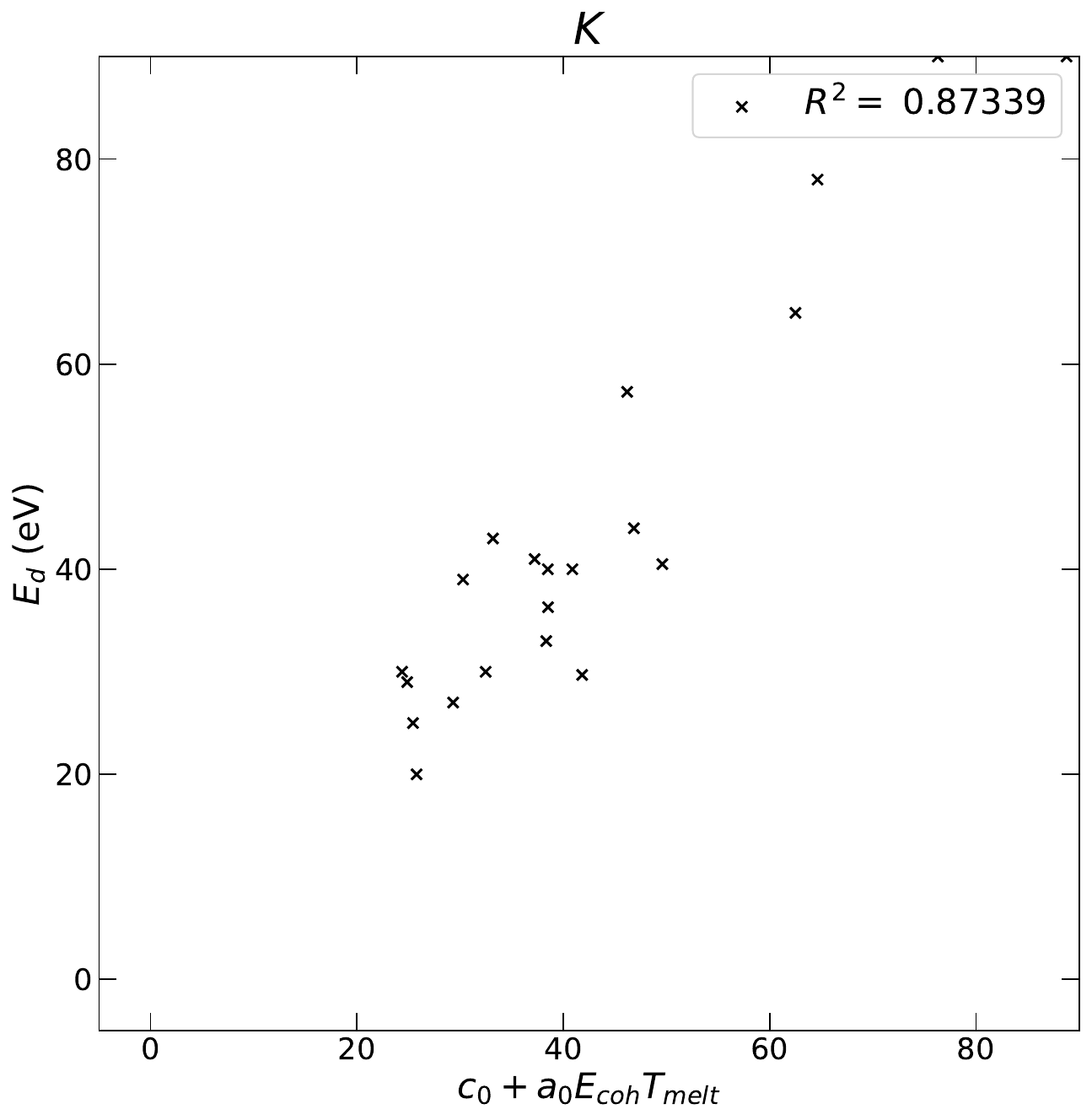}
        \includegraphics[width=5.7cm]{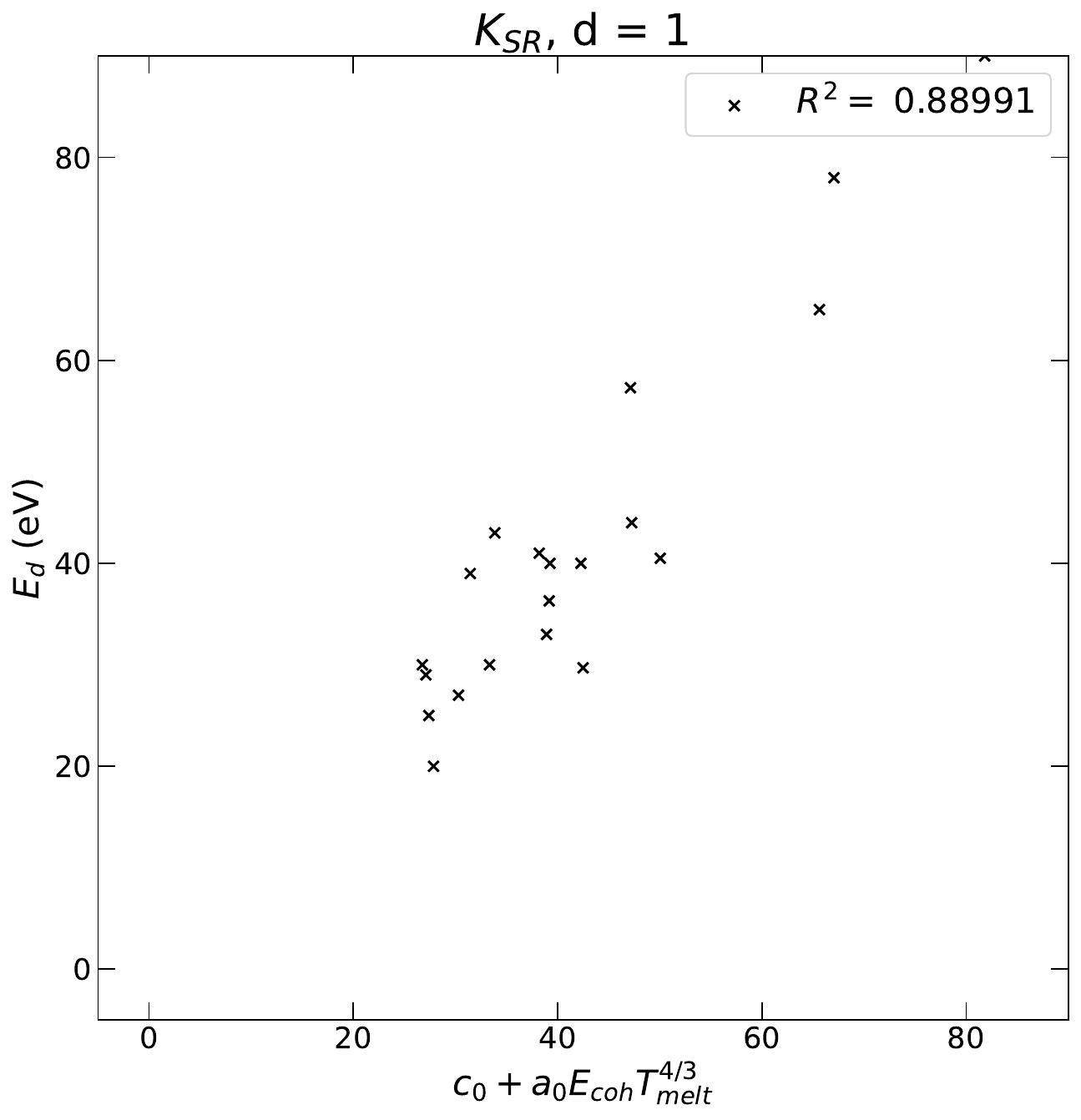}
            \includegraphics[width=5.7cm]{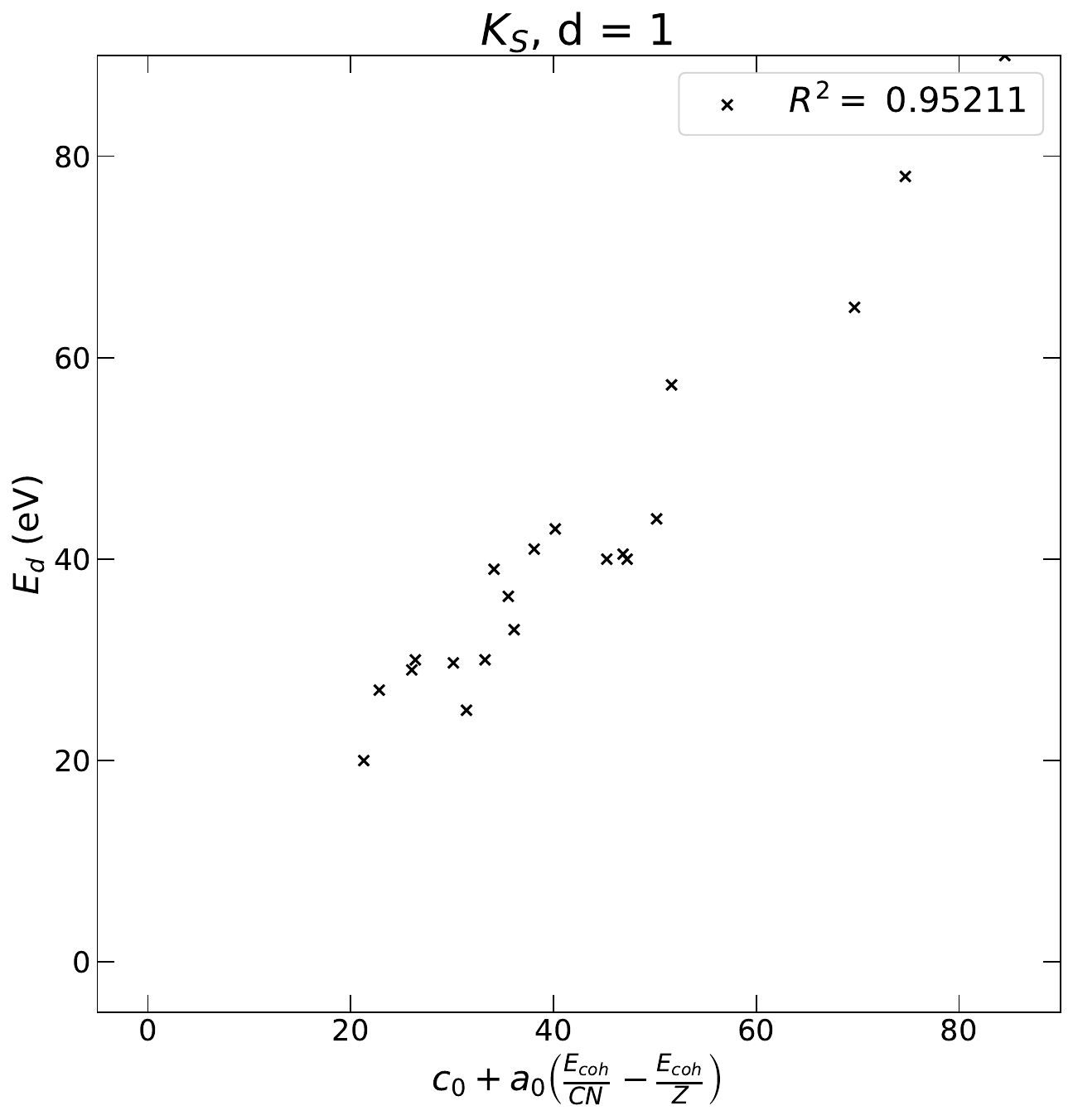}
    \caption{Plots of $E_d$ data versus $E_d$ $K$ models for each corresponding subset and dimension.}
    \label{R_squared_figure_K}
\end{figure*}

Figs.~\ref{R_squared_figure}~and~\ref{R_squared_figure_K} shows all the $R^2$ plots for the different models with SISSO dimension d.
Similarly as before, for Fig.~\ref{R_squared_figure_K} we extract Konobeyev et al.'s experimental $E_d$ data from Fig. 4 in their report.

\subsection*{Model $M_0$}

\begin{figure*}[t]
    \includegraphics[width = 17cm]{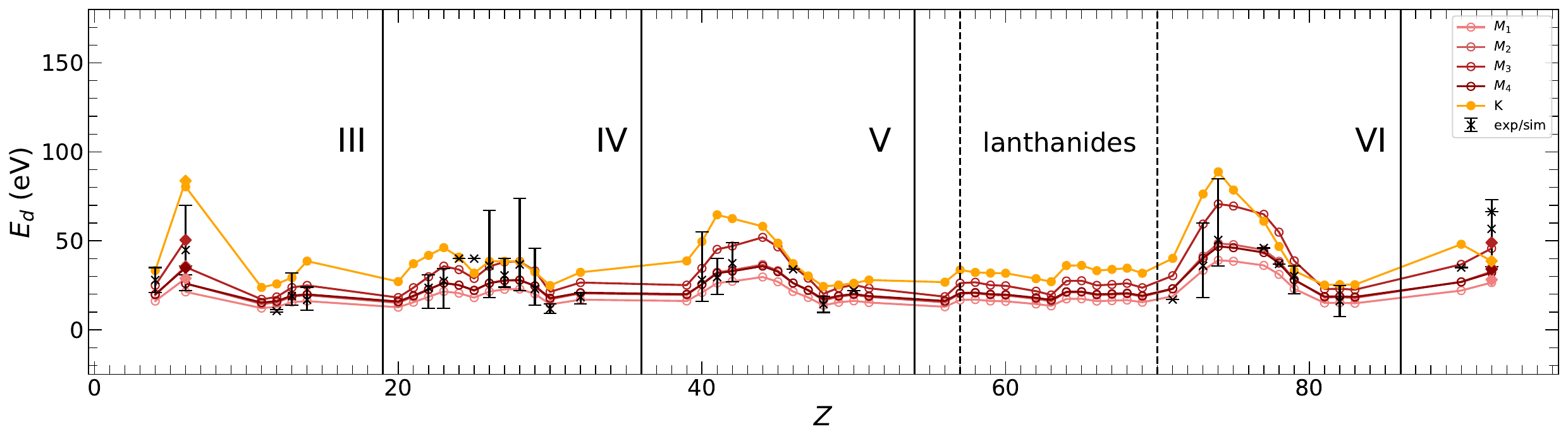}
    \caption{Predicted (connected points) and measured (cross markers) average threshold displacement energy ($E_d$) values as a function of the atomic number ($Z$), corresponding to $M_0$ for the different constant values of each model $M_i$.
    Model K is shown as a reference.
    All specifications for the plot remain consistent with those outlined in Fig.~\ref{Monoatom_Ed_Z_Figure}.
    }
    \label{Monoatom_Ed_Z_Figure_M0}
\end{figure*}

The performance of model $M_0$ is presented in Fig~\ref{Monoatom_Ed_Z_Figure_M0}, where we plot $E_d$ against $Z$ similar to Fig.~\ref{Monoatom_Ed_Z_Figure}.
The labels ($M_{0M_i}$) in Fig~\ref{Monoatom_Ed_Z_Figure_M0} follow the same function ($M_0$) from Table~\ref{r_squared_table} with different constant values for $a_1$ and $c_1$ as shown in Table~\ref{constants_table_M0}.
Different values should be adopted depending on the nature of the calculation the model is trying to replicate; however, they all exhibit similar performance.

\subsection*{Models with SISSO dimensionality set to two}

\begin{figure*}[t]
    \includegraphics[width = 17cm]{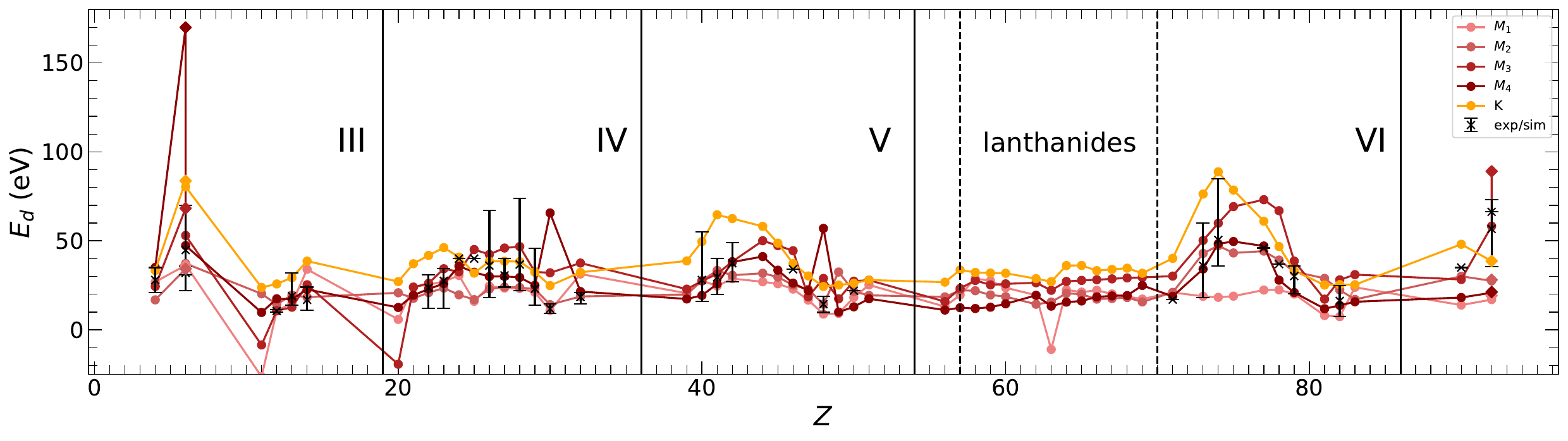}
    \caption{Predicted (connected points) and measured (cross markers) average threshold displacement energy ($E_d$) values as a function of the atomic number ($Z$), corresponding to the predictive functions in Table~\ref{r_squared_table} when $\text{d} = 2$.
    Model K is shown as a reference.
    All specifications for the plot remain consistent with those outlined in Fig.~\ref{Monoatom_Ed_Z_Figure}.
    }
    \label{Monoatom_Ed_Z_Figure_d2}
\end{figure*}

Regarding the performance of the models $M_i$ (excluding $M_0$), by setting SISSO's dimension to two, we obtain the models labeled with $\text{d}=2$ in Table~\ref{r_squared_table}.
These models are plotted in Figure~\ref{Monoatom_Ed_Z_Figure_d2} as a function of $Z$.
We observe that the performance is not significantly better than when $\text{d}=1$ (Fig.\ref{Monoatom_Ed_Z_Figure}).
Furthermore, the model $M_3$ yields negative values, which are not allowed; therefore, we can state that its performance has worsened, even though this model has a higher $R^2$ value.

\section*{Appendix C: Temperature considerations}

We apply the same approach explained in the main manuscript for monoatomic materials, including $T$ as a feature and excluding data from reports that do not specify the measuring or simulation temperature.
If multiple reports exist for the same material, using the same method and at the same temperature, we consider a single value by taking their average.
The intention behind this approach is to capture a potential temperature dependence in the $E_d$. However, only the models for subsets $M_0$ and $M_2$ display any dependency on $T$ when $\text{d}=2$.
Additionally, these results generally lead to similar models but with a lower $R^2$, likely due to the reduced amount of data per subdivision.
The temperature-dependent models are presented in Table~\ref{r_squared_table_T} and their constants in Table~\ref{constants_table_T}.
n general, from Table~\ref{r_squared_table_T}, we observe that model $M_0$ includes a term that increases with $T$, whereas, in contrast, model $M_2$ includes a term that decreases as $T$ increases.
Nonetheless, these results are not conclusive enough, and not presented in the main manuscript of this work.

\begin{table}[t]
    \centering
    \setlength{\tabcolsep}{4pt}
    \setlength\extrarowheight{9pt}
    \begin{tabular}{c c c c}
    \Xhline{1pt}     
       Label & d & $E_d$ & $R^2$  \\ \hline
               $M_0$ & 2 & $c_0 + a_0 \ell_{\text{bond}}^4T + a_1 \frac{E_{\text{coh}}\sqrt{\rho}}{\ell_{\text{bond}}}$ & 0.62 \\
        $M_2$ & 2 & $c_0 + a_0 \frac{E_{\text{ioniz}}\left(T_{\text{melt}}-T\right)}{T_{\text{melt}}} + a_1 \sqrt[3]{\rho}\sqrt{E_{\text{coh}}}$ & 0.74 \\       
        \vspace{-3ex}  
        \\
     \Xhline{1pt}         
    \end{tabular}
    \caption{Threshold displacement energy ($E_d$) models, dataset subdivision labels ($M_i$, $K$, $K_{\text{SR}}$, $K_{\text{S}}$), SISSO dimension (d), and coefficient of determination ($R^2$) for models dependent on $T$.
    The constants $a_i$ and $c_0$ are unique for each function and are reported in Table~\ref{constants_table_T}. 
    }
    \label{r_squared_table_T}
\end{table}
\begin{table}[t]
    \centering
    \setlength\extrarowheight{7pt}
    \begin{tabular}{c c c c c}
    \Xhline{1pt}     
       Label & d & $c_0$ & $a_0$ & $a_1$  \\ \hline
              $M_2$ & 2 & -9.87877510 & 1.87112033 & 5.03143901
       \\
       \hline
            $M_{0M_1}$ & 2 & $3.6760343\cdot10$ & $-8.2028199\cdot 10^{-4}$ & $-8.0706561\cdot10^{-2}$
       \\
            $M_{0M_2}$ & 2 & $1.4993954\cdot10$ & $-2.1801895\cdot 10^{-4}$ & $2.2623755$
       \\
            $M_{0M_3}$ & 2 & $1.2968922\cdot10$ & $3.0633423\cdot 10^{-4}$ & $3.5860318$
       \\      
            $M_{0M_4}$ & 2 & $1.8840480\cdot10$ & $-8.1067802\cdot 10^{-3}$ & $2.3535802$
       \\    
     \Xhline{1pt}         
    \end{tabular}
    \caption{Constants $c_0$ and $a_i$ for models dependent on $T$.
    }
    \label{constants_table_T}
\end{table}

\section*{Appendix D: Features as a function of the atomic number}

To better understand the behavior of the main results in Fig.~\ref{Monoatom_Ed_Z_Figure}, we plot e values of the considered features in monoatomic materials as the atomic number ($Z$) increases in Fig.~\ref{Z_features_figure}, illustrating how these properties evolve across the periodic table.
Since the resulting models are functions of these parameters, Fig.~\ref{Monoatom_Ed_Z_Figure} follow a similar trend to some of the curves in Fig.~\ref{Z_features_figure}.
This is particularly evident for $E_{\text{ioniz}}$, $E$, $r$, $\rho$, and $T_{\text{melt}}$, where ``bumps'' appear for each row of the periodic table (separated by vertical lines) and a linear trend is observed in the lanthanides section.

\begin{figure*}[h]
    \includegraphics[width=17cm]{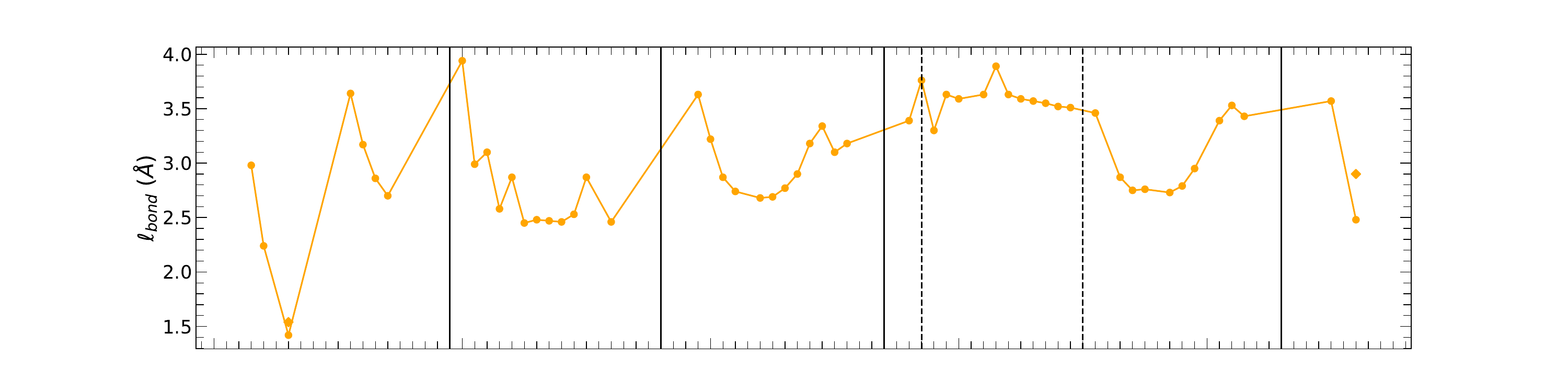}
    \includegraphics[width=17cm]{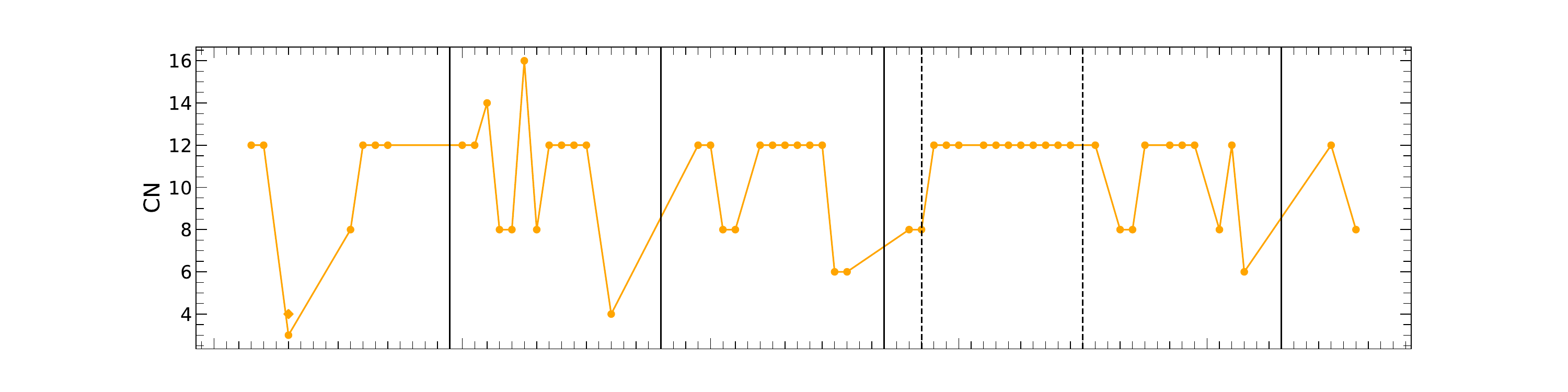}
    \includegraphics[width=17cm]{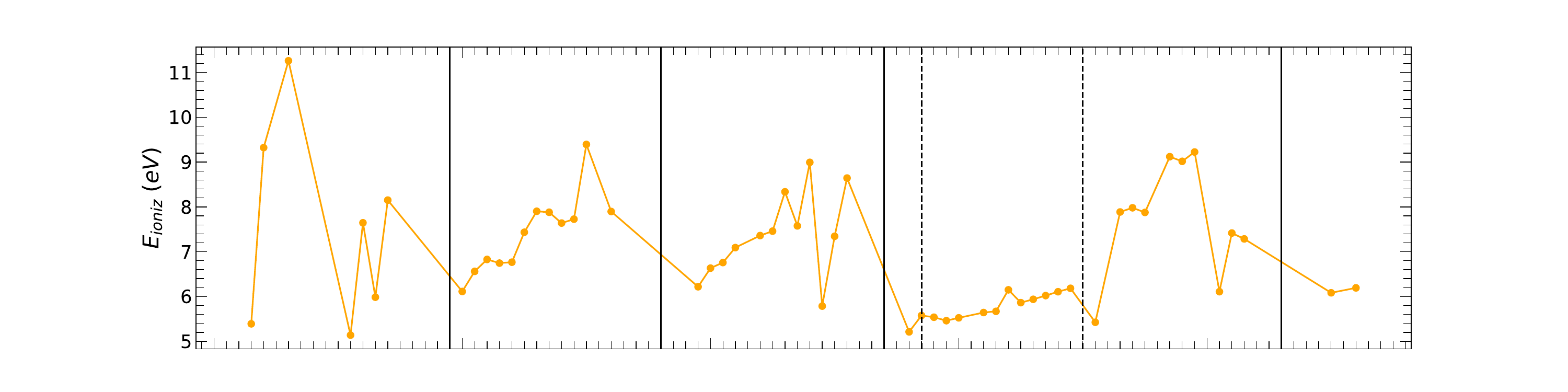}    
    \includegraphics[width=17cm]{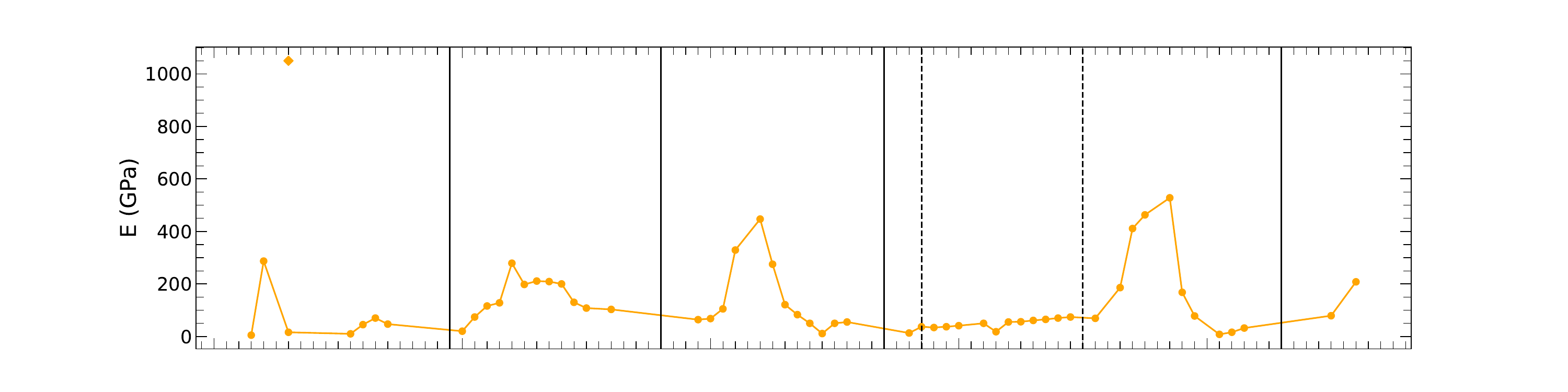}          
        \includegraphics[width=17cm]{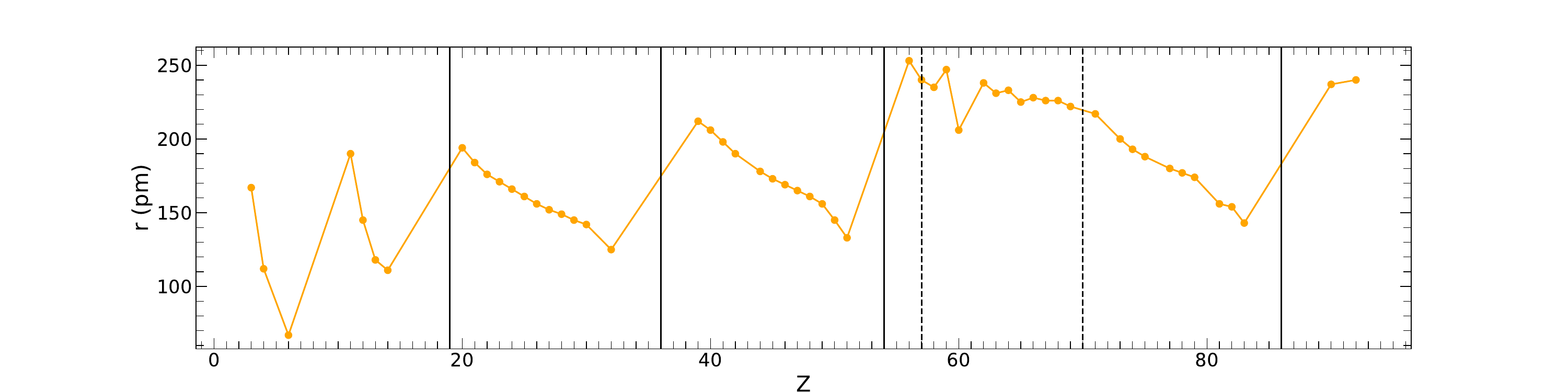}            
\end{figure*}
\begin{figure*}
    \includegraphics[width=17cm]{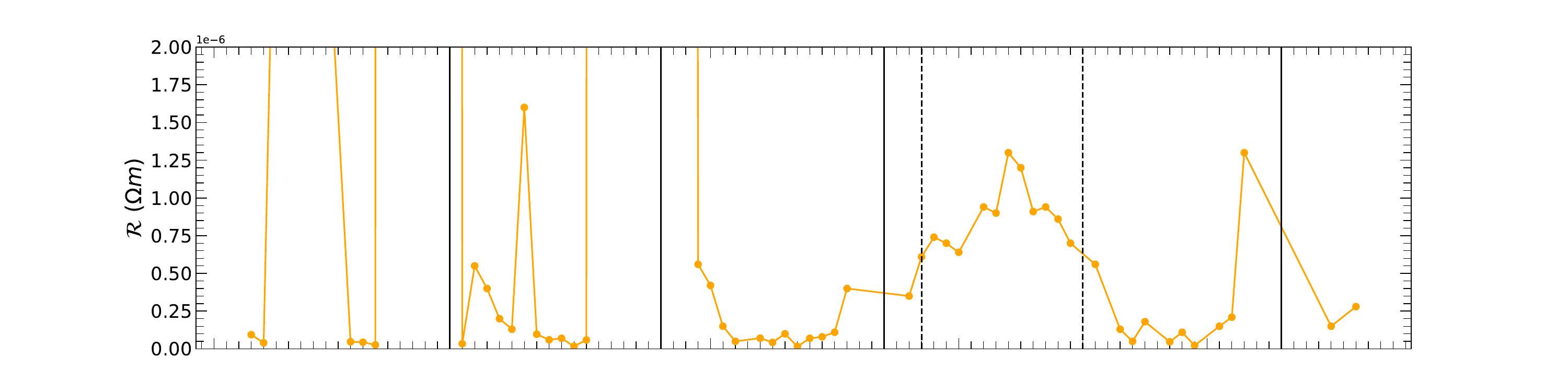}  
    \includegraphics[width=17cm]{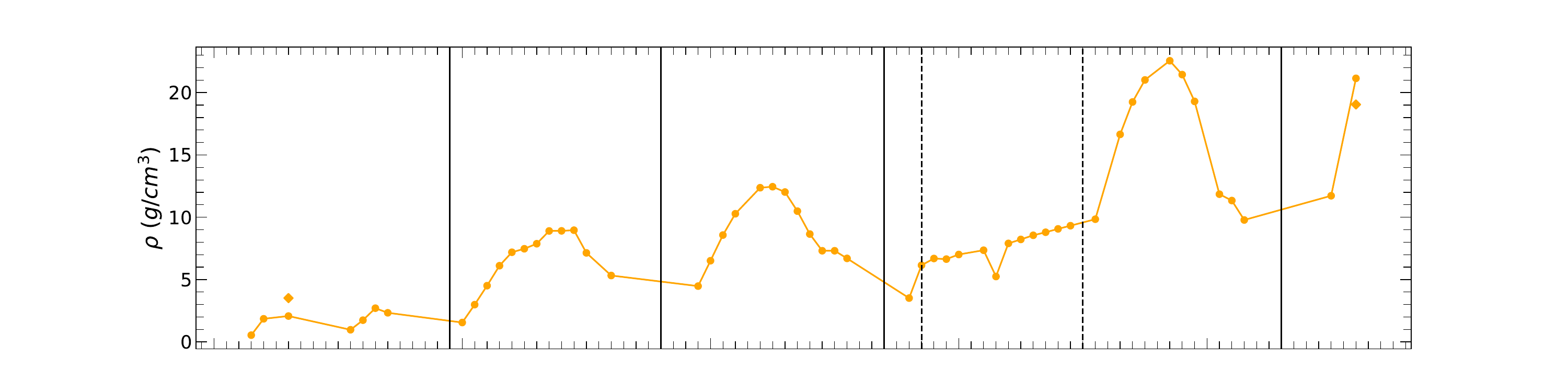} 
    \includegraphics[width=17cm]{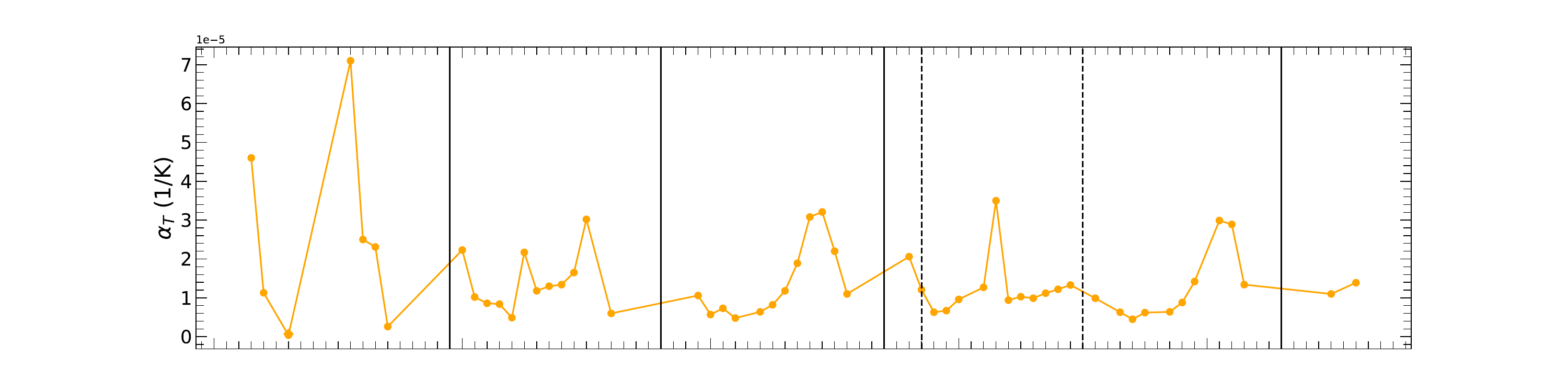}
    \includegraphics[width=17cm]{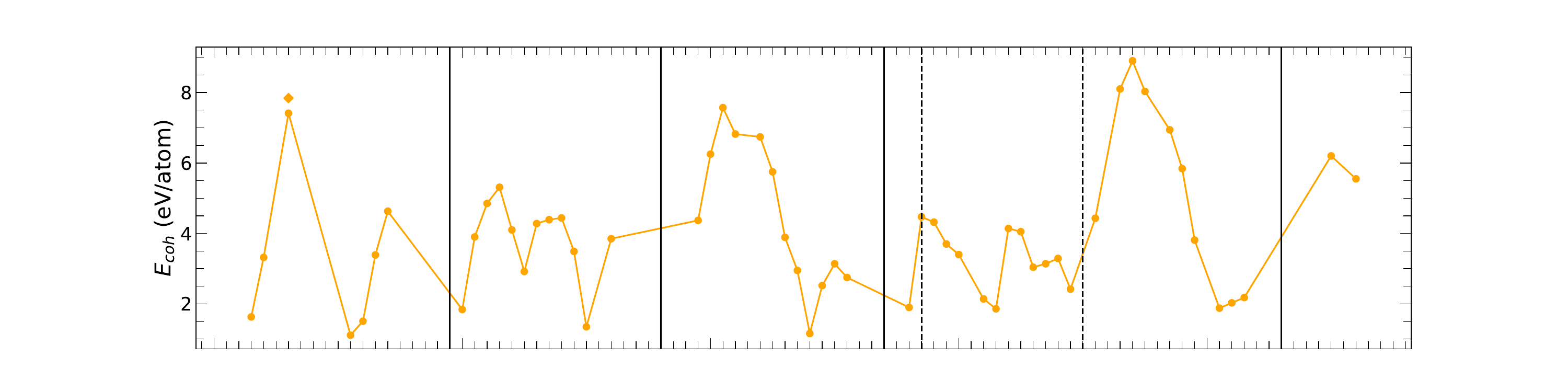}
    \includegraphics[width=17cm]{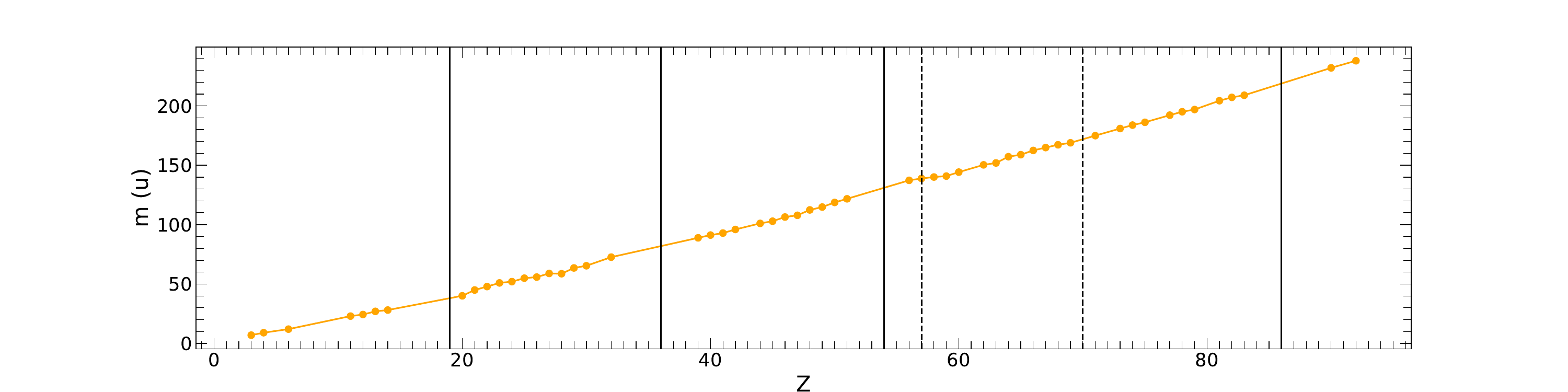} 
    \label{Z_features_figure}
\end{figure*}
\begin{figure*}
    \includegraphics[width=17cm]{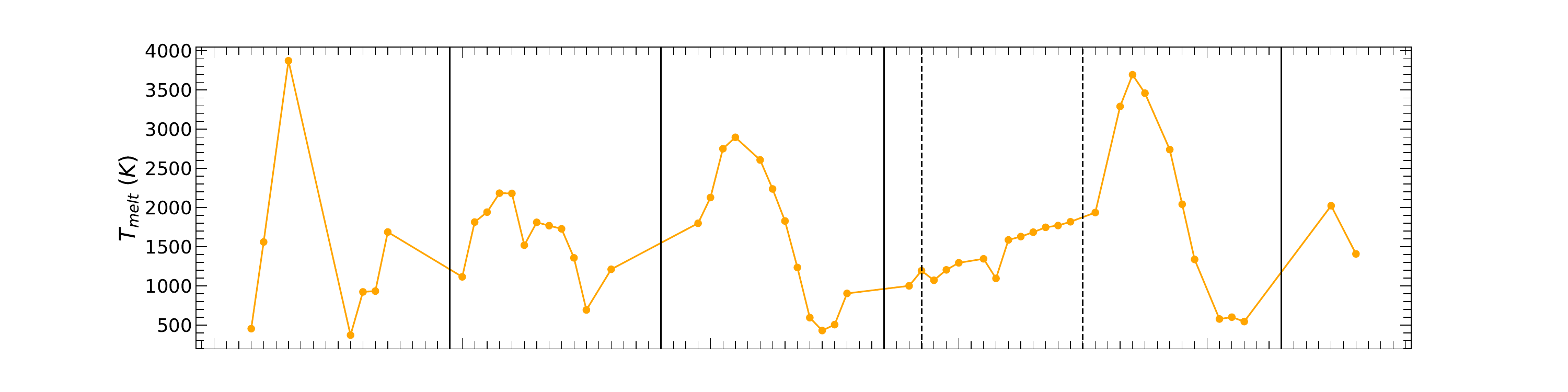}  
    \caption{Different used parameters plotted as a function of the atomic number ($Z$).
     We use diamond markers to represent materials with the same $Z$ value, such as diamond and gamma uranium (diamond marker), in contrast to graphite and alpha uranium (point marker).
     Some values are out of scale and therefore not shown in the plots.
     }
    \label{Z_features_figure}
\end{figure*}

\section*{Appendix E: Effective Threshold Displacement Energy}

\begin{figure*}
    \centering
    \includegraphics[height=0.23\textheight]{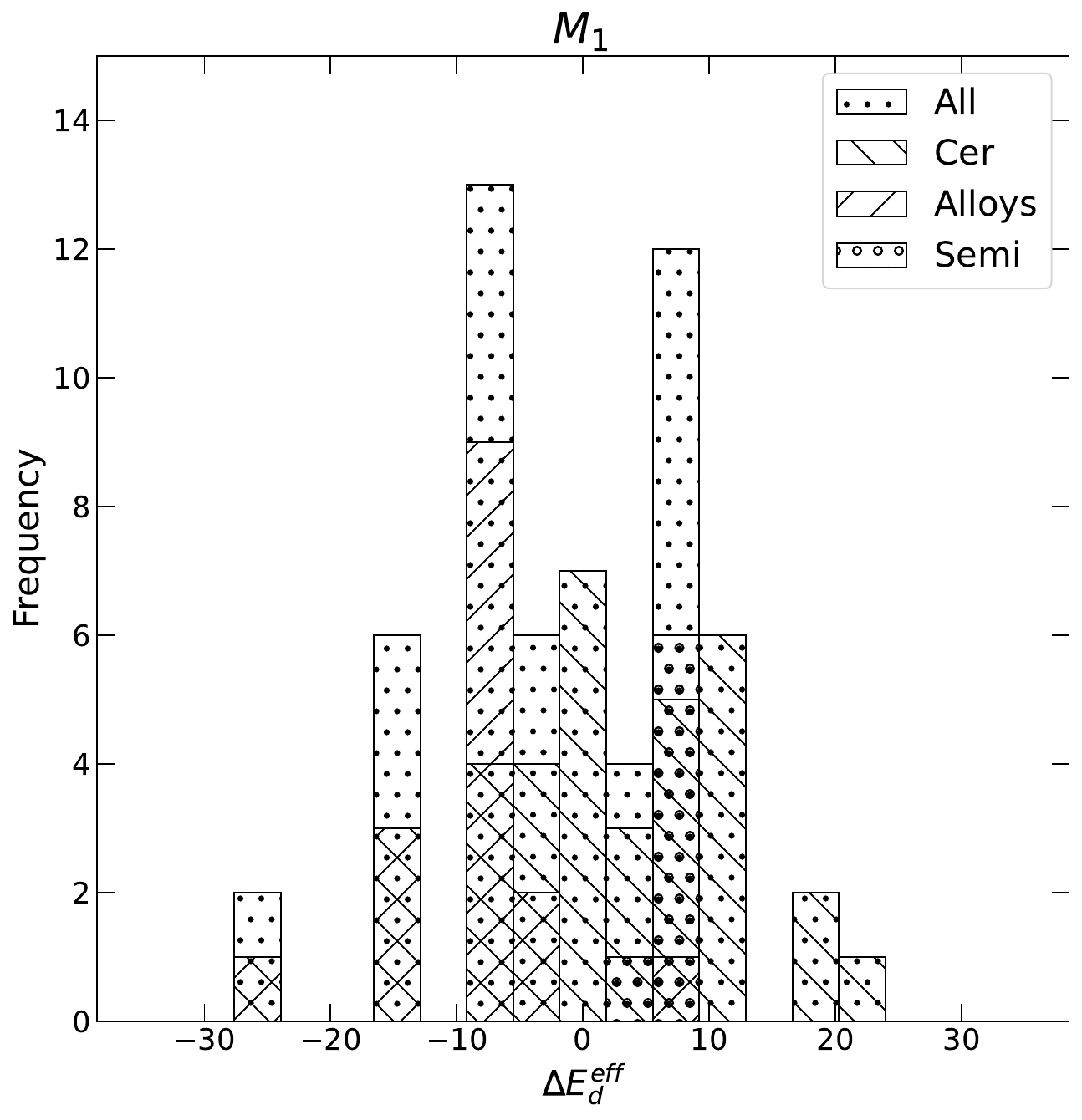}
        \includegraphics[ height=0.23\textheight]{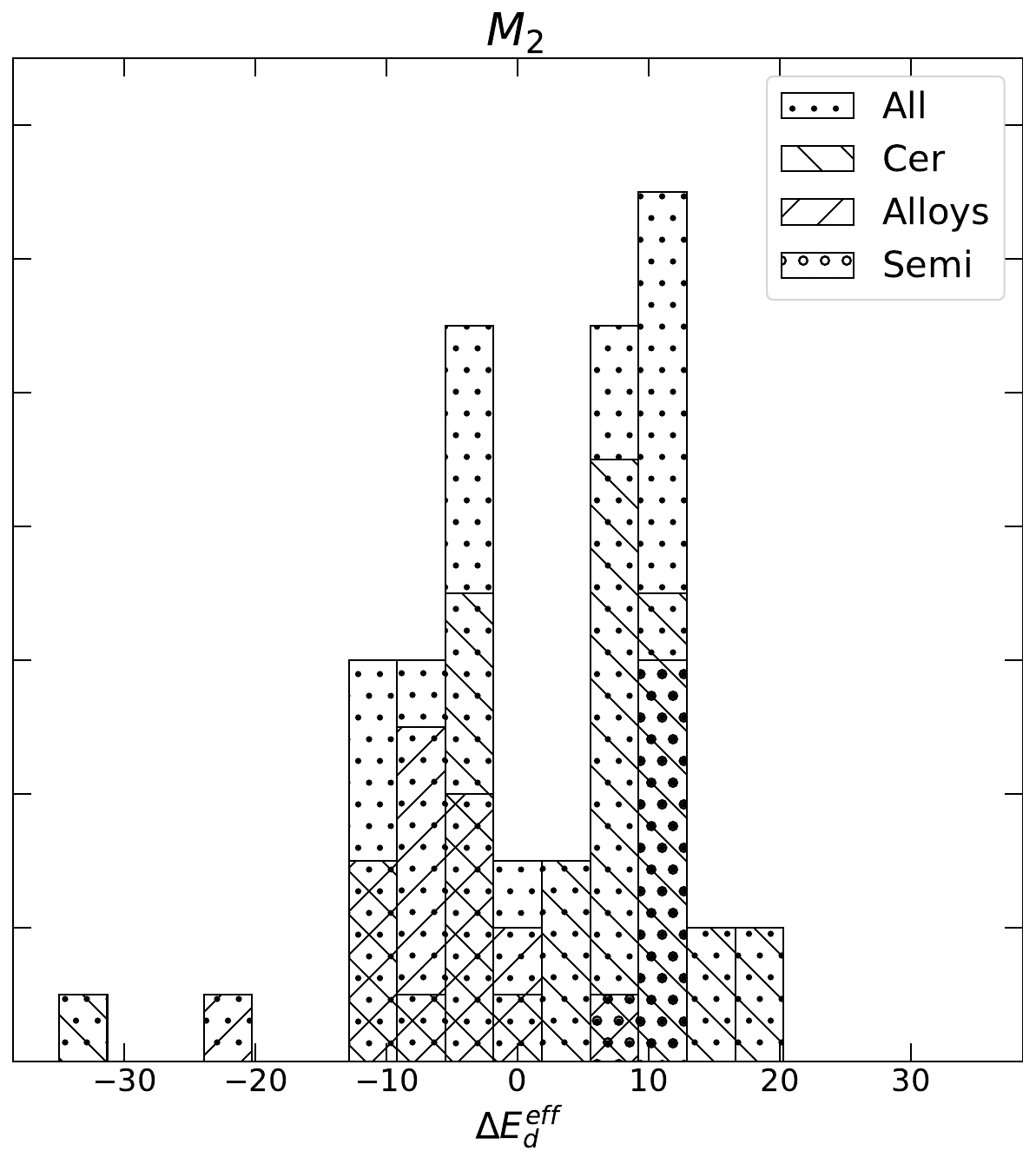}
            \includegraphics[ height=0.23\textheight]{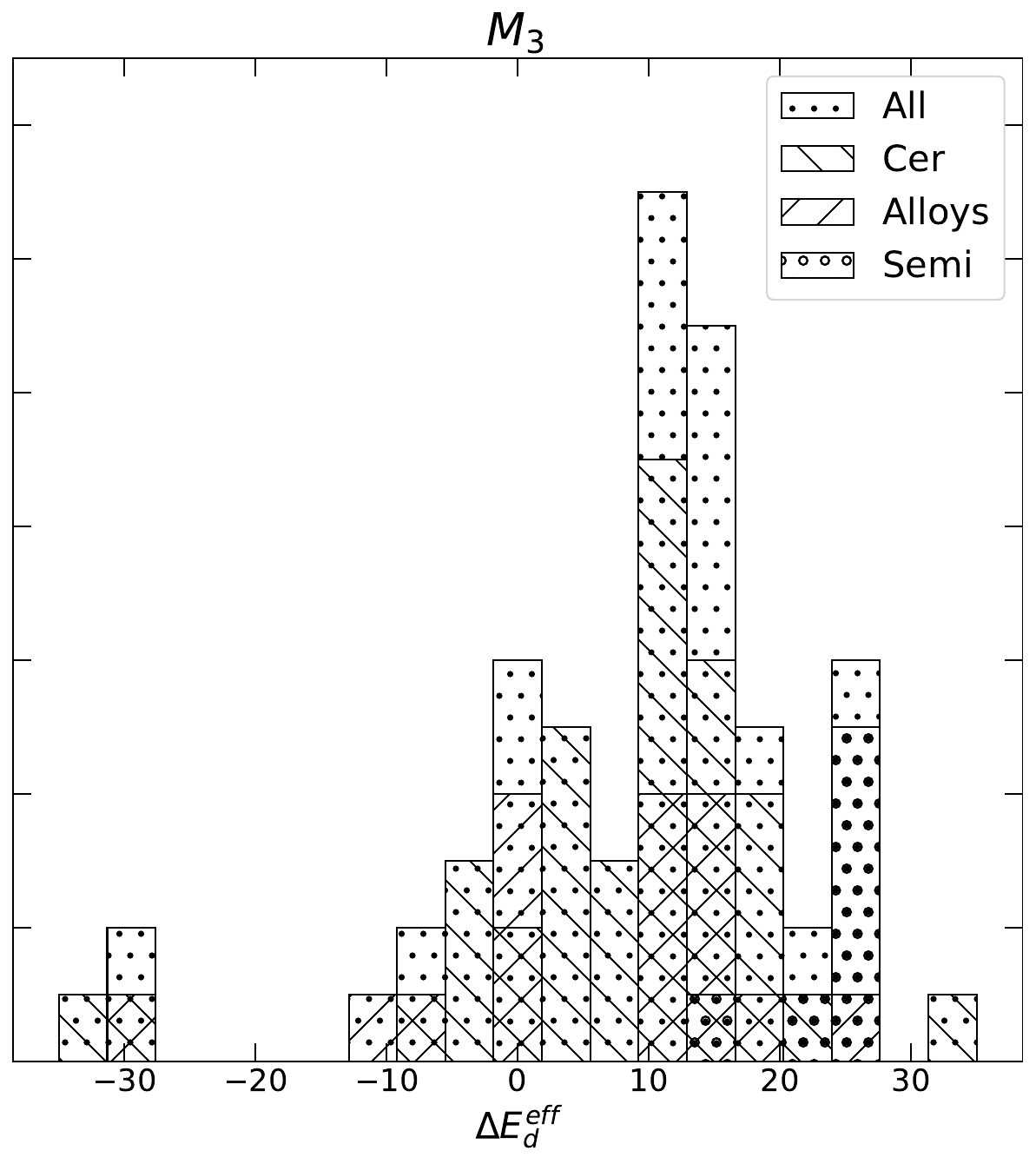}
                \includegraphics[height=0.23\textheight]{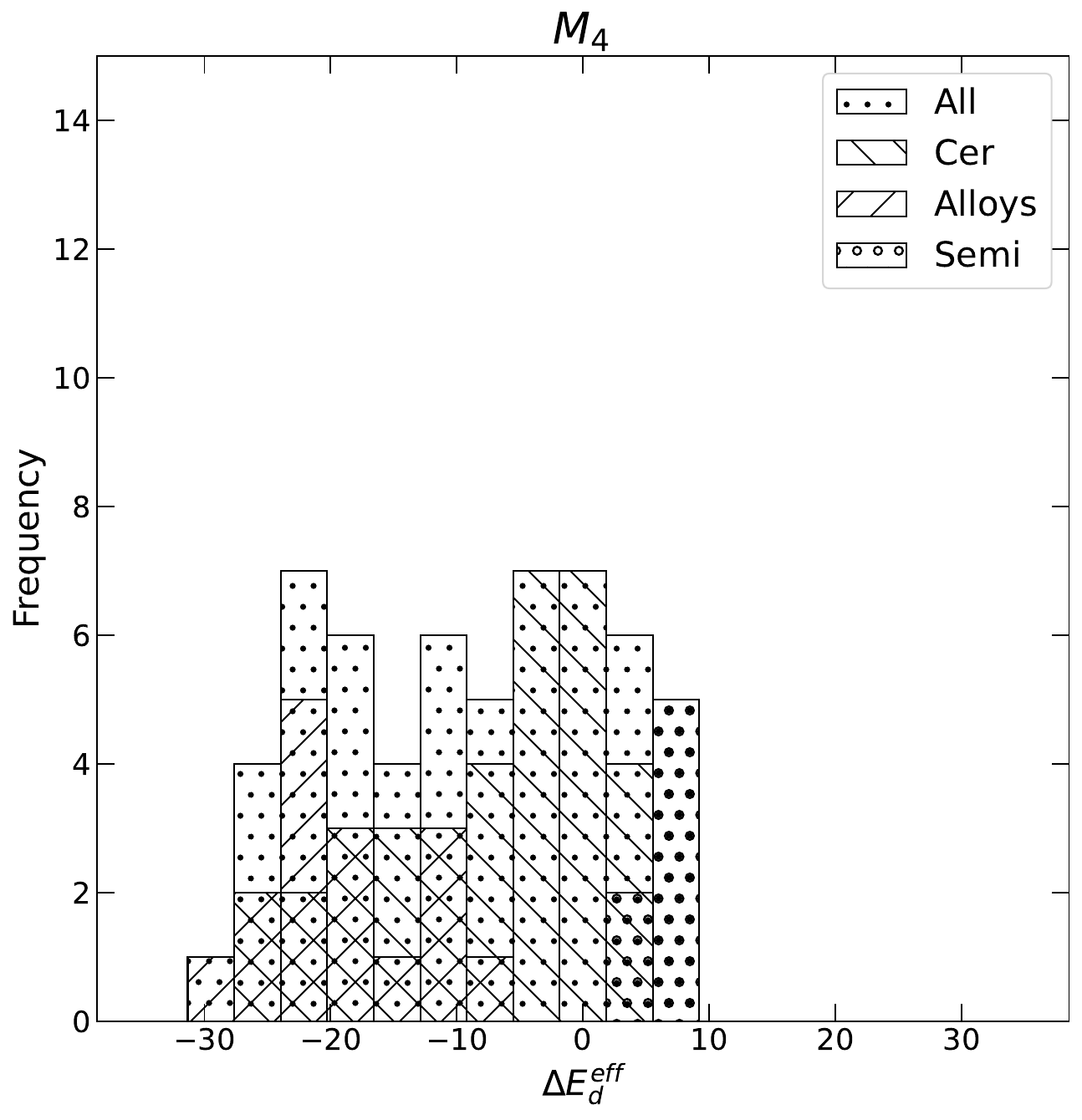}
                    \includegraphics[height=0.23\textheight]{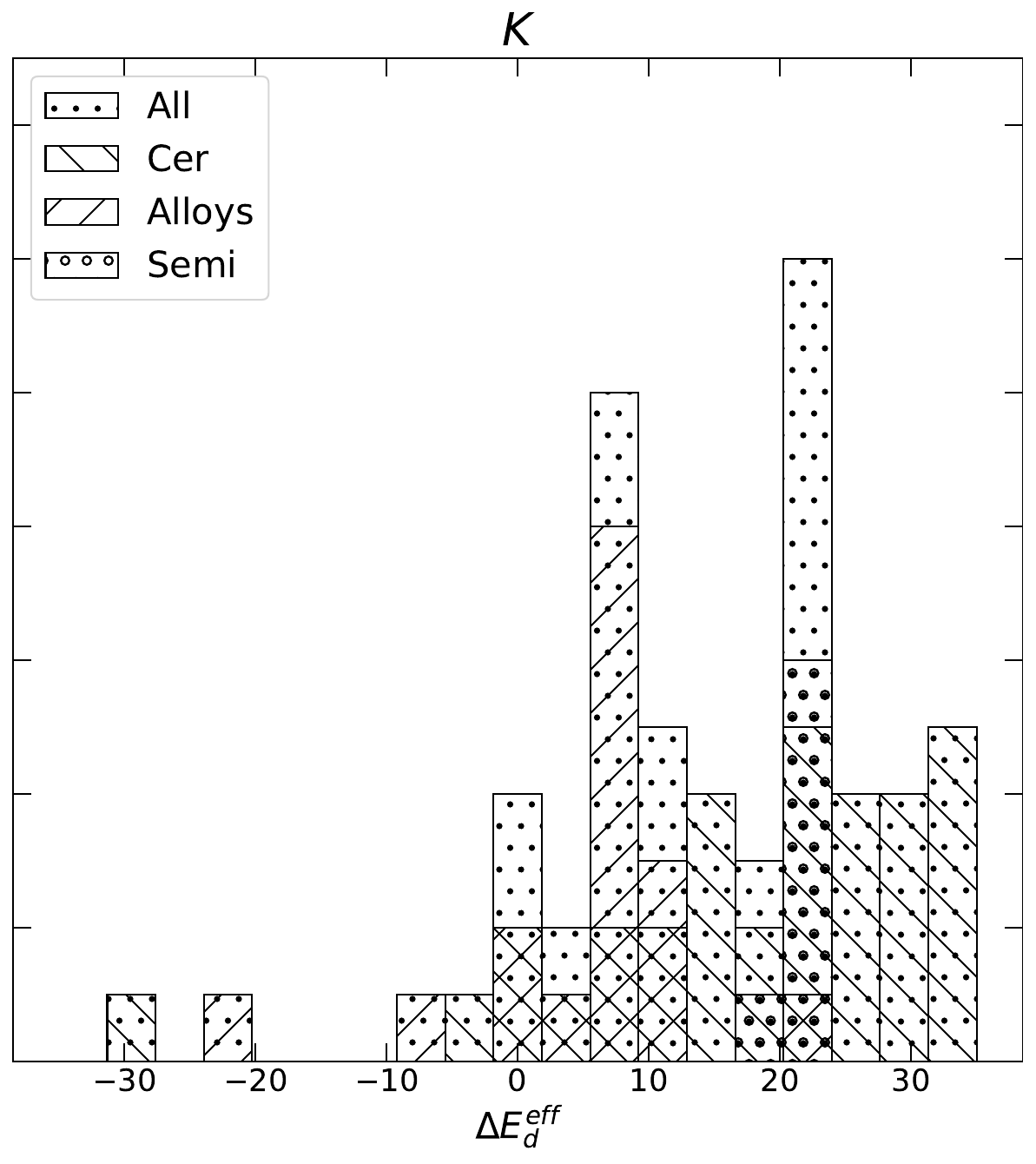}
    \caption{Histograms of the difference between the effective threshold displacement energy from models and the average experimental/simulation data ($\overline{\Delta E_d^{\text{eff}}}$) for all polyatomic data and for subsets specific to alloys and ceramics.}
    \label{ghoniem_hists}
\end{figure*}

To visually represent the results in Section~\ref{section_edeff}, in Fig~\ref{ghoniem_hists}, we present histograms depicting the differences between the effective threshold displacement energy predicted by the models and the average experimental and simulation data ($\overline{\Delta E_d^{\text{eff}}}$) for all categories: the entire dataset (All), alloys (Alloys), ceramics (Cer), and non-ceramic semiconductors (Semi).
These results, along with specific values such as the data mean and standard deviation, are presented and discussed in the main manuscript.

\section*{Appendix F: Polyatomic Materials}

\begin{figure*}
\includegraphics[width=5.7cm]{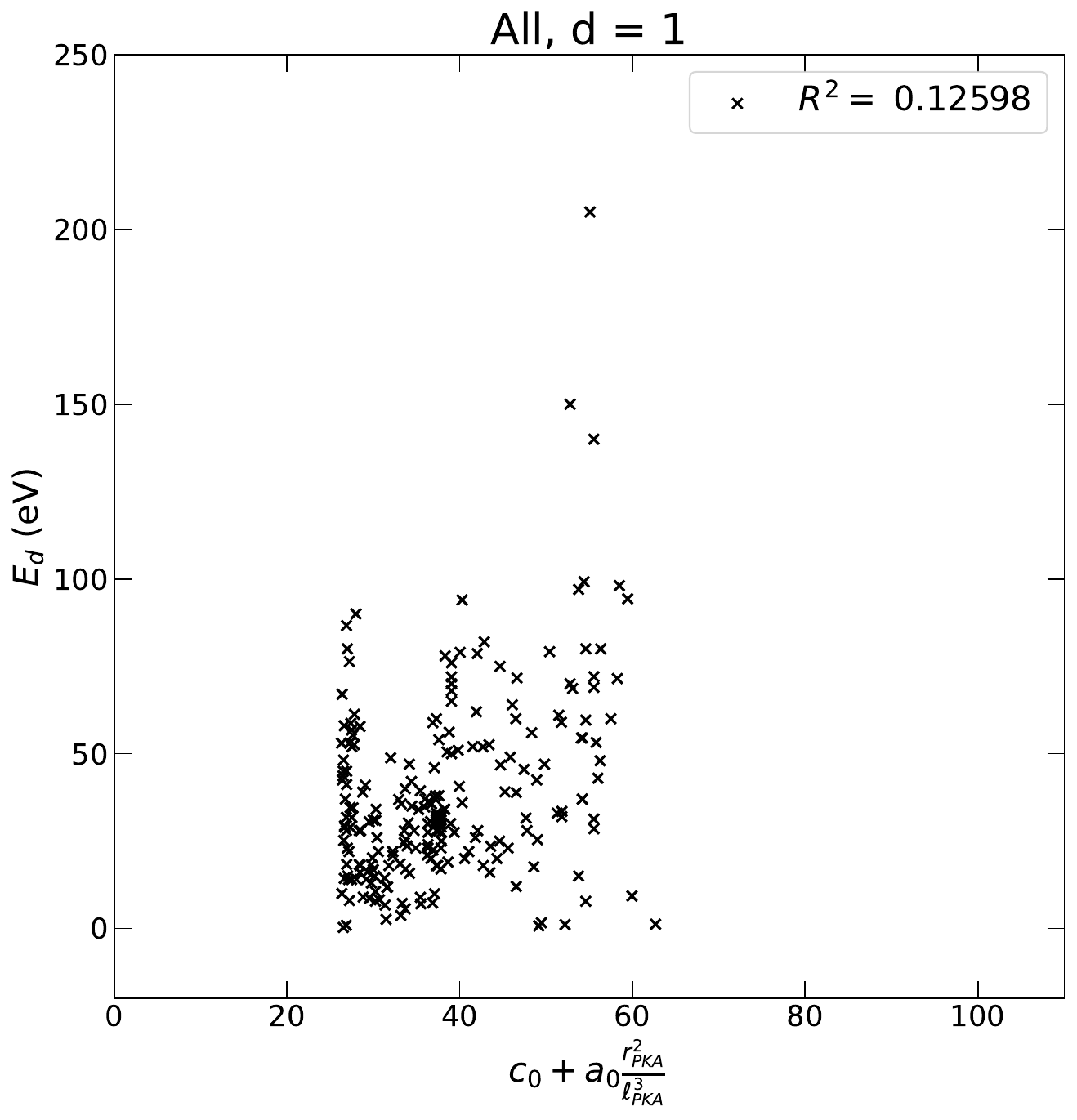}
        \includegraphics[width=5.7cm]{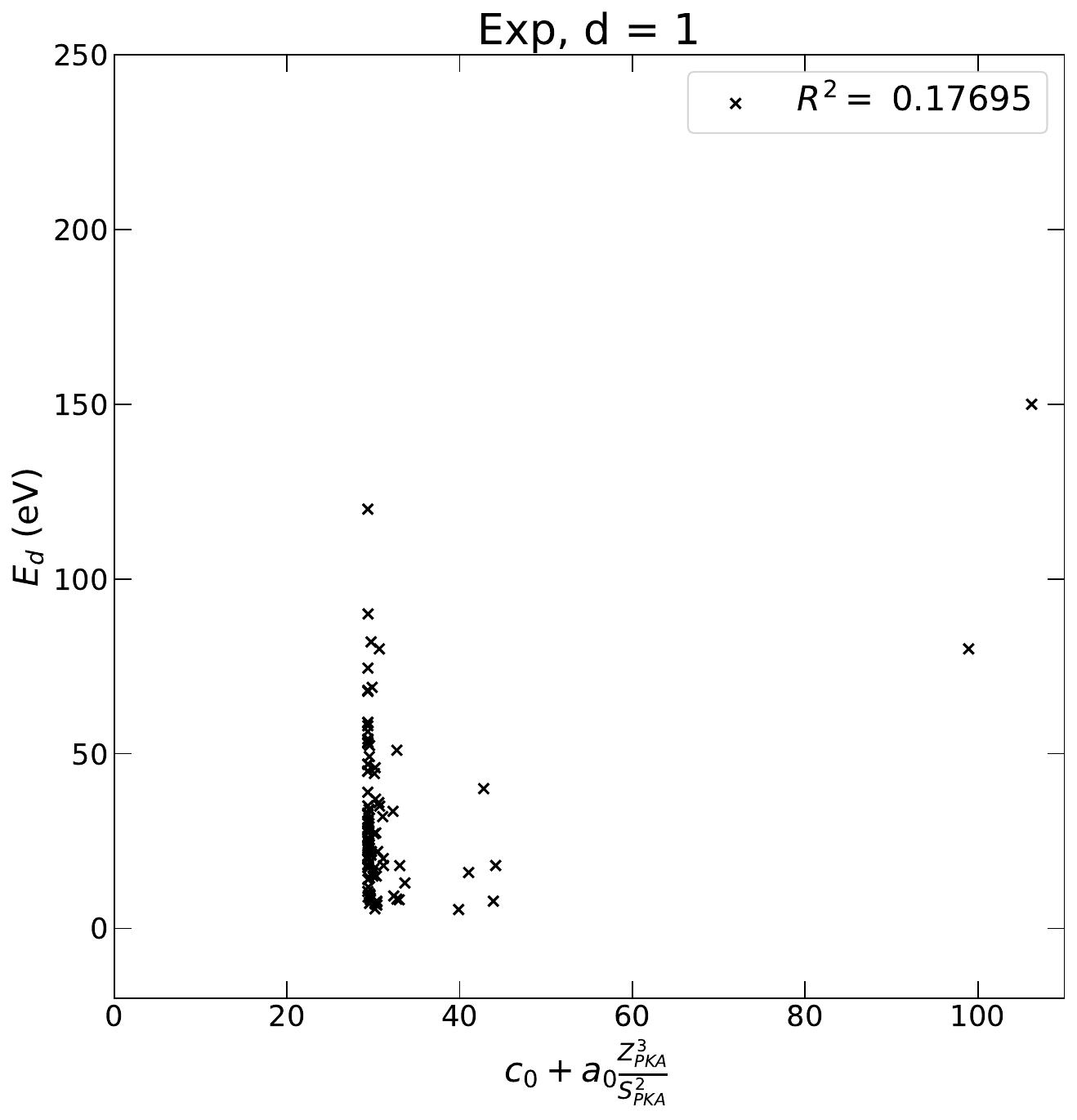}
            \includegraphics[width=5.7cm]{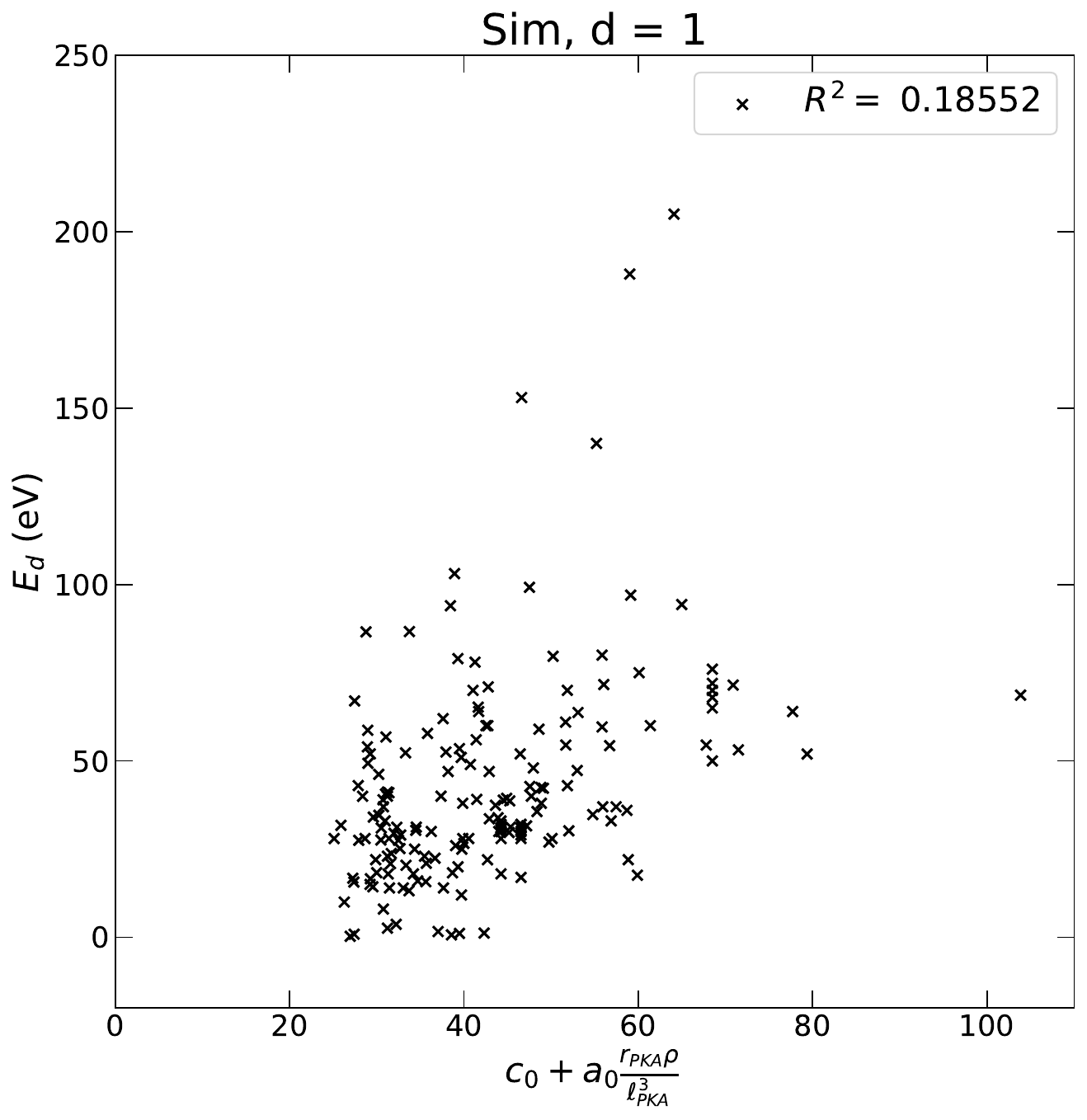}
\includegraphics[width=5.7cm]{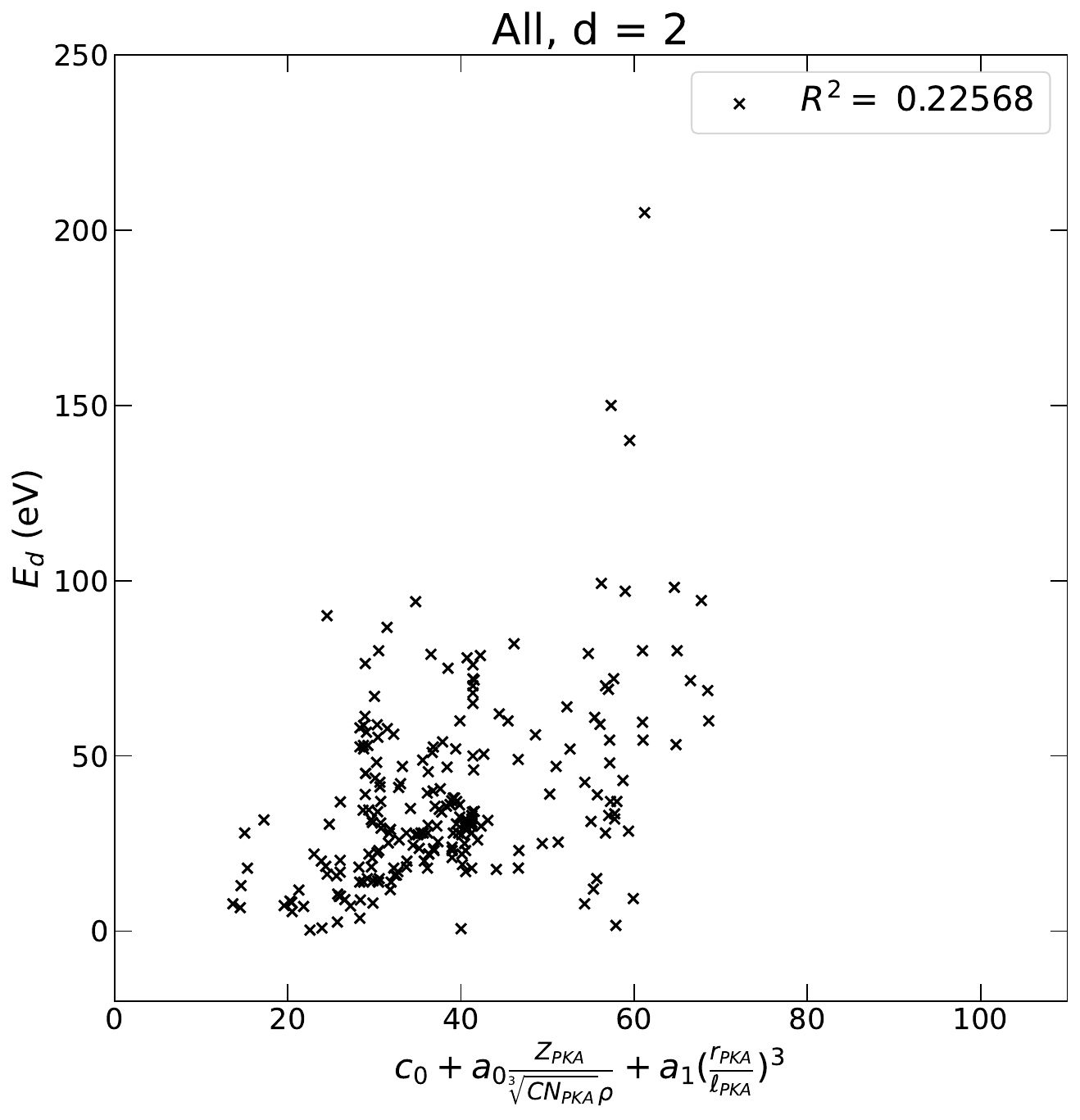}
        \includegraphics[width=5.7cm]{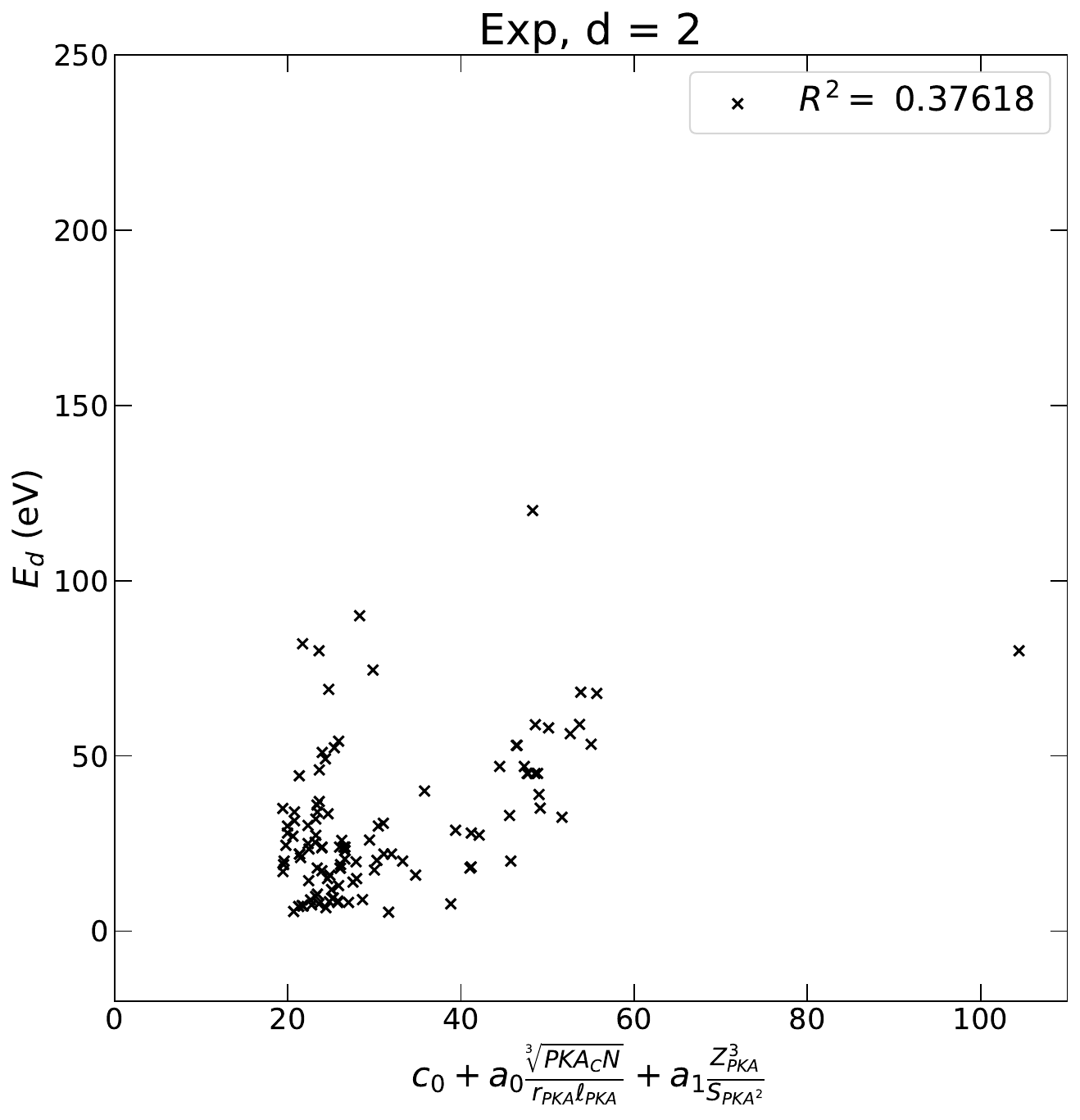}
            \includegraphics[width=5.7cm]{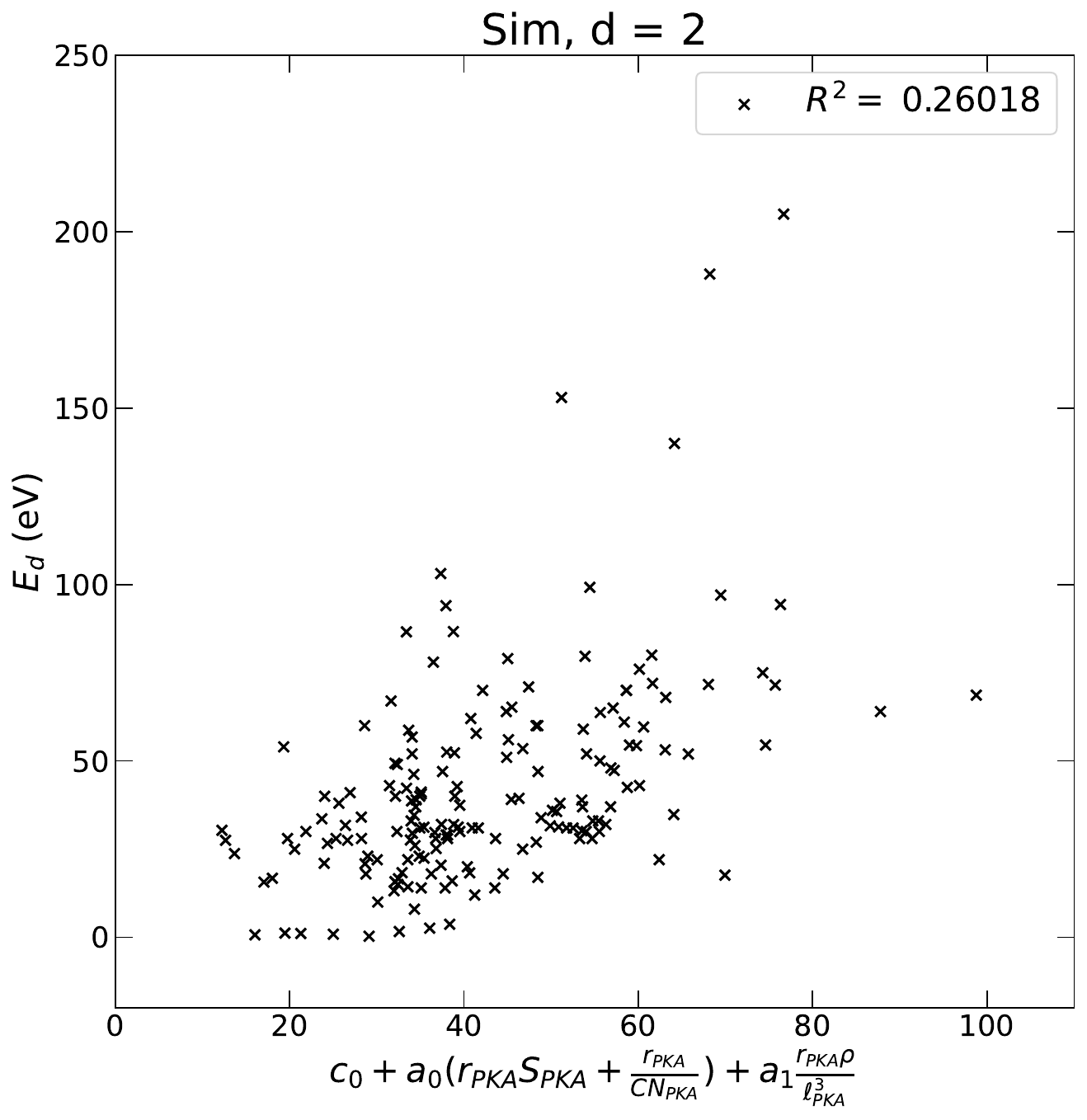}          
\includegraphics[width=5.7cm]{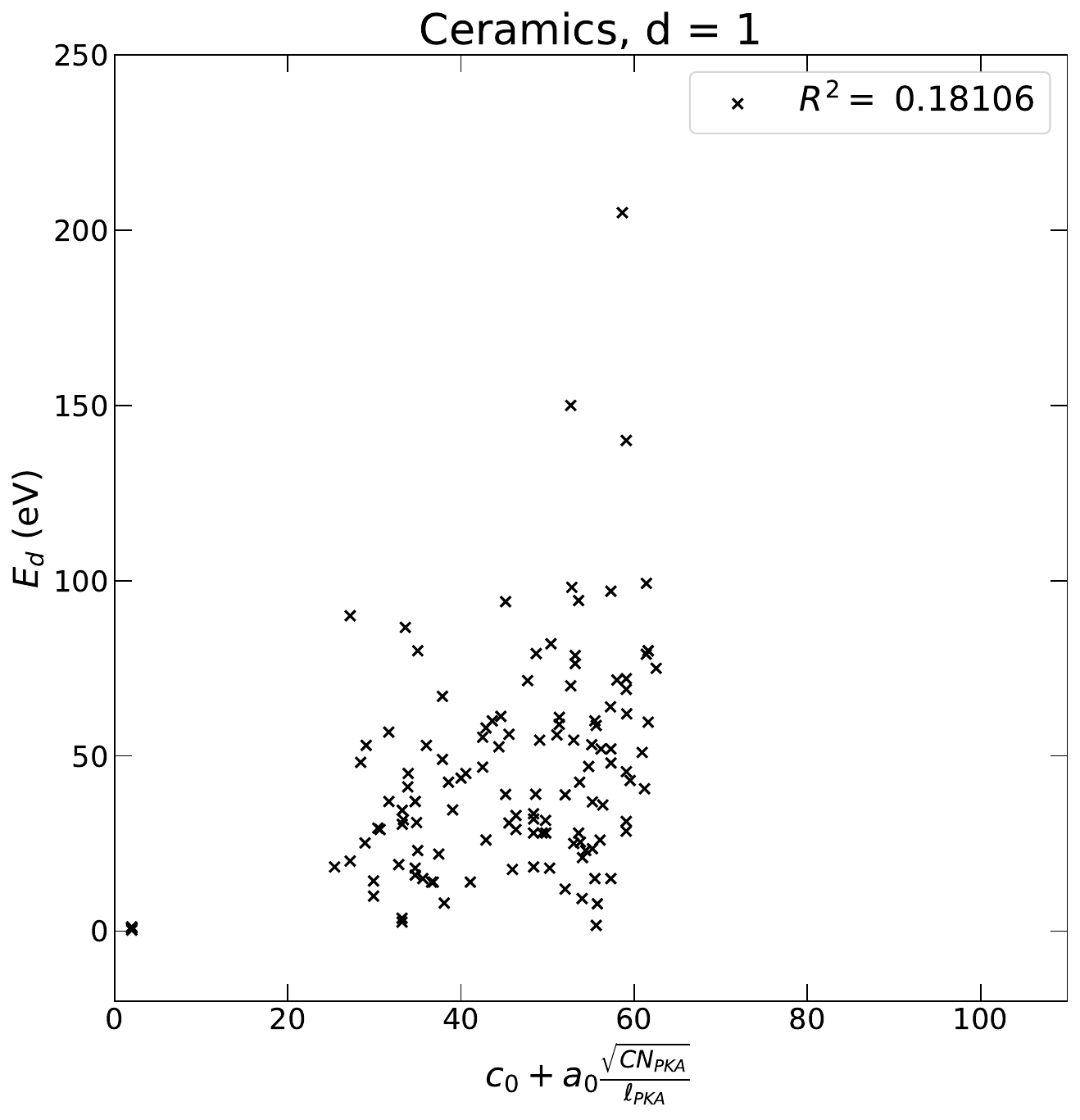}
    \includegraphics[width=5.7cm]{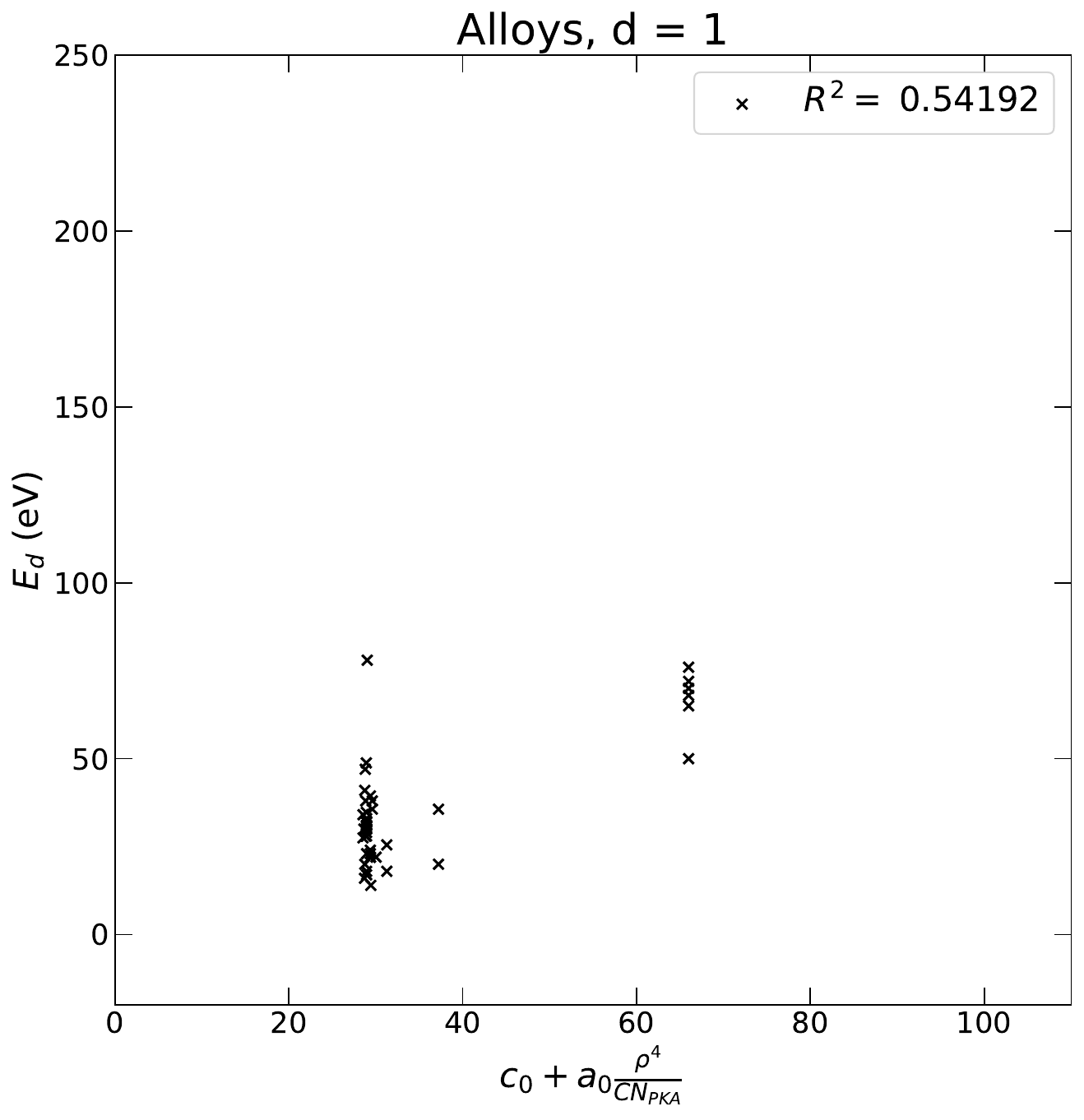}
        \includegraphics[width=5.7cm]{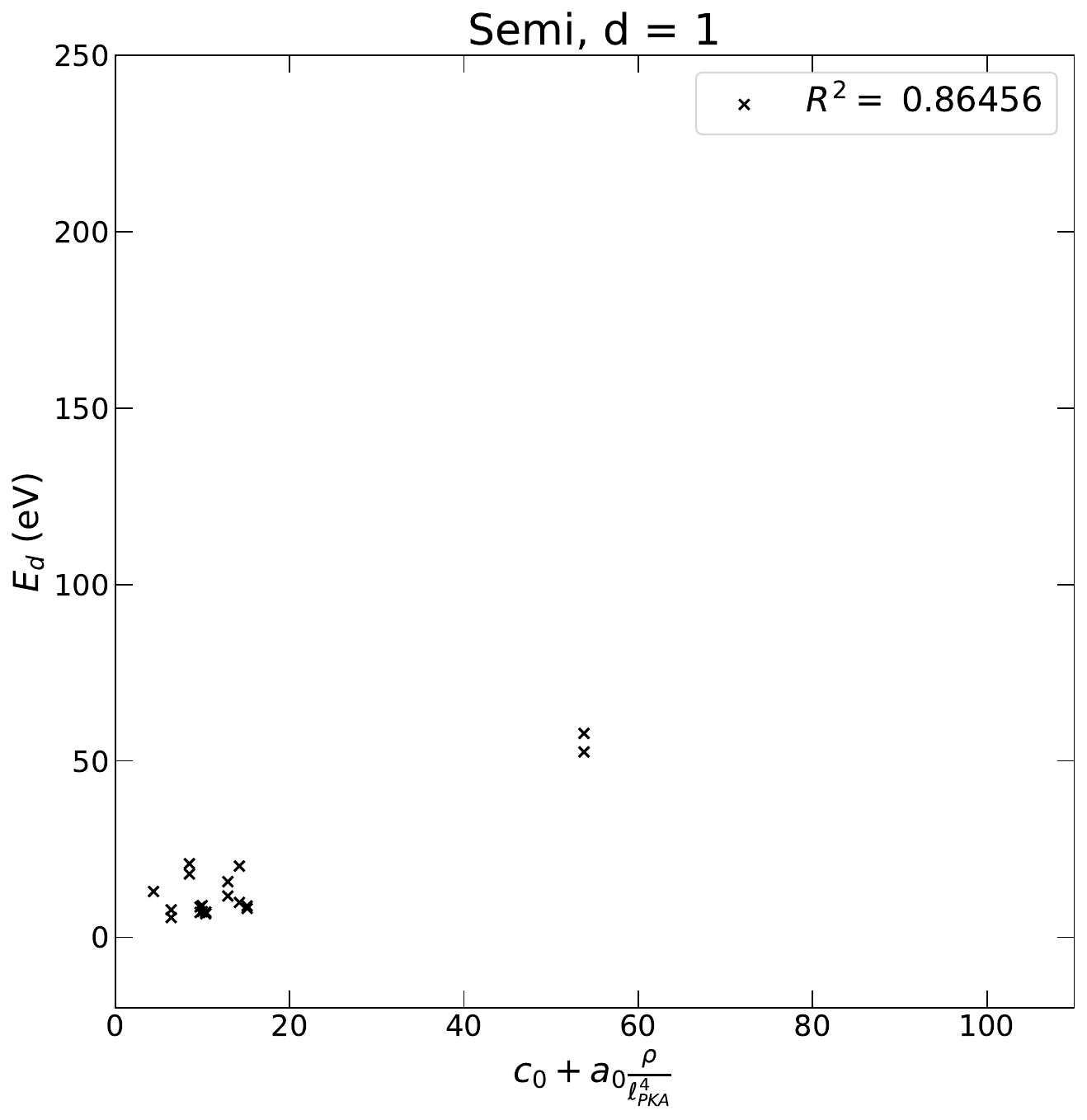}    
\includegraphics[width=5.7cm]{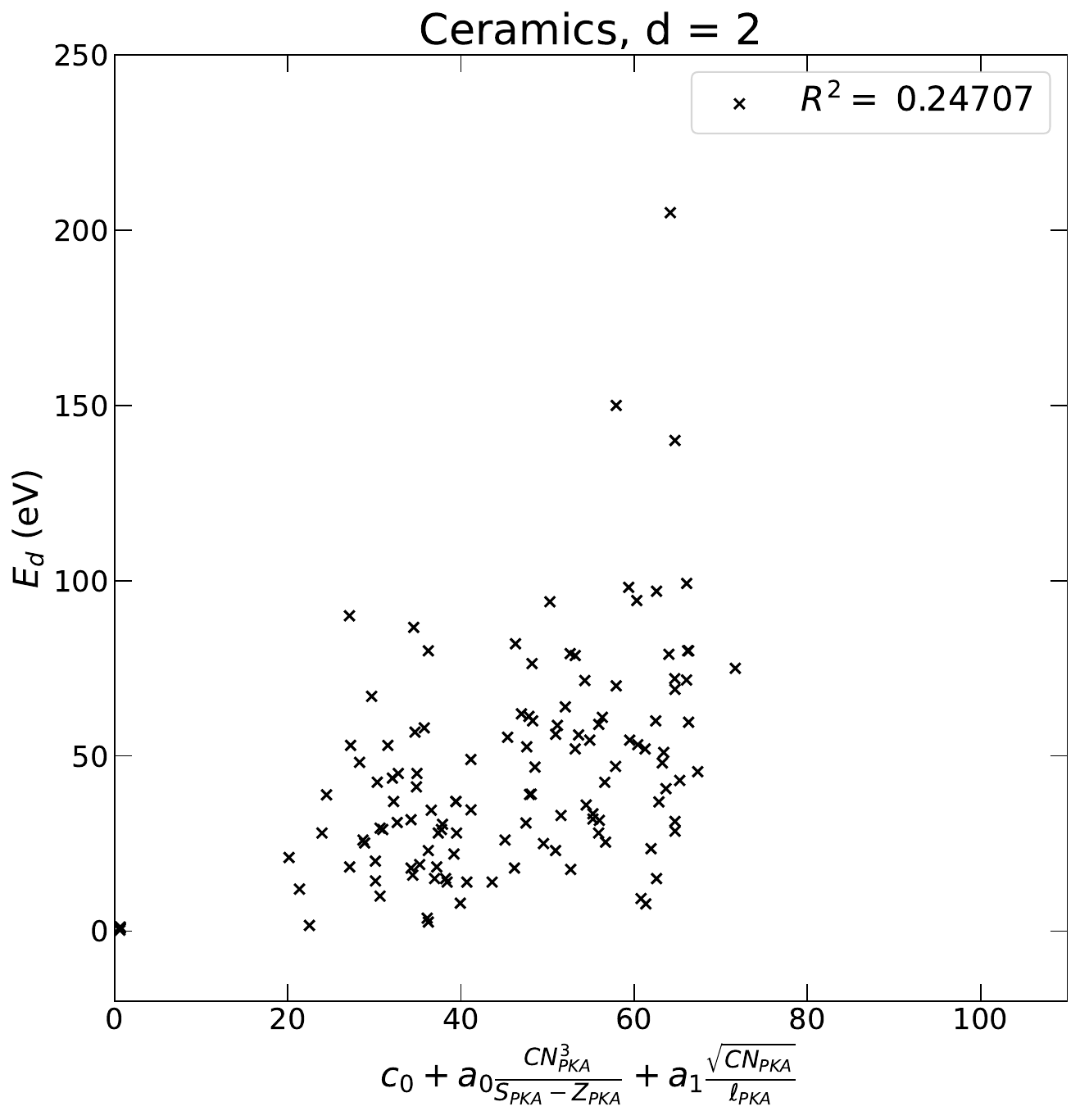}
    \includegraphics[width=5.7cm]{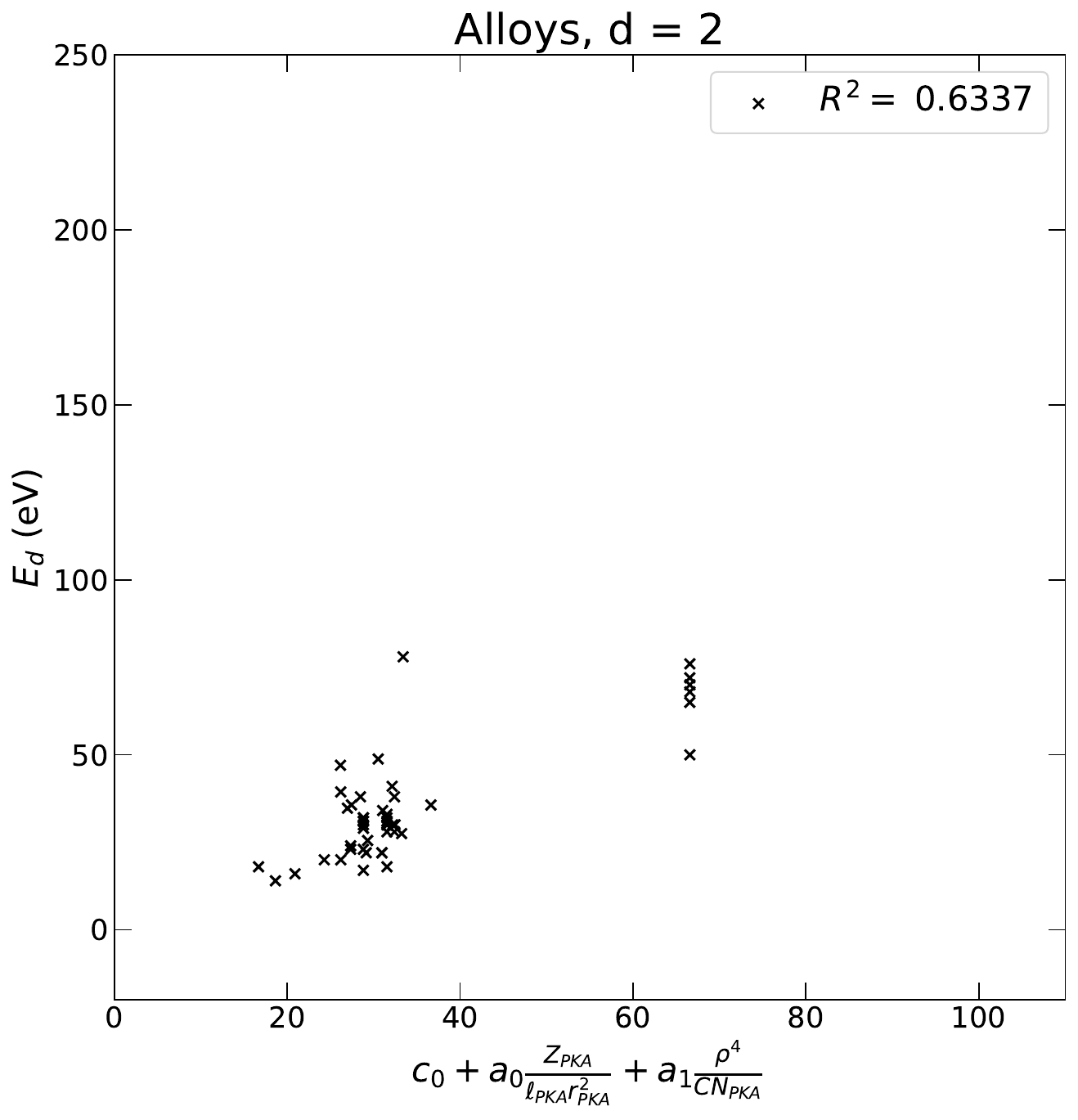}
        \includegraphics[width=5.7cm]{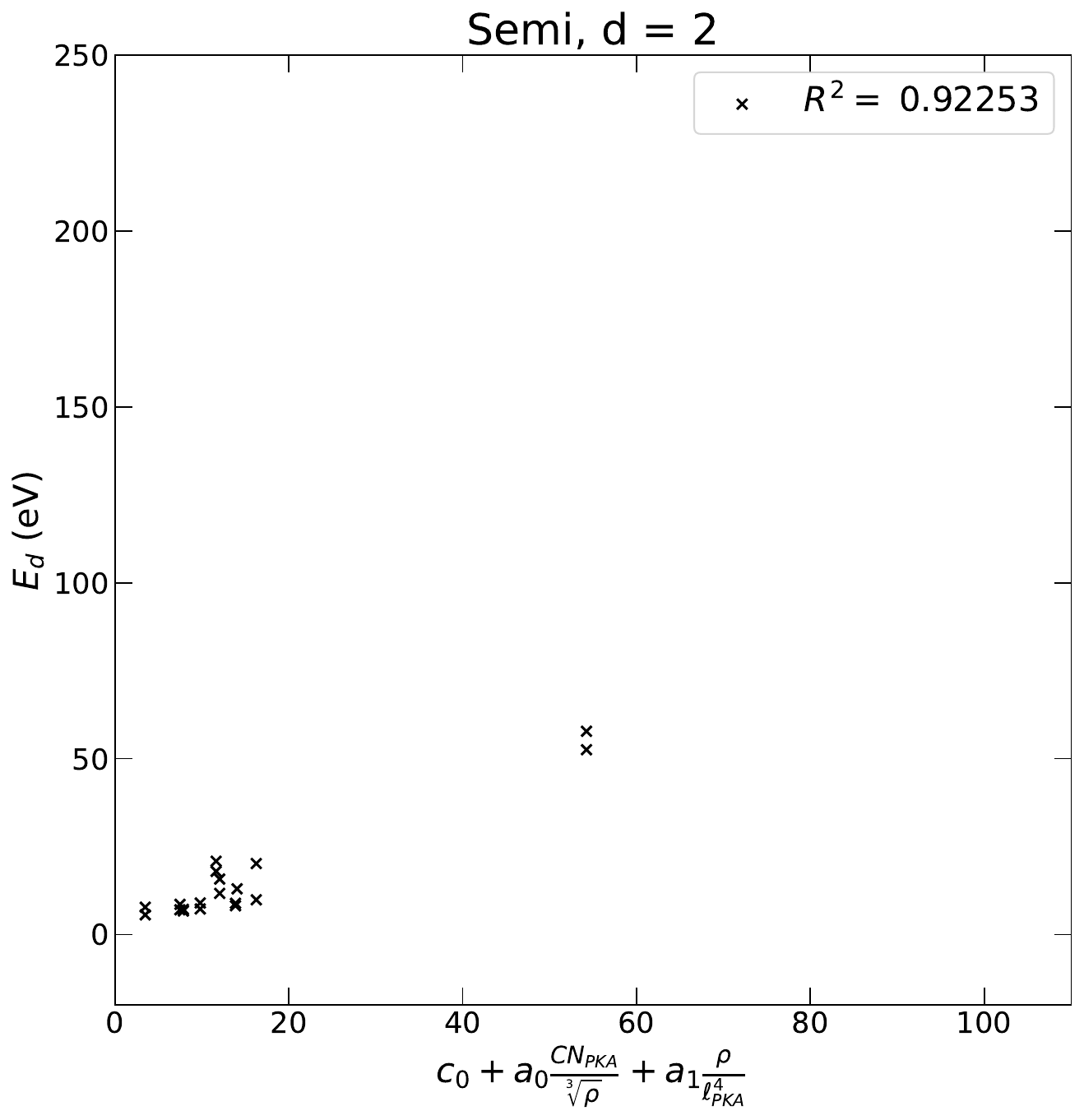}  
    \caption{Plots of $E_d$ data versus $E_d$ $K$ models for each corresponding subset and dimension for polyatomic materials.}
    \label{R_squared_figure_K_poly}
\end{figure*}

Similar to Appendix B, this section presents the constant values in Table~\ref{constants_table_poly} and $R^2$  plots for the polyatomic results in Fig~\ref{R_squared_figure_K_poly}.
The corresponding functions for each model with $\text{d}=2$ are indicated on the horizontal axis of Fig.~\ref{R_squared_figure_K_poly}.
\begin{table}[t]
    \centering
    \setlength\extrarowheight{7pt}
    \begin{tabular}{c c c c c}
    \Xhline{1pt}     
       Label & d & $c_0$ & $a_0$ & $a_1$  \\ \hline
              All & 1 & $2.5405299\cdot10$ & $7.3247623\cdot10^{-3}$ & 0
       \\
            All & 2 & $3.3290184\cdot10$ & $-4.0750820$ & $4.9693342\cdot10^{-5}$
       \\
            Exp & 1 & $2.9356494\cdot10$ & $1.8767521\cdot10^{-6}$ & 0
       \\
             Exp & 2 & $1.1104710\cdot10$ & $2.4725480\cdot10^{3}$ & $2.2314192\cdot10^{-6}$
       \\
            Sim & 1 & $2.3727270\cdot10$ & $2.6732752\cdot10^{-1}$ & 0
       \\    
            Sim & 2 & $3.1431970\cdot10$ & $-1.5094931\cdot10^{-1}$ & $3.6064510\cdot10^{-1}$
       \\    
            Cer & 1 & $-2.9235180\cdot10$ & $7.0664236\cdot10$ & 0
       \\   
            Cer & 2 & $-3.7037252\cdot10$ & $4.9722505\cdot10^{-1}$ & $8.5358349\cdot10$
       \\  
             Alloys & 1 & $2.8488816\cdot10$ & $1.4321412\cdot10^{-3}$ & 0
       \\       
            Alloys & 2 & $4.0181193\cdot10$ & $-2.7090534\cdot10^{4}$ & $1.6870308\cdot10^{-3}$
       \\   
            Semi & 1 & $-6.2188096$ & $1.4562305\cdot10^{2}$ & 0
       \\  
            Semi & 2 & $1.7316019\cdot10$ & $1.4562305\cdot10^{2}$ & $1.5899731\cdot10^{2}$
       \\ 
     \Xhline{1pt}         
    \end{tabular}
    \caption{Constants $c_0$ and $a_i$ for SISSO models' for polyatomic materials with different dimension.
    }
    \label{constants_table_poly}
\end{table}

\section*{Appendix G: Minimum Threshold Displacement Energy in Monoatomic Materials}

\begin{figure*}[b]
    \includegraphics[width = 17cm]{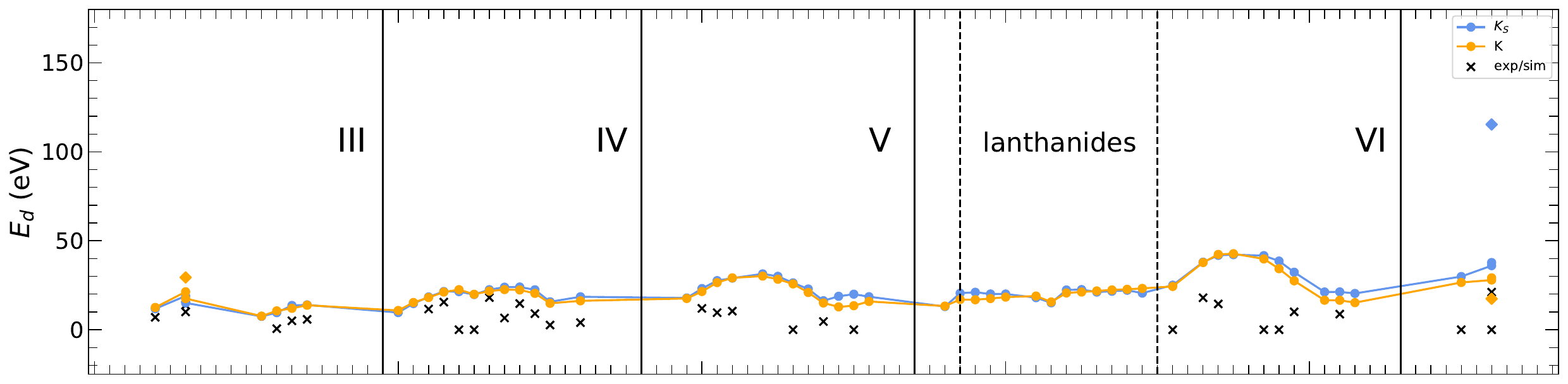}
    \caption{Predicted (connected point markers) and measured from experiments and simulations (cross markers) minimum threshold displacement energy ($E_d$) values as a function of the atomic number ($Z$), corresponding to the predictive functions in Table~\ref{r_squared_table_Edmin}.
    The diamond-shaped marker represents different materials that share the same $Z$, specifically, graphite and diamond (diamond marker), as well as gamma uranium and alpha uranium (diamond marker).
    The points for the different models are connected for better visibility.
    Vertical lines serve as references to denote each row in the periodic table, labeled with Roman numerals.
    The dashed lines serve as references to denote the lanthanides section in the periodic table. 
    }
    \label{Monoatom_Edmin_Z_Figure}
\end{figure*}

Similarly as for the average $E_d$, we apply SISSO to identify optimal functions for describing the minimum $E_d$ by using the minimum values from our datasets. 
Once again, the SISSO model $K_\text{S}$ achieved a slightly higher $R^2$ compared to Konobeyev’s et al model $K$, as shwon in Table~\ref{r_squared_table_Edmin}. 
On the other hand, Fig.~\ref{Monoatom_Edmin_Z_Figure} shows the minimum $E_d$ values as function of $Z$.
Model $K_\text{SR}$ is omitted since $K_\text{S}$ includes only variables from the reduced feature set.
\begin{table}[t]
    \centering
    \setlength{\tabcolsep}{4pt}
    \setlength\extrarowheight{9pt}
    \begin{tabular}{c c c}
    \Xhline{1pt}     
       Label  & $E_d$ & $R^2$  \\ \hline
                      $K$  & $c_0 + a_0 \sqrt{\rho T_{\text{melt}}} $ & 0.74 \\
                 $K_\text{S}$  & $c_0 + a_0 \sqrt{E_{\text{coh}}} \sqrt[3]{\rho} $ & 0.78 \\
        \vspace{-3ex}  
        \\
     \Xhline{1pt}         
    \end{tabular}
    \caption{Minium threshold displacement energy ($E_d$) models, dataset subdivision labels ($M_i$, $K$, $K_{\text{S}}$), and coefficient of determination ($R^2$) for models.
    The constants $a_i$ and $c_0$ are unique for each function and are reported in Table~\ref{constants_table_Edmin}. 
    }
    \label{r_squared_table_Edmin}
\end{table}
\begin{table}[t]
    \centering
    \setlength\extrarowheight{7pt}
    \begin{tabular}{c c c}
    \Xhline{1pt}     
       Label & $c_0$ & $a_0$ \\ \hline
              $K$ &  5 & $0.14$ 
       \\
                    $K_{\text{SR}}$ & 4.2496767 & $3.15969724$
       \\
     \Xhline{1pt}         
    \end{tabular}
    \caption{Constants $c_0$ and $a_0$ for the minimum $E_d$ models' in Table~\ref{r_squared_table_Edmin}.
    }
    \label{constants_table_Edmin}
\end{table}
\clearpage
\bibliography{Bibliography}

\begin{thebibliography}{96}%
\makeatletter
\providecommand \@ifxundefined [1]{%
 \@ifx{#1\undefined}
}%
\providecommand \@ifnum [1]{%
 \ifnum #1\expandafter \@firstoftwo
 \else \expandafter \@secondoftwo
 \fi
}%
\providecommand \@ifx [1]{%
 \ifx #1\expandafter \@firstoftwo
 \else \expandafter \@secondoftwo
 \fi
}%
\providecommand \natexlab [1]{#1}%
\providecommand \enquote  [1]{``#1''}%
\providecommand \bibnamefont  [1]{#1}%
\providecommand \bibfnamefont [1]{#1}%
\providecommand \citenamefont [1]{#1}%
\providecommand \href@noop [0]{\@secondoftwo}%
\providecommand \href [0]{\begingroup \@sanitize@url \@href}%
\providecommand \@href[1]{\@@startlink{#1}\@@href}%
\providecommand \@@href[1]{\endgroup#1\@@endlink}%
\providecommand \@sanitize@url [0]{\catcode `\\12\catcode `\$12\catcode `\&12\catcode `\#12\catcode `\^12\catcode `\_12\catcode `\%12\relax}%
\providecommand \@@startlink[1]{}%
\providecommand \@@endlink[0]{}%
\providecommand \url  [0]{\begingroup\@sanitize@url \@url }%
\providecommand \@url [1]{\endgroup\@href {#1}{\urlprefix }}%
\providecommand \urlprefix  [0]{URL }%
\providecommand \Eprint [0]{\href }%
\providecommand \doibase [0]{https://doi.org/}%
\providecommand \selectlanguage [0]{\@gobble}%
\providecommand \bibinfo  [0]{\@secondoftwo}%
\providecommand \bibfield  [0]{\@secondoftwo}%
\providecommand \translation [1]{[#1]}%
\providecommand \BibitemOpen [0]{}%
\providecommand \bibitemStop [0]{}%
\providecommand \bibitemNoStop [0]{.\EOS\space}%
\providecommand \EOS [0]{\spacefactor3000\relax}%
\providecommand \BibitemShut  [1]{\csname bibitem#1\endcsname}%
\let\auto@bib@innerbib\@empty
\bibitem [{\citenamefont {Dacus}\ \emph {et~al.}(2019)\citenamefont {Dacus}, \citenamefont {Beeler},\ and\ \citenamefont {Schwen}}]{dacus2019calculation}%
  \BibitemOpen
  \bibfield  {author} {\bibinfo {author} {\bibfnamefont {B.}~\bibnamefont {Dacus}}, \bibinfo {author} {\bibfnamefont {B.}~\bibnamefont {Beeler}},\ and\ \bibinfo {author} {\bibfnamefont {D.}~\bibnamefont {Schwen}},\ }\bibfield  {title} {\bibinfo {title} {Calculation of threshold displacement energies in $\text{UO}_2$},\ }\href@noop {} {\bibfield  {journal} {\bibinfo  {journal} {J. Nucl. Mater.}\ }\textbf {\bibinfo {volume} {520}},\ \bibinfo {pages} {152} (\bibinfo {year} {2019})}\BibitemShut {NoStop}%
\bibitem [{\citenamefont {Chen}\ \emph {et~al.}(2021)\citenamefont {Chen}, \citenamefont {Bernard}, \citenamefont {Tamagno}, \citenamefont {Litaize},\ and\ \citenamefont {Blaise}}]{chen2021uncertainty}%
  \BibitemOpen
  \bibfield  {author} {\bibinfo {author} {\bibfnamefont {S.}~\bibnamefont {Chen}}, \bibinfo {author} {\bibfnamefont {D.}~\bibnamefont {Bernard}}, \bibinfo {author} {\bibfnamefont {L.}~\bibnamefont {Tamagno}}, \bibinfo {author} {\bibfnamefont {O.}~\bibnamefont {Litaize}},\ and\ \bibinfo {author} {\bibfnamefont {P.}~\bibnamefont {Blaise}},\ }\bibfield  {title} {\bibinfo {title} {Uncertainty assessment for the displacement damage of a pressurized water reactor vessel},\ }\href@noop {} {\bibfield  {journal} {\bibinfo  {journal} {Nucl. Mater. Energy}\ }\textbf {\bibinfo {volume} {28}},\ \bibinfo {pages} {101017} (\bibinfo {year} {2021})}\BibitemShut {NoStop}%
\bibitem [{\citenamefont {Bany~Salman}\ \emph {et~al.}(2023)\citenamefont {Bany~Salman}, \citenamefont {Park},\ and\ \citenamefont {Banisalman}}]{bany2023atomistic}%
  \BibitemOpen
  \bibfield  {author} {\bibinfo {author} {\bibfnamefont {M.}~\bibnamefont {Bany~Salman}}, \bibinfo {author} {\bibfnamefont {M.}~\bibnamefont {Park}},\ and\ \bibinfo {author} {\bibfnamefont {M.~J.}\ \bibnamefont {Banisalman}},\ }\bibfield  {title} {\bibinfo {title} {Atomistic study for the tantalum and tantalum--tungsten alloy threshold displacement energy under local strain},\ }\href@noop {} {\bibfield  {journal} {\bibinfo  {journal} {Int. J. Mol. Sci.}\ }\textbf {\bibinfo {volume} {24}},\ \bibinfo {pages} {3289} (\bibinfo {year} {2023})}\BibitemShut {NoStop}%
\bibitem [{\citenamefont {Fang}\ \emph {et~al.}(2022)\citenamefont {Fang}, \citenamefont {Zhan}, \citenamefont {Xie}, \citenamefont {Du}, \citenamefont {Chen}, \citenamefont {Zeng}, \citenamefont {Zhang}, \citenamefont {Chen}, \citenamefont {Wang}, \citenamefont {Liang} \emph {et~al.}}]{fang2022integration}%
  \BibitemOpen
  \bibfield  {author} {\bibinfo {author} {\bibfnamefont {Y.}~\bibnamefont {Fang}}, \bibinfo {author} {\bibfnamefont {Y.}~\bibnamefont {Zhan}}, \bibinfo {author} {\bibfnamefont {Y.}~\bibnamefont {Xie}}, \bibinfo {author} {\bibfnamefont {S.}~\bibnamefont {Du}}, \bibinfo {author} {\bibfnamefont {Y.}~\bibnamefont {Chen}}, \bibinfo {author} {\bibfnamefont {Z.}~\bibnamefont {Zeng}}, \bibinfo {author} {\bibfnamefont {Y.}~\bibnamefont {Zhang}}, \bibinfo {author} {\bibfnamefont {K.}~\bibnamefont {Chen}}, \bibinfo {author} {\bibfnamefont {Y.}~\bibnamefont {Wang}}, \bibinfo {author} {\bibfnamefont {L.}~\bibnamefont {Liang}}, \emph {et~al.},\ }\bibfield  {title} {\bibinfo {title} {Integration of glucose and cardiolipin anabolism confers radiation resistance of hcc},\ }\href@noop {} {\bibfield  {journal} {\bibinfo  {journal} {Hepatology}\ }\textbf {\bibinfo {volume} {75}},\ \bibinfo {pages} {1386} (\bibinfo {year} {2022})}\BibitemShut {NoStop}%
\bibitem [{\citenamefont {Lu}\ \emph {et~al.}(2021)\citenamefont {Lu}, \citenamefont {Sun}, \citenamefont {Lu}, \citenamefont {Wu},\ and\ \citenamefont {Huang}}]{lu2021high}%
  \BibitemOpen
  \bibfield  {author} {\bibinfo {author} {\bibfnamefont {L.}~\bibnamefont {Lu}}, \bibinfo {author} {\bibfnamefont {M.}~\bibnamefont {Sun}}, \bibinfo {author} {\bibfnamefont {Q.}~\bibnamefont {Lu}}, \bibinfo {author} {\bibfnamefont {T.}~\bibnamefont {Wu}},\ and\ \bibinfo {author} {\bibfnamefont {B.}~\bibnamefont {Huang}},\ }\bibfield  {title} {\bibinfo {title} {High energy x-ray radiation sensitive scintillating materials for medical imaging, cancer diagnosis and therapy},\ }\href@noop {} {\bibfield  {journal} {\bibinfo  {journal} {Nano Energy}\ }\textbf {\bibinfo {volume} {79}},\ \bibinfo {pages} {105437} (\bibinfo {year} {2021})}\BibitemShut {NoStop}%
\bibitem [{\citenamefont {De~Cesare}\ \emph {et~al.}(2020)\citenamefont {De~Cesare}, \citenamefont {Savino}, \citenamefont {Ceglia}, \citenamefont {Alfano}, \citenamefont {Di~Carolo}, \citenamefont {French}, \citenamefont {Rapagnani}, \citenamefont {Gravina}, \citenamefont {Cipullo}, \citenamefont {Del~Vecchio} \emph {et~al.}}]{de2020applied}%
  \BibitemOpen
  \bibfield  {author} {\bibinfo {author} {\bibfnamefont {M.}~\bibnamefont {De~Cesare}}, \bibinfo {author} {\bibfnamefont {L.}~\bibnamefont {Savino}}, \bibinfo {author} {\bibfnamefont {G.}~\bibnamefont {Ceglia}}, \bibinfo {author} {\bibfnamefont {D.}~\bibnamefont {Alfano}}, \bibinfo {author} {\bibfnamefont {F.}~\bibnamefont {Di~Carolo}}, \bibinfo {author} {\bibfnamefont {A.}~\bibnamefont {French}}, \bibinfo {author} {\bibfnamefont {D.}~\bibnamefont {Rapagnani}}, \bibinfo {author} {\bibfnamefont {S.}~\bibnamefont {Gravina}}, \bibinfo {author} {\bibfnamefont {A.}~\bibnamefont {Cipullo}}, \bibinfo {author} {\bibfnamefont {A.}~\bibnamefont {Del~Vecchio}}, \emph {et~al.},\ }\bibfield  {title} {\bibinfo {title} {Applied radiation physics techniques for diagnostic evaluation of the plasma wind and thermal protection system critical parameters in aerospace re-entry},\ }\href@noop {} {\bibfield  {journal} {\bibinfo  {journal} {Prog. Aerosp. Sci.}\ }\textbf {\bibinfo {volume} {112}},\ \bibinfo {pages} {100550} (\bibinfo
  {year} {2020})}\BibitemShut {NoStop}%
\bibitem [{\citenamefont {Benton}\ and\ \citenamefont {Benton}(2001)}]{benton2001space}%
  \BibitemOpen
  \bibfield  {author} {\bibinfo {author} {\bibfnamefont {E.~R.}\ \bibnamefont {Benton}}\ and\ \bibinfo {author} {\bibfnamefont {E.}~\bibnamefont {Benton}},\ }\bibfield  {title} {\bibinfo {title} {Space radiation dosimetry in low-earth orbit and beyond},\ }\href@noop {} {\bibfield  {journal} {\bibinfo  {journal} {Nucl. Instrum. Methods Phys. Rev. B}\ }\textbf {\bibinfo {volume} {184}},\ \bibinfo {pages} {255} (\bibinfo {year} {2001})}\BibitemShut {NoStop}%
\bibitem [{\citenamefont {Miyazawa}\ \emph {et~al.}(2018)\citenamefont {Miyazawa}, \citenamefont {Ikegami}, \citenamefont {Chen}, \citenamefont {Ohshima}, \citenamefont {Imaizumi}, \citenamefont {Hirose},\ and\ \citenamefont {Miyasaka}}]{miyazawa2018tolerance}%
  \BibitemOpen
  \bibfield  {author} {\bibinfo {author} {\bibfnamefont {Y.}~\bibnamefont {Miyazawa}}, \bibinfo {author} {\bibfnamefont {M.}~\bibnamefont {Ikegami}}, \bibinfo {author} {\bibfnamefont {H.-W.}\ \bibnamefont {Chen}}, \bibinfo {author} {\bibfnamefont {T.}~\bibnamefont {Ohshima}}, \bibinfo {author} {\bibfnamefont {M.}~\bibnamefont {Imaizumi}}, \bibinfo {author} {\bibfnamefont {K.}~\bibnamefont {Hirose}},\ and\ \bibinfo {author} {\bibfnamefont {T.}~\bibnamefont {Miyasaka}},\ }\bibfield  {title} {\bibinfo {title} {Tolerance of perovskite solar cell to high-energy particle irradiations in space environment},\ }\href@noop {} {\bibfield  {journal} {\bibinfo  {journal} {IScience}\ }\textbf {\bibinfo {volume} {2}},\ \bibinfo {pages} {148} (\bibinfo {year} {2018})}\BibitemShut {NoStop}%
\bibitem [{\citenamefont {Kirmani}\ \emph {et~al.}(2022)\citenamefont {Kirmani}, \citenamefont {Durant}, \citenamefont {Grandidier}, \citenamefont {Haegel}, \citenamefont {Kelzenberg}, \citenamefont {Lao}, \citenamefont {McGehee}, \citenamefont {McMillon-Brown}, \citenamefont {Ostrowski}, \citenamefont {Peshek} \emph {et~al.}}]{kirmani2022countdown}%
  \BibitemOpen
  \bibfield  {author} {\bibinfo {author} {\bibfnamefont {A.~R.}\ \bibnamefont {Kirmani}}, \bibinfo {author} {\bibfnamefont {B.~K.}\ \bibnamefont {Durant}}, \bibinfo {author} {\bibfnamefont {J.}~\bibnamefont {Grandidier}}, \bibinfo {author} {\bibfnamefont {N.~M.}\ \bibnamefont {Haegel}}, \bibinfo {author} {\bibfnamefont {M.~D.}\ \bibnamefont {Kelzenberg}}, \bibinfo {author} {\bibfnamefont {Y.~M.}\ \bibnamefont {Lao}}, \bibinfo {author} {\bibfnamefont {M.~D.}\ \bibnamefont {McGehee}}, \bibinfo {author} {\bibfnamefont {L.}~\bibnamefont {McMillon-Brown}}, \bibinfo {author} {\bibfnamefont {D.~P.}\ \bibnamefont {Ostrowski}}, \bibinfo {author} {\bibfnamefont {T.~J.}\ \bibnamefont {Peshek}}, \emph {et~al.},\ }\bibfield  {title} {\bibinfo {title} {Countdown to perovskite space launch: Guidelines to performing relevant radiation-hardness experiments},\ }\href@noop {} {\bibfield  {journal} {\bibinfo  {journal} {Joule}\ } (\bibinfo {year} {2022})}\BibitemShut {NoStop}%
\bibitem [{\citenamefont {Kirmani}\ \emph {et~al.}(2024)\citenamefont {Kirmani}, \citenamefont {Byers}, \citenamefont {Ni}, \citenamefont {VanSant}, \citenamefont {Saini}, \citenamefont {Scheidt}, \citenamefont {Zheng}, \citenamefont {Kum}, \citenamefont {Sellers}, \citenamefont {McMillon-Brown} \emph {et~al.}}]{kirmani2024unraveling}%
  \BibitemOpen
  \bibfield  {author} {\bibinfo {author} {\bibfnamefont {A.~R.}\ \bibnamefont {Kirmani}}, \bibinfo {author} {\bibfnamefont {T.~A.}\ \bibnamefont {Byers}}, \bibinfo {author} {\bibfnamefont {Z.}~\bibnamefont {Ni}}, \bibinfo {author} {\bibfnamefont {K.}~\bibnamefont {VanSant}}, \bibinfo {author} {\bibfnamefont {D.~K.}\ \bibnamefont {Saini}}, \bibinfo {author} {\bibfnamefont {R.}~\bibnamefont {Scheidt}}, \bibinfo {author} {\bibfnamefont {X.}~\bibnamefont {Zheng}}, \bibinfo {author} {\bibfnamefont {T.~B.}\ \bibnamefont {Kum}}, \bibinfo {author} {\bibfnamefont {I.~R.}\ \bibnamefont {Sellers}}, \bibinfo {author} {\bibfnamefont {L.}~\bibnamefont {McMillon-Brown}}, \emph {et~al.},\ }\bibfield  {title} {\bibinfo {title} {Unraveling radiation damage and healing mechanisms in halide perovskites using energy-tuned dual irradiation dosing},\ }\href@noop {} {\bibfield  {journal} {\bibinfo  {journal} {Nat. Commun.}\ }\textbf {\bibinfo {volume} {15}},\ \bibinfo {pages} {696} (\bibinfo {year} {2024})}\BibitemShut {NoStop}%
\bibitem [{\citenamefont {Lang}\ \emph {et~al.}(2020)\citenamefont {Lang}, \citenamefont {Jo{\v{s}}t}, \citenamefont {Frohna}, \citenamefont {K{\"o}hnen}, \citenamefont {Al-Ashouri}, \citenamefont {Bowman}, \citenamefont {Bertram}, \citenamefont {Morales-Vilches}, \citenamefont {Koushik}, \citenamefont {Tennyson} \emph {et~al.}}]{lang2020proton}%
  \BibitemOpen
  \bibfield  {author} {\bibinfo {author} {\bibfnamefont {F.}~\bibnamefont {Lang}}, \bibinfo {author} {\bibfnamefont {M.}~\bibnamefont {Jo{\v{s}}t}}, \bibinfo {author} {\bibfnamefont {K.}~\bibnamefont {Frohna}}, \bibinfo {author} {\bibfnamefont {E.}~\bibnamefont {K{\"o}hnen}}, \bibinfo {author} {\bibfnamefont {A.}~\bibnamefont {Al-Ashouri}}, \bibinfo {author} {\bibfnamefont {A.~R.}\ \bibnamefont {Bowman}}, \bibinfo {author} {\bibfnamefont {T.}~\bibnamefont {Bertram}}, \bibinfo {author} {\bibfnamefont {A.~B.}\ \bibnamefont {Morales-Vilches}}, \bibinfo {author} {\bibfnamefont {D.}~\bibnamefont {Koushik}}, \bibinfo {author} {\bibfnamefont {E.~M.}\ \bibnamefont {Tennyson}}, \emph {et~al.},\ }\bibfield  {title} {\bibinfo {title} {Proton radiation hardness of perovskite tandem photovoltaics},\ }\href@noop {} {\bibfield  {journal} {\bibinfo  {journal} {Joule}\ } (\bibinfo {year} {2020})}\BibitemShut {NoStop}%
\bibitem [{\citenamefont {Nordlund}\ \emph {et~al.}(2015)\citenamefont {Nordlund}, \citenamefont {Sand}, \citenamefont {Granberg}, \citenamefont {Zinkle}, \citenamefont {Stoller}, \citenamefont {Averback}, \citenamefont {Suzudo}, \citenamefont {Malerba}, \citenamefont {Banhart}, \citenamefont {Weber} \emph {et~al.}}]{nordlund2015primary}%
  \BibitemOpen
  \bibfield  {author} {\bibinfo {author} {\bibfnamefont {K.}~\bibnamefont {Nordlund}}, \bibinfo {author} {\bibfnamefont {A.~E.}\ \bibnamefont {Sand}}, \bibinfo {author} {\bibfnamefont {F.}~\bibnamefont {Granberg}}, \bibinfo {author} {\bibfnamefont {S.~J.}\ \bibnamefont {Zinkle}}, \bibinfo {author} {\bibfnamefont {R.}~\bibnamefont {Stoller}}, \bibinfo {author} {\bibfnamefont {R.~S.}\ \bibnamefont {Averback}}, \bibinfo {author} {\bibfnamefont {T.}~\bibnamefont {Suzudo}}, \bibinfo {author} {\bibfnamefont {L.}~\bibnamefont {Malerba}}, \bibinfo {author} {\bibfnamefont {F.}~\bibnamefont {Banhart}}, \bibinfo {author} {\bibfnamefont {W.~J.}\ \bibnamefont {Weber}}, \emph {et~al.},\ }\bibfield  {title} {\bibinfo {title} {Primary radiation damage in materials. review of current understanding and proposed new standard displacement damage model to incorporate in cascade defect production efficiency and mixing effects},\ }\href@noop {} {\bibfield  {journal} {\bibinfo  {journal} {OECD}\ } (\bibinfo {year}
  {2015})}\BibitemShut {NoStop}%
\bibitem [{\citenamefont {Nordlund}\ \emph {et~al.}(2018{\natexlab{a}})\citenamefont {Nordlund}, \citenamefont {Zinkle}, \citenamefont {Sand}, \citenamefont {Granberg}, \citenamefont {Averback}, \citenamefont {Stoller}, \citenamefont {Suzudo}, \citenamefont {Malerba}, \citenamefont {Banhart}, \citenamefont {Weber} \emph {et~al.}}]{nordlund2018improving}%
  \BibitemOpen
  \bibfield  {author} {\bibinfo {author} {\bibfnamefont {K.}~\bibnamefont {Nordlund}}, \bibinfo {author} {\bibfnamefont {S.~J.}\ \bibnamefont {Zinkle}}, \bibinfo {author} {\bibfnamefont {A.~E.}\ \bibnamefont {Sand}}, \bibinfo {author} {\bibfnamefont {F.}~\bibnamefont {Granberg}}, \bibinfo {author} {\bibfnamefont {R.~S.}\ \bibnamefont {Averback}}, \bibinfo {author} {\bibfnamefont {R.}~\bibnamefont {Stoller}}, \bibinfo {author} {\bibfnamefont {T.}~\bibnamefont {Suzudo}}, \bibinfo {author} {\bibfnamefont {L.}~\bibnamefont {Malerba}}, \bibinfo {author} {\bibfnamefont {F.}~\bibnamefont {Banhart}}, \bibinfo {author} {\bibfnamefont {W.~J.}\ \bibnamefont {Weber}}, \emph {et~al.},\ }\bibfield  {title} {\bibinfo {title} {Improving atomic displacement and replacement calculations with physically realistic damage models},\ }\href@noop {} {\bibfield  {journal} {\bibinfo  {journal} {Nat. Commun.}\ }\textbf {\bibinfo {volume} {9}},\ \bibinfo {pages} {1084} (\bibinfo {year} {2018}{\natexlab{a}})}\BibitemShut {NoStop}%
\bibitem [{\citenamefont {Ziegler}\ \emph {et~al.}(2010)\citenamefont {Ziegler}, \citenamefont {Ziegler},\ and\ \citenamefont {Biersack}}]{ziegler2010srim}%
  \BibitemOpen
  \bibfield  {author} {\bibinfo {author} {\bibfnamefont {J.~F.}\ \bibnamefont {Ziegler}}, \bibinfo {author} {\bibfnamefont {M.~D.}\ \bibnamefont {Ziegler}},\ and\ \bibinfo {author} {\bibfnamefont {J.~P.}\ \bibnamefont {Biersack}},\ }\bibfield  {title} {\bibinfo {title} {\text{SRIM}--the stopping and range of ions in matter (2010)},\ }\href@noop {} {\bibfield  {journal} {\bibinfo  {journal} {Nucl. Instrum. Methods Phys. Res. B}\ }\textbf {\bibinfo {volume} {268}},\ \bibinfo {pages} {1818} (\bibinfo {year} {2010})}\BibitemShut {NoStop}%
\bibitem [{\citenamefont {Norgett}\ \emph {et~al.}(1975)\citenamefont {Norgett}, \citenamefont {Robinson},\ and\ \citenamefont {Torrens}}]{norgett1975proposed}%
  \BibitemOpen
  \bibfield  {author} {\bibinfo {author} {\bibfnamefont {M.}~\bibnamefont {Norgett}}, \bibinfo {author} {\bibfnamefont {M.}~\bibnamefont {Robinson}},\ and\ \bibinfo {author} {\bibfnamefont {I.~M.}\ \bibnamefont {Torrens}},\ }\bibfield  {title} {\bibinfo {title} {A proposed method of calculating displacement dose rates},\ }\href@noop {} {\bibfield  {journal} {\bibinfo  {journal} {Nucl. Eng. Des.}\ }\textbf {\bibinfo {volume} {33}},\ \bibinfo {pages} {50} (\bibinfo {year} {1975})}\BibitemShut {NoStop}%
\bibitem [{\citenamefont {Kinchin}\ and\ \citenamefont {Pease}(1955)}]{kinchin1955displacement}%
  \BibitemOpen
  \bibfield  {author} {\bibinfo {author} {\bibfnamefont {G.}~\bibnamefont {Kinchin}}\ and\ \bibinfo {author} {\bibfnamefont {R.}~\bibnamefont {Pease}},\ }\bibfield  {title} {\bibinfo {title} {The displacement of atoms in solids by radiation},\ }\href@noop {} {\bibfield  {journal} {\bibinfo  {journal} {Rep. Prog. Phys.}\ }\textbf {\bibinfo {volume} {18}},\ \bibinfo {pages} {1} (\bibinfo {year} {1955})}\BibitemShut {NoStop}%
\bibitem [{\citenamefont {Ziegler}\ \emph {et~al.}(2008)\citenamefont {Ziegler}, \citenamefont {Biersack},\ and\ \citenamefont {Ziegler}}]{ziegler2008srim}%
  \BibitemOpen
  \bibfield  {author} {\bibinfo {author} {\bibfnamefont {J.}~\bibnamefont {Ziegler}}, \bibinfo {author} {\bibfnamefont {J.}~\bibnamefont {Biersack}},\ and\ \bibinfo {author} {\bibfnamefont {M.}~\bibnamefont {Ziegler}},\ }\href {https://books.google.com/books?id=Ox73wAEACAAJ} {\emph {\bibinfo {title} {\text{SRIM}, the Stopping and Range of Ions in Matter}}}\ (\bibinfo  {publisher} {SRIM Company},\ \bibinfo {year} {2008})\BibitemShut {NoStop}%
\bibitem [{\citenamefont {Robinson}(1990)}]{robinson1990temporal}%
  \BibitemOpen
  \bibfield  {author} {\bibinfo {author} {\bibfnamefont {M.~T.}\ \bibnamefont {Robinson}},\ }\bibfield  {title} {\bibinfo {title} {The temporal development of collision cascades in the binary-collision approximation},\ }\href@noop {} {\bibfield  {journal} {\bibinfo  {journal} {Nucl. Instrum. Methods Phys. Res. B}\ }\textbf {\bibinfo {volume} {48}},\ \bibinfo {pages} {408} (\bibinfo {year} {1990})}\BibitemShut {NoStop}%
\bibitem [{\citenamefont {Was}\ \emph {et~al.}(2017)\citenamefont {Was} \emph {et~al.}}]{was2016fundamentals}%
  \BibitemOpen
  \bibfield  {author} {\bibinfo {author} {\bibfnamefont {G.~S.}\ \bibnamefont {Was}} \emph {et~al.},\ }\href@noop {} {\emph {\bibinfo {title} {Fundamentals of radiation materials science: metals and alloys}}},\ Vol.\ \bibinfo {volume} {510}\ (\bibinfo  {publisher} {Springer},\ \bibinfo {year} {2017})\BibitemShut {NoStop}%
\bibitem [{\citenamefont {Stoller}\ \emph {et~al.}(2013)\citenamefont {Stoller}, \citenamefont {Toloczko}, \citenamefont {Was}, \citenamefont {Certain}, \citenamefont {Dwaraknath},\ and\ \citenamefont {Garner}}]{stoller2013use}%
  \BibitemOpen
  \bibfield  {author} {\bibinfo {author} {\bibfnamefont {R.~E.}\ \bibnamefont {Stoller}}, \bibinfo {author} {\bibfnamefont {M.~B.}\ \bibnamefont {Toloczko}}, \bibinfo {author} {\bibfnamefont {G.~S.}\ \bibnamefont {Was}}, \bibinfo {author} {\bibfnamefont {A.~G.}\ \bibnamefont {Certain}}, \bibinfo {author} {\bibfnamefont {S.}~\bibnamefont {Dwaraknath}},\ and\ \bibinfo {author} {\bibfnamefont {F.~A.}\ \bibnamefont {Garner}},\ }\bibfield  {title} {\bibinfo {title} {On the use of \text{SRIM} for computing radiation damage exposure},\ }\href@noop {} {\bibfield  {journal} {\bibinfo  {journal} {Nucl. Instrum. Methods Phys. Rev., B}\ }\textbf {\bibinfo {volume} {310}},\ \bibinfo {pages} {75} (\bibinfo {year} {2013})}\BibitemShut {NoStop}%
\bibitem [{\citenamefont {Durant}\ \emph {et~al.}(2021)\citenamefont {Durant}, \citenamefont {Afshari}, \citenamefont {Sourabh}, \citenamefont {Yeddu}, \citenamefont {Bamidele}, \citenamefont {Singh}, \citenamefont {Rout}, \citenamefont {Eperon}, \citenamefont {Sellers} \emph {et~al.}}]{durant2021radiation}%
  \BibitemOpen
  \bibfield  {author} {\bibinfo {author} {\bibfnamefont {B.~K.}\ \bibnamefont {Durant}}, \bibinfo {author} {\bibfnamefont {H.}~\bibnamefont {Afshari}}, \bibinfo {author} {\bibfnamefont {S.}~\bibnamefont {Sourabh}}, \bibinfo {author} {\bibfnamefont {V.}~\bibnamefont {Yeddu}}, \bibinfo {author} {\bibfnamefont {M.~T.}\ \bibnamefont {Bamidele}}, \bibinfo {author} {\bibfnamefont {S.}~\bibnamefont {Singh}}, \bibinfo {author} {\bibfnamefont {B.}~\bibnamefont {Rout}}, \bibinfo {author} {\bibfnamefont {G.~E.}\ \bibnamefont {Eperon}}, \bibinfo {author} {\bibfnamefont {I.~R.}\ \bibnamefont {Sellers}}, \emph {et~al.},\ }\bibfield  {title} {\bibinfo {title} {Radiation stability of mixed tin--lead halide perovskites: Implications for space applications},\ }\href@noop {} {\bibfield  {journal} {\bibinfo  {journal} {Sol. Energy Mater Sol. Cells}\ }\textbf {\bibinfo {volume} {230}},\ \bibinfo {pages} {111232} (\bibinfo {year} {2021})}\BibitemShut {NoStop}%
\bibitem [{\citenamefont {Seitz}(1949)}]{seitz1949disordering}%
  \BibitemOpen
  \bibfield  {author} {\bibinfo {author} {\bibfnamefont {F.}~\bibnamefont {Seitz}},\ }\bibfield  {title} {\bibinfo {title} {On the disordering of solids by action of fast massive particles},\ }\href@noop {} {\bibfield  {journal} {\bibinfo  {journal} {Discuss. Faraday Soc.}\ }\textbf {\bibinfo {volume} {5}},\ \bibinfo {pages} {271} (\bibinfo {year} {1949})}\BibitemShut {NoStop}%
\bibitem [{\citenamefont {Zag}\ and\ \citenamefont {Urban}(1983)}]{zag1983temperature}%
  \BibitemOpen
  \bibfield  {author} {\bibinfo {author} {\bibfnamefont {W.}~\bibnamefont {Zag}}\ and\ \bibinfo {author} {\bibfnamefont {K.}~\bibnamefont {Urban}},\ }\bibfield  {title} {\bibinfo {title} {Temperature dependence of the threshold energy for atom displacement in irradiated molybdenum},\ }\href@noop {} {\bibfield  {journal} {\bibinfo  {journal} {Phys. Status Solidi A}\ }\textbf {\bibinfo {volume} {76}},\ \bibinfo {pages} {285} (\bibinfo {year} {1983})}\BibitemShut {NoStop}%
\bibitem [{\citenamefont {Saile}(1985)}]{saile1985temperature}%
  \BibitemOpen
  \bibfield  {author} {\bibinfo {author} {\bibfnamefont {B.}~\bibnamefont {Saile}},\ }\bibfield  {title} {\bibinfo {title} {The temperature dependence of the effective threshold energy for atom displacement in tantalum},\ }\href@noop {} {\bibfield  {journal} {\bibinfo  {journal} {Phys. Status Solidi A Appl. Res.}\ }\textbf {\bibinfo {volume} {89}},\ \bibinfo {pages} {K143} (\bibinfo {year} {1985})}\BibitemShut {NoStop}%
\bibitem [{\citenamefont {Chen}\ \emph {et~al.}(2019)\citenamefont {Chen}, \citenamefont {Deo},\ and\ \citenamefont {Dingreville}}]{chen2019atomistic}%
  \BibitemOpen
  \bibfield  {author} {\bibinfo {author} {\bibfnamefont {E.~Y.}\ \bibnamefont {Chen}}, \bibinfo {author} {\bibfnamefont {C.}~\bibnamefont {Deo}},\ and\ \bibinfo {author} {\bibfnamefont {R.}~\bibnamefont {Dingreville}},\ }\bibfield  {title} {\bibinfo {title} {Atomistic simulations of temperature and direction dependent threshold displacement energies in $\alpha$-and $\gamma$-uranium},\ }\href@noop {} {\bibfield  {journal} {\bibinfo  {journal} {Comput. Mater. Sci.}\ }\textbf {\bibinfo {volume} {157}},\ \bibinfo {pages} {75} (\bibinfo {year} {2019})}\BibitemShut {NoStop}%
\bibitem [{\citenamefont {Beeler}\ \emph {et~al.}(2016)\citenamefont {Beeler}, \citenamefont {Asta}, \citenamefont {Hosemann},\ and\ \citenamefont {Gr{\o}nbech-Jensen}}]{beeler2016effect}%
  \BibitemOpen
  \bibfield  {author} {\bibinfo {author} {\bibfnamefont {B.}~\bibnamefont {Beeler}}, \bibinfo {author} {\bibfnamefont {M.}~\bibnamefont {Asta}}, \bibinfo {author} {\bibfnamefont {P.}~\bibnamefont {Hosemann}},\ and\ \bibinfo {author} {\bibfnamefont {N.}~\bibnamefont {Gr{\o}nbech-Jensen}},\ }\bibfield  {title} {\bibinfo {title} {Effect of strain and temperature on the threshold displacement energy in body-centered cubic iron},\ }\href@noop {} {\bibfield  {journal} {\bibinfo  {journal} {J. Nucl. Mater.}\ }\textbf {\bibinfo {volume} {474}},\ \bibinfo {pages} {113} (\bibinfo {year} {2016})}\BibitemShut {NoStop}%
\bibitem [{\citenamefont {Banisalman}\ and\ \citenamefont {Oda}(2019)}]{banisalman2019atomistic}%
  \BibitemOpen
  \bibfield  {author} {\bibinfo {author} {\bibfnamefont {M.~J.}\ \bibnamefont {Banisalman}}\ and\ \bibinfo {author} {\bibfnamefont {T.}~\bibnamefont {Oda}},\ }\bibfield  {title} {\bibinfo {title} {Atomistic simulation for strain effects on threshold displacement energies in refractory metals},\ }\href@noop {} {\bibfield  {journal} {\bibinfo  {journal} {Comput. Mater. Sci.}\ }\textbf {\bibinfo {volume} {158}},\ \bibinfo {pages} {346} (\bibinfo {year} {2019})}\BibitemShut {NoStop}%
\bibitem [{\citenamefont {Wang}\ \emph {et~al.}(2016)\citenamefont {Wang}, \citenamefont {Gao}, \citenamefont {Wang}, \citenamefont {Gao}, \citenamefont {He}, \citenamefont {Cui}, \citenamefont {Pang},\ and\ \citenamefont {Zhu}}]{wang2016effect}%
  \BibitemOpen
  \bibfield  {author} {\bibinfo {author} {\bibfnamefont {D.}~\bibnamefont {Wang}}, \bibinfo {author} {\bibfnamefont {N.}~\bibnamefont {Gao}}, \bibinfo {author} {\bibfnamefont {Z.}~\bibnamefont {Wang}}, \bibinfo {author} {\bibfnamefont {X.}~\bibnamefont {Gao}}, \bibinfo {author} {\bibfnamefont {W.}~\bibnamefont {He}}, \bibinfo {author} {\bibfnamefont {M.}~\bibnamefont {Cui}}, \bibinfo {author} {\bibfnamefont {L.}~\bibnamefont {Pang}},\ and\ \bibinfo {author} {\bibfnamefont {Y.}~\bibnamefont {Zhu}},\ }\bibfield  {title} {\bibinfo {title} {Effect of strain field on displacement cascade in tungsten studied by molecular dynamics simulation},\ }\href@noop {} {\bibfield  {journal} {\bibinfo  {journal} {Nucl. Instrum. Methods Phys. Res., Sect. B}\ }\textbf {\bibinfo {volume} {384}},\ \bibinfo {pages} {68} (\bibinfo {year} {2016})}\BibitemShut {NoStop}%
\bibitem [{\citenamefont {Tikhonchev}\ and\ \citenamefont {Svetukhin}(2017)}]{tikhonchev2017threshold}%
  \BibitemOpen
  \bibfield  {author} {\bibinfo {author} {\bibfnamefont {M.~Y.}\ \bibnamefont {Tikhonchev}}\ and\ \bibinfo {author} {\bibfnamefont {V.~V.}\ \bibnamefont {Svetukhin}},\ }\bibfield  {title} {\bibinfo {title} {Threshold energies of atomic displacements in $\alpha$-$\text{Fe}$ under deformation: Molecular dynamics simulation},\ }\href@noop {} {\bibfield  {journal} {\bibinfo  {journal} {Tech. Phys. Lett.}\ }\textbf {\bibinfo {volume} {43}},\ \bibinfo {pages} {348} (\bibinfo {year} {2017})}\BibitemShut {NoStop}%
\bibitem [{\citenamefont {Zhao}\ \emph {et~al.}(2012)\citenamefont {Zhao}, \citenamefont {Xue}, \citenamefont {Lan}, \citenamefont {Sun}, \citenamefont {Wang},\ and\ \citenamefont {Yan}}]{zhao2012influence}%
  \BibitemOpen
  \bibfield  {author} {\bibinfo {author} {\bibfnamefont {S.}~\bibnamefont {Zhao}}, \bibinfo {author} {\bibfnamefont {J.}~\bibnamefont {Xue}}, \bibinfo {author} {\bibfnamefont {C.}~\bibnamefont {Lan}}, \bibinfo {author} {\bibfnamefont {L.}~\bibnamefont {Sun}}, \bibinfo {author} {\bibfnamefont {Y.}~\bibnamefont {Wang}},\ and\ \bibinfo {author} {\bibfnamefont {S.}~\bibnamefont {Yan}},\ }\bibfield  {title} {\bibinfo {title} {Influence of high pressure on the threshold displacement energies in silicon carbide: A car--parrinello molecular dynamics approach},\ }\href@noop {} {\bibfield  {journal} {\bibinfo  {journal} {Nucl. Instrum. Methods Phys. Res. Sect. B Beam Interact. Mater. Atoms}\ }\textbf {\bibinfo {volume} {286}},\ \bibinfo {pages} {119} (\bibinfo {year} {2012})}\BibitemShut {NoStop}%
\bibitem [{\citenamefont {Uglov}\ \emph {et~al.}(2015)\citenamefont {Uglov}, \citenamefont {Kvasov}, \citenamefont {Remnev},\ and\ \citenamefont {Polikarpov}}]{uglov2015physical}%
  \BibitemOpen
  \bibfield  {author} {\bibinfo {author} {\bibfnamefont {V.}~\bibnamefont {Uglov}}, \bibinfo {author} {\bibfnamefont {N.}~\bibnamefont {Kvasov}}, \bibinfo {author} {\bibfnamefont {G.}~\bibnamefont {Remnev}},\ and\ \bibinfo {author} {\bibfnamefont {R.}~\bibnamefont {Polikarpov}},\ }\bibfield  {title} {\bibinfo {title} {On the physical nature of the threshold displacement energy in radiation physics},\ }\href@noop {} {\bibfield  {journal} {\bibinfo  {journal} {J. Surf. Investig.}\ }\textbf {\bibinfo {volume} {9}},\ \bibinfo {pages} {1206} (\bibinfo {year} {2015})}\BibitemShut {NoStop}%
\bibitem [{\citenamefont {Zinkle}\ and\ \citenamefont {Kinoshita}(1997)}]{zinkle1997defect}%
  \BibitemOpen
  \bibfield  {author} {\bibinfo {author} {\bibfnamefont {S.}~\bibnamefont {Zinkle}}\ and\ \bibinfo {author} {\bibfnamefont {C.}~\bibnamefont {Kinoshita}},\ }\bibfield  {title} {\bibinfo {title} {Defect production in ceramics},\ }\href@noop {} {\bibfield  {journal} {\bibinfo  {journal} {J. Nucl. Mater.}\ }\textbf {\bibinfo {volume} {251}},\ \bibinfo {pages} {200} (\bibinfo {year} {1997})}\BibitemShut {NoStop}%
\bibitem [{\citenamefont {Nordlund}\ \emph {et~al.}(2018{\natexlab{b}})\citenamefont {Nordlund}, \citenamefont {Zinkle}, \citenamefont {Sand}, \citenamefont {Granberg}, \citenamefont {Averback}, \citenamefont {Stoller}, \citenamefont {Suzudo}, \citenamefont {Malerba}, \citenamefont {Banhart}, \citenamefont {Weber} \emph {et~al.}}]{nordlund2018primary}%
  \BibitemOpen
  \bibfield  {author} {\bibinfo {author} {\bibfnamefont {K.}~\bibnamefont {Nordlund}}, \bibinfo {author} {\bibfnamefont {S.~J.}\ \bibnamefont {Zinkle}}, \bibinfo {author} {\bibfnamefont {A.~E.}\ \bibnamefont {Sand}}, \bibinfo {author} {\bibfnamefont {F.}~\bibnamefont {Granberg}}, \bibinfo {author} {\bibfnamefont {R.~S.}\ \bibnamefont {Averback}}, \bibinfo {author} {\bibfnamefont {R.~E.}\ \bibnamefont {Stoller}}, \bibinfo {author} {\bibfnamefont {T.}~\bibnamefont {Suzudo}}, \bibinfo {author} {\bibfnamefont {L.}~\bibnamefont {Malerba}}, \bibinfo {author} {\bibfnamefont {F.}~\bibnamefont {Banhart}}, \bibinfo {author} {\bibfnamefont {W.~J.}\ \bibnamefont {Weber}}, \emph {et~al.},\ }\bibfield  {title} {\bibinfo {title} {Primary radiation damage: A review of current understanding and models},\ }\href@noop {} {\bibfield  {journal} {\bibinfo  {journal} {J. Nucl. Mater.}\ }\textbf {\bibinfo {volume} {512}},\ \bibinfo {pages} {450} (\bibinfo {year} {2018}{\natexlab{b}})}\BibitemShut {NoStop}%
\bibitem [{\citenamefont {Ghoniem}\ and\ \citenamefont {Chou}(1988)}]{ghoniem1988binary}%
  \BibitemOpen
  \bibfield  {author} {\bibinfo {author} {\bibfnamefont {N.~M.}\ \bibnamefont {Ghoniem}}\ and\ \bibinfo {author} {\bibfnamefont {S.}~\bibnamefont {Chou}},\ }\bibfield  {title} {\bibinfo {title} {Binary collision monte carlo simulations of cascades in polyatomic ceramics},\ }\href@noop {} {\bibfield  {journal} {\bibinfo  {journal} {J. Nucl. Mater.}\ }\textbf {\bibinfo {volume} {155}},\ \bibinfo {pages} {1263} (\bibinfo {year} {1988})}\BibitemShut {NoStop}%
\bibitem [{\citenamefont {Gibson}\ \emph {et~al.}(1960)\citenamefont {Gibson}, \citenamefont {Goland}, \citenamefont {Milgram},\ and\ \citenamefont {Vineyard}}]{gibson1960dynamics}%
  \BibitemOpen
  \bibfield  {author} {\bibinfo {author} {\bibfnamefont {J.}~\bibnamefont {Gibson}}, \bibinfo {author} {\bibfnamefont {A.~N.}\ \bibnamefont {Goland}}, \bibinfo {author} {\bibfnamefont {M.}~\bibnamefont {Milgram}},\ and\ \bibinfo {author} {\bibfnamefont {G.}~\bibnamefont {Vineyard}},\ }\bibfield  {title} {\bibinfo {title} {Dynamics of radiation damage},\ }\href@noop {} {\bibfield  {journal} {\bibinfo  {journal} {Phys. Rev.}\ }\textbf {\bibinfo {volume} {120}},\ \bibinfo {pages} {1229} (\bibinfo {year} {1960})}\BibitemShut {NoStop}%
\bibitem [{\citenamefont {Corbett}\ \emph {et~al.}(1957)\citenamefont {Corbett}, \citenamefont {Denney}, \citenamefont {Fiske},\ and\ \citenamefont {Walker}}]{corbett1957electron}%
  \BibitemOpen
  \bibfield  {author} {\bibinfo {author} {\bibfnamefont {J.}~\bibnamefont {Corbett}}, \bibinfo {author} {\bibfnamefont {J.}~\bibnamefont {Denney}}, \bibinfo {author} {\bibfnamefont {M.}~\bibnamefont {Fiske}},\ and\ \bibinfo {author} {\bibfnamefont {R.}~\bibnamefont {Walker}},\ }\bibfield  {title} {\bibinfo {title} {Electron irradiation of copper near 10 k},\ }\href@noop {} {\bibfield  {journal} {\bibinfo  {journal} {Phys. Rev.}\ }\textbf {\bibinfo {volume} {108}},\ \bibinfo {pages} {954} (\bibinfo {year} {1957})}\BibitemShut {NoStop}%
\bibitem [{\citenamefont {Arnold}\ and\ \citenamefont {Compton}(1960)}]{arnold1960threshold}%
  \BibitemOpen
  \bibfield  {author} {\bibinfo {author} {\bibfnamefont {G.~W.}\ \bibnamefont {Arnold}}\ and\ \bibinfo {author} {\bibfnamefont {W.~D.}\ \bibnamefont {Compton}},\ }\bibfield  {title} {\bibinfo {title} {Threshold energy for lattice displacement in $\alpha$-$\text{Al}_2\text{O}_3$},\ }\href@noop {} {\bibfield  {journal} {\bibinfo  {journal} {Phys. Rev. Lett.}\ }\textbf {\bibinfo {volume} {4}},\ \bibinfo {pages} {66} (\bibinfo {year} {1960})}\BibitemShut {NoStop}%
\bibitem [{\citenamefont {Sosin}(1962)}]{sosin1962energy}%
  \BibitemOpen
  \bibfield  {author} {\bibinfo {author} {\bibfnamefont {A.}~\bibnamefont {Sosin}},\ }\bibfield  {title} {\bibinfo {title} {Energy dependence of electron damage in copper},\ }\href@noop {} {\bibfield  {journal} {\bibinfo  {journal} {Phys. Rev.}\ }\textbf {\bibinfo {volume} {126}},\ \bibinfo {pages} {1698} (\bibinfo {year} {1962})}\BibitemShut {NoStop}%
\bibitem [{\citenamefont {Lucasson}\ and\ \citenamefont {Walker}(1962)}]{lucasson1962production}%
  \BibitemOpen
  \bibfield  {author} {\bibinfo {author} {\bibfnamefont {P.}~\bibnamefont {Lucasson}}\ and\ \bibinfo {author} {\bibfnamefont {R.}~\bibnamefont {Walker}},\ }\bibfield  {title} {\bibinfo {title} {Production and recovery of electron-induced radiation damage in a number of metals},\ }\href@noop {} {\bibfield  {journal} {\bibinfo  {journal} {Phys. Rev.}\ }\textbf {\bibinfo {volume} {127}},\ \bibinfo {pages} {485} (\bibinfo {year} {1962})}\BibitemShut {NoStop}%
\bibitem [{\citenamefont {Makin}(1968)}]{makin1968electron}%
  \BibitemOpen
  \bibfield  {author} {\bibinfo {author} {\bibfnamefont {M.}~\bibnamefont {Makin}},\ }\bibfield  {title} {\bibinfo {title} {Electron displacement damage in copper and aluminium in a high voltage electron microscope},\ }\href@noop {} {\bibfield  {journal} {\bibinfo  {journal} {Philos. Mag.}\ }\textbf {\bibinfo {volume} {18}},\ \bibinfo {pages} {637} (\bibinfo {year} {1968})}\BibitemShut {NoStop}%
\bibitem [{\citenamefont {Sharp}\ and\ \citenamefont {Rumsby}(1973)}]{sharp1973electron}%
  \BibitemOpen
  \bibfield  {author} {\bibinfo {author} {\bibfnamefont {J.}~\bibnamefont {Sharp}}\ and\ \bibinfo {author} {\bibfnamefont {D.}~\bibnamefont {Rumsby}},\ }\bibfield  {title} {\bibinfo {title} {Electron irradjation damage in magnesium oxide},\ }\href@noop {} {\bibfield  {journal} {\bibinfo  {journal} {Radiation Effects}\ }\textbf {\bibinfo {volume} {17}},\ \bibinfo {pages} {65} (\bibinfo {year} {1973})}\BibitemShut {NoStop}%
\bibitem [{\citenamefont {Kenik}\ and\ \citenamefont {Mitchell}(1975)}]{kenik1975orientation}%
  \BibitemOpen
  \bibfield  {author} {\bibinfo {author} {\bibfnamefont {E.}~\bibnamefont {Kenik}}\ and\ \bibinfo {author} {\bibfnamefont {T.}~\bibnamefont {Mitchell}},\ }\bibfield  {title} {\bibinfo {title} {Orientation dependence of the threshold displacement energy in copper and vanadium},\ }\href@noop {} {\bibfield  {journal} {\bibinfo  {journal} {Philos. Mag.}\ }\textbf {\bibinfo {volume} {32}},\ \bibinfo {pages} {815} (\bibinfo {year} {1975})}\BibitemShut {NoStop}%
\bibitem [{\citenamefont {Urban}\ and\ \citenamefont {Seeger}(1974)}]{urban1974radiation}%
  \BibitemOpen
  \bibfield  {author} {\bibinfo {author} {\bibfnamefont {K.}~\bibnamefont {Urban}}\ and\ \bibinfo {author} {\bibfnamefont {A.}~\bibnamefont {Seeger}},\ }\bibfield  {title} {\bibinfo {title} {Radiation-induced diffusion of point-defects during low-temperature electron irradiation},\ }\href@noop {} {\bibfield  {journal} {\bibinfo  {journal} {Philos. Mag.}\ }\textbf {\bibinfo {volume} {30}},\ \bibinfo {pages} {1395} (\bibinfo {year} {1974})}\BibitemShut {NoStop}%
\bibitem [{\citenamefont {Yoshiie}\ \emph {et~al.}(1979)\citenamefont {Yoshiie}, \citenamefont {Iwanaga}, \citenamefont {Shibata}, \citenamefont {Ichihara},\ and\ \citenamefont {Takeuchi}}]{yoshiie1979orientation}%
  \BibitemOpen
  \bibfield  {author} {\bibinfo {author} {\bibfnamefont {T.}~\bibnamefont {Yoshiie}}, \bibinfo {author} {\bibfnamefont {H.}~\bibnamefont {Iwanaga}}, \bibinfo {author} {\bibfnamefont {N.}~\bibnamefont {Shibata}}, \bibinfo {author} {\bibfnamefont {M.}~\bibnamefont {Ichihara}},\ and\ \bibinfo {author} {\bibfnamefont {S.}~\bibnamefont {Takeuchi}},\ }\bibfield  {title} {\bibinfo {title} {Orientation dependence of electron--irradiation damage in zinc oxide},\ }\href@noop {} {\bibfield  {journal} {\bibinfo  {journal} {Philos. Mag. A}\ }\textbf {\bibinfo {volume} {40}},\ \bibinfo {pages} {297} (\bibinfo {year} {1979})}\BibitemShut {NoStop}%
\bibitem [{\citenamefont {Caulfield}\ \emph {et~al.}(1995)\citenamefont {Caulfield}, \citenamefont {Cooper},\ and\ \citenamefont {Boas}}]{caulfield1995point}%
  \BibitemOpen
  \bibfield  {author} {\bibinfo {author} {\bibfnamefont {K.~J.}\ \bibnamefont {Caulfield}}, \bibinfo {author} {\bibfnamefont {R.}~\bibnamefont {Cooper}},\ and\ \bibinfo {author} {\bibfnamefont {J.~E.}\ \bibnamefont {Boas}},\ }\bibfield  {title} {\bibinfo {title} {Point defects in electron-irradiated oxide single crystals},\ }\href@noop {} {\bibfield  {journal} {\bibinfo  {journal} {J. Am. Ceram. Soc.}\ }\textbf {\bibinfo {volume} {78}},\ \bibinfo {pages} {1054} (\bibinfo {year} {1995})}\BibitemShut {NoStop}%
\bibitem [{\citenamefont {Smith}\ \emph {et~al.}(2000)\citenamefont {Smith}, \citenamefont {Cooper}, \citenamefont {Colella},\ and\ \citenamefont {Vance}}]{smith2000measured}%
  \BibitemOpen
  \bibfield  {author} {\bibinfo {author} {\bibfnamefont {K.~L.}\ \bibnamefont {Smith}}, \bibinfo {author} {\bibfnamefont {R.}~\bibnamefont {Cooper}}, \bibinfo {author} {\bibfnamefont {M.}~\bibnamefont {Colella}},\ and\ \bibinfo {author} {\bibfnamefont {E.~R.}\ \bibnamefont {Vance}},\ }\bibfield  {title} {\bibinfo {title} {Measured displacement energies of oxygen ions in zirconolite and rutile},\ }\href@noop {} {\bibfield  {journal} {\bibinfo  {journal} {MRS Online Proc. Libr.}\ }\textbf {\bibinfo {volume} {663}},\ \bibinfo {pages} {373} (\bibinfo {year} {2000})}\BibitemShut {NoStop}%
\bibitem [{\citenamefont {Smith}\ and\ \citenamefont {Zaluzec}(2005)}]{smith2005displacement}%
  \BibitemOpen
  \bibfield  {author} {\bibinfo {author} {\bibfnamefont {K.~L.}\ \bibnamefont {Smith}}\ and\ \bibinfo {author} {\bibfnamefont {N.~J.}\ \bibnamefont {Zaluzec}},\ }\bibfield  {title} {\bibinfo {title} {The displacement energies of cations in perovskite ($\text{CaTiO}_3$)},\ }\href@noop {} {\bibfield  {journal} {\bibinfo  {journal} {J. Nucl. Mater.}\ }\textbf {\bibinfo {volume} {336}},\ \bibinfo {pages} {261} (\bibinfo {year} {2005})}\BibitemShut {NoStop}%
\bibitem [{\citenamefont {Cooper}\ \emph {et~al.}(2001)\citenamefont {Cooper}, \citenamefont {Smith}, \citenamefont {Colella}, \citenamefont {Vance},\ and\ \citenamefont {Phillips}}]{cooper2001optical}%
  \BibitemOpen
  \bibfield  {author} {\bibinfo {author} {\bibfnamefont {R.}~\bibnamefont {Cooper}}, \bibinfo {author} {\bibfnamefont {K.}~\bibnamefont {Smith}}, \bibinfo {author} {\bibfnamefont {M.}~\bibnamefont {Colella}}, \bibinfo {author} {\bibfnamefont {E.}~\bibnamefont {Vance}},\ and\ \bibinfo {author} {\bibfnamefont {M.}~\bibnamefont {Phillips}},\ }\bibfield  {title} {\bibinfo {title} {Optical emission due to ionic displacements in alkaline earth titanates},\ }\href@noop {} {\bibfield  {journal} {\bibinfo  {journal} {J. Nucl. Mater.}\ }\textbf {\bibinfo {volume} {289}},\ \bibinfo {pages} {199} (\bibinfo {year} {2001})}\BibitemShut {NoStop}%
\bibitem [{\citenamefont {Torrens}\ \emph {et~al.}(1966)\citenamefont {Torrens}, \citenamefont {Chadderton},\ and\ \citenamefont {Morgan}}]{torrens1966ionic}%
  \BibitemOpen
  \bibfield  {author} {\bibinfo {author} {\bibfnamefont {I.~M.}\ \bibnamefont {Torrens}}, \bibinfo {author} {\bibfnamefont {L.~T.}\ \bibnamefont {Chadderton}},\ and\ \bibinfo {author} {\bibfnamefont {D.~V.}\ \bibnamefont {Morgan}},\ }\bibfield  {title} {\bibinfo {title} {Ionic displacement in the alkali halides},\ }\href@noop {} {\bibfield  {journal} {\bibinfo  {journal} {J. Appl. Phys.}\ }\textbf {\bibinfo {volume} {37}},\ \bibinfo {pages} {2395} (\bibinfo {year} {1966})}\BibitemShut {NoStop}%
\bibitem [{\citenamefont {Devanathan}\ and\ \citenamefont {Weber}(2000)}]{devanathan2000displacement}%
  \BibitemOpen
  \bibfield  {author} {\bibinfo {author} {\bibfnamefont {R.}~\bibnamefont {Devanathan}}\ and\ \bibinfo {author} {\bibfnamefont {W.~J.}\ \bibnamefont {Weber}},\ }\bibfield  {title} {\bibinfo {title} {Displacement energy surface in $3\text{C}$ and $6\text{H}$ $\text{SiC}$},\ }\href@noop {} {\bibfield  {journal} {\bibinfo  {journal} {J. Nucl. Mater.}\ }\textbf {\bibinfo {volume} {278}},\ \bibinfo {pages} {258} (\bibinfo {year} {2000})}\BibitemShut {NoStop}%
\bibitem [{\citenamefont {Lucas}\ and\ \citenamefont {Pizzagalli}(2005)}]{lucas2005ab}%
  \BibitemOpen
  \bibfield  {author} {\bibinfo {author} {\bibfnamefont {G.}~\bibnamefont {Lucas}}\ and\ \bibinfo {author} {\bibfnamefont {L.}~\bibnamefont {Pizzagalli}},\ }\bibfield  {title} {\bibinfo {title} {Ab initio molecular dynamics calculations of threshold displacement energies in silicon carbide},\ }\href@noop {} {\bibfield  {journal} {\bibinfo  {journal} {Phys. Rev. B}\ }\textbf {\bibinfo {volume} {72}},\ \bibinfo {pages} {161202} (\bibinfo {year} {2005})}\BibitemShut {NoStop}%
\bibitem [{\citenamefont {Liu}\ \emph {et~al.}(2015)\citenamefont {Liu}, \citenamefont {Yuan}, \citenamefont {Jin}, \citenamefont {Zhang},\ and\ \citenamefont {Weber}}]{liu2015ab}%
  \BibitemOpen
  \bibfield  {author} {\bibinfo {author} {\bibfnamefont {B.}~\bibnamefont {Liu}}, \bibinfo {author} {\bibfnamefont {F.}~\bibnamefont {Yuan}}, \bibinfo {author} {\bibfnamefont {K.}~\bibnamefont {Jin}}, \bibinfo {author} {\bibfnamefont {Y.}~\bibnamefont {Zhang}},\ and\ \bibinfo {author} {\bibfnamefont {W.~J.}\ \bibnamefont {Weber}},\ }\bibfield  {title} {\bibinfo {title} {Ab initio molecular dynamics investigations of low-energy recoil events in \text{Ni} and \text{NiCo}},\ }\href@noop {} {\bibfield  {journal} {\bibinfo  {journal} {J. Phys.: Condens. Matter}\ }\textbf {\bibinfo {volume} {27}},\ \bibinfo {pages} {435006} (\bibinfo {year} {2015})}\BibitemShut {NoStop}%
\bibitem [{\citenamefont {Yang}\ and\ \citenamefont {Olsson}(2021)}]{yang2021full}%
  \BibitemOpen
  \bibfield  {author} {\bibinfo {author} {\bibfnamefont {Q.}~\bibnamefont {Yang}}\ and\ \bibinfo {author} {\bibfnamefont {P.}~\bibnamefont {Olsson}},\ }\bibfield  {title} {\bibinfo {title} {Full energy range primary radiation damage model},\ }\href@noop {} {\bibfield  {journal} {\bibinfo  {journal} {Phys. Rev. Mater.}\ }\textbf {\bibinfo {volume} {5}},\ \bibinfo {pages} {073602} (\bibinfo {year} {2021})}\BibitemShut {NoStop}%
\bibitem [{\citenamefont {{ASTM International}}(1996)}]{ASTM_E521}%
  \BibitemOpen
  \bibfield  {author} {\bibinfo {author} {\bibnamefont {{ASTM International}}},\ }\href@noop {} {\bibinfo {title} {{Standard Practice for Neutron Radiation Damage Simulation by Charged-Particle Irradiation}}},\ \bibinfo {howpublished} {Available from: \url{https://www.astm.org/}} (\bibinfo {year} {1996}),\ \bibinfo {note} {accessed: 2024-05-10}\BibitemShut {NoStop}%
\bibitem [{\citenamefont {Merrill}\ \emph {et~al.}(2015)\citenamefont {Merrill}, \citenamefont {Cress}, \citenamefont {Rossi}, \citenamefont {Cox},\ and\ \citenamefont {Landi}}]{merrill2015threshold}%
  \BibitemOpen
  \bibfield  {author} {\bibinfo {author} {\bibfnamefont {A.}~\bibnamefont {Merrill}}, \bibinfo {author} {\bibfnamefont {C.~D.}\ \bibnamefont {Cress}}, \bibinfo {author} {\bibfnamefont {J.~E.}\ \bibnamefont {Rossi}}, \bibinfo {author} {\bibfnamefont {N.~D.}\ \bibnamefont {Cox}},\ and\ \bibinfo {author} {\bibfnamefont {B.~J.}\ \bibnamefont {Landi}},\ }\bibfield  {title} {\bibinfo {title} {Threshold displacement energies in graphene and single-walled carbon nanotubes},\ }\href@noop {} {\bibfield  {journal} {\bibinfo  {journal} {Phys. Rev. B}\ }\textbf {\bibinfo {volume} {92}},\ \bibinfo {pages} {075404} (\bibinfo {year} {2015})}\BibitemShut {NoStop}%
\bibitem [{\citenamefont {Hossain}\ and\ \citenamefont {Brown}(1977)}]{hossain1977electron}%
  \BibitemOpen
  \bibfield  {author} {\bibinfo {author} {\bibfnamefont {M.}~\bibnamefont {Hossain}}\ and\ \bibinfo {author} {\bibfnamefont {L.}~\bibnamefont {Brown}},\ }\bibfield  {title} {\bibinfo {title} {Electron irradiation damage in magnesium},\ }\href@noop {} {\bibfield  {journal} {\bibinfo  {journal} {Acta Metall.}\ }\textbf {\bibinfo {volume} {25}},\ \bibinfo {pages} {257} (\bibinfo {year} {1977})}\BibitemShut {NoStop}%
\bibitem [{\citenamefont {Faust}\ \emph {et~al.}(1969)\citenamefont {Faust}, \citenamefont {O'Neal},\ and\ \citenamefont {Chaplin}}]{faust1969measurements}%
  \BibitemOpen
  \bibfield  {author} {\bibinfo {author} {\bibfnamefont {W.}~\bibnamefont {Faust}}, \bibinfo {author} {\bibfnamefont {T.}~\bibnamefont {O'Neal}},\ and\ \bibinfo {author} {\bibfnamefont {R.}~\bibnamefont {Chaplin}},\ }\bibfield  {title} {\bibinfo {title} {Measurements of the electron-irradiation damage rates in magnesium},\ }\href@noop {} {\bibfield  {journal} {\bibinfo  {journal} {Phys. Rev.}\ }\textbf {\bibinfo {volume} {183}},\ \bibinfo {pages} {609} (\bibinfo {year} {1969})}\BibitemShut {NoStop}%
\bibitem [{\citenamefont {Iseler}\ \emph {et~al.}(1966)\citenamefont {Iseler}, \citenamefont {Dawson}, \citenamefont {Mehner},\ and\ \citenamefont {Kauffman}}]{iseler1966production}%
  \BibitemOpen
  \bibfield  {author} {\bibinfo {author} {\bibfnamefont {G.}~\bibnamefont {Iseler}}, \bibinfo {author} {\bibfnamefont {H.}~\bibnamefont {Dawson}}, \bibinfo {author} {\bibfnamefont {A.}~\bibnamefont {Mehner}},\ and\ \bibinfo {author} {\bibfnamefont {J.}~\bibnamefont {Kauffman}},\ }\bibfield  {title} {\bibinfo {title} {Production rates of electrical resistivity in copper and aluminum induced by electron irradiation},\ }\href@noop {} {\bibfield  {journal} {\bibinfo  {journal} {Phys. Rev.}\ }\textbf {\bibinfo {volume} {146}},\ \bibinfo {pages} {468} (\bibinfo {year} {1966})}\BibitemShut {NoStop}%
\bibitem [{\citenamefont {Konobeyev}\ \emph {et~al.}(2017)\citenamefont {Konobeyev}, \citenamefont {Fischer}, \citenamefont {Korovin},\ and\ \citenamefont {Simakov}}]{konobeyev2017evaluation}%
  \BibitemOpen
  \bibfield  {author} {\bibinfo {author} {\bibfnamefont {A.~Y.}\ \bibnamefont {Konobeyev}}, \bibinfo {author} {\bibfnamefont {U.}~\bibnamefont {Fischer}}, \bibinfo {author} {\bibfnamefont {Y.~A.}\ \bibnamefont {Korovin}},\ and\ \bibinfo {author} {\bibfnamefont {S.}~\bibnamefont {Simakov}},\ }\bibfield  {title} {\bibinfo {title} {Evaluation of effective threshold displacement energies and other data required for the calculation of advanced atomic displacement cross-sections},\ }\href@noop {} {\bibfield  {journal} {\bibinfo  {journal} {Nucl. Eng. Technol.}\ }\textbf {\bibinfo {volume} {3}},\ \bibinfo {pages} {169} (\bibinfo {year} {2017})}\BibitemShut {NoStop}%
\bibitem [{\citenamefont {Ouyang}\ \emph {et~al.}(2018)\citenamefont {Ouyang}, \citenamefont {Curtarolo}, \citenamefont {Ahmetcik}, \citenamefont {Scheffler},\ and\ \citenamefont {Ghiringhelli}}]{ouyang2018sisso}%
  \BibitemOpen
  \bibfield  {author} {\bibinfo {author} {\bibfnamefont {R.}~\bibnamefont {Ouyang}}, \bibinfo {author} {\bibfnamefont {S.}~\bibnamefont {Curtarolo}}, \bibinfo {author} {\bibfnamefont {E.}~\bibnamefont {Ahmetcik}}, \bibinfo {author} {\bibfnamefont {M.}~\bibnamefont {Scheffler}},\ and\ \bibinfo {author} {\bibfnamefont {L.~M.}\ \bibnamefont {Ghiringhelli}},\ }\bibfield  {title} {\bibinfo {title} {\text{SISSO}: A compressed-sensing method for identifying the best low-dimensional descriptor in an immensity of offered candidates},\ }\href@noop {} {\bibfield  {journal} {\bibinfo  {journal} {Phys. Rev. Mater.}\ }\textbf {\bibinfo {volume} {2}},\ \bibinfo {pages} {083802} (\bibinfo {year} {2018})}\BibitemShut {NoStop}%
\bibitem [{\citenamefont {Purcell}\ \emph {et~al.}(2023)\citenamefont {Purcell}, \citenamefont {Scheffler},\ and\ \citenamefont {Ghiringhelli}}]{purcell2023recent}%
  \BibitemOpen
  \bibfield  {author} {\bibinfo {author} {\bibfnamefont {T.~A.}\ \bibnamefont {Purcell}}, \bibinfo {author} {\bibfnamefont {M.}~\bibnamefont {Scheffler}},\ and\ \bibinfo {author} {\bibfnamefont {L.~M.}\ \bibnamefont {Ghiringhelli}},\ }\bibfield  {title} {\bibinfo {title} {Recent advances in the \text{SISSO} method and their implementation in the \text{SISSO}++ code},\ }\href@noop {} {\bibfield  {journal} {\bibinfo  {journal} {J. Chem. Phys.}\ }\textbf {\bibinfo {volume} {159}} (\bibinfo {year} {2023})}\BibitemShut {NoStop}%
\bibitem [{\citenamefont {Bartel}\ \emph {et~al.}(2019)\citenamefont {Bartel}, \citenamefont {Sutton}, \citenamefont {Goldsmith}, \citenamefont {Ouyang}, \citenamefont {Musgrave}, \citenamefont {Ghiringhelli},\ and\ \citenamefont {Scheffler}}]{bartel2019new}%
  \BibitemOpen
  \bibfield  {author} {\bibinfo {author} {\bibfnamefont {C.~J.}\ \bibnamefont {Bartel}}, \bibinfo {author} {\bibfnamefont {C.}~\bibnamefont {Sutton}}, \bibinfo {author} {\bibfnamefont {B.~R.}\ \bibnamefont {Goldsmith}}, \bibinfo {author} {\bibfnamefont {R.}~\bibnamefont {Ouyang}}, \bibinfo {author} {\bibfnamefont {C.~B.}\ \bibnamefont {Musgrave}}, \bibinfo {author} {\bibfnamefont {L.~M.}\ \bibnamefont {Ghiringhelli}},\ and\ \bibinfo {author} {\bibfnamefont {M.}~\bibnamefont {Scheffler}},\ }\bibfield  {title} {\bibinfo {title} {New tolerance factor to predict the stability of perovskite oxides and halides},\ }\href@noop {} {\bibfield  {journal} {\bibinfo  {journal} {Sci. Adv.}\ }\textbf {\bibinfo {volume} {5}},\ \bibinfo {pages} {eaav0693} (\bibinfo {year} {2019})}\BibitemShut {NoStop}%
\bibitem [{\citenamefont {Ouyang}\ \emph {et~al.}(2019)\citenamefont {Ouyang}, \citenamefont {Ahmetcik}, \citenamefont {Carbogno}, \citenamefont {Scheffler},\ and\ \citenamefont {Ghiringhelli}}]{ouyang2019simultaneous}%
  \BibitemOpen
  \bibfield  {author} {\bibinfo {author} {\bibfnamefont {R.}~\bibnamefont {Ouyang}}, \bibinfo {author} {\bibfnamefont {E.}~\bibnamefont {Ahmetcik}}, \bibinfo {author} {\bibfnamefont {C.}~\bibnamefont {Carbogno}}, \bibinfo {author} {\bibfnamefont {M.}~\bibnamefont {Scheffler}},\ and\ \bibinfo {author} {\bibfnamefont {L.~M.}\ \bibnamefont {Ghiringhelli}},\ }\bibfield  {title} {\bibinfo {title} {Simultaneous learning of several materials properties from incomplete databases with multi-task \text{SISSO}},\ }\href@noop {} {\bibfield  {journal} {\bibinfo  {journal} {J. Phys. Mater.}\ }\textbf {\bibinfo {volume} {2}},\ \bibinfo {pages} {024002} (\bibinfo {year} {2019})}\BibitemShut {NoStop}%
\bibitem [{\citenamefont {Liu}\ \emph {et~al.}(2022)\citenamefont {Liu}, \citenamefont {Wang}, \citenamefont {Gao}, \citenamefont {Chang}, \citenamefont {Tom}, \citenamefont {Yu}, \citenamefont {Ghiringhelli},\ and\ \citenamefont {Marom}}]{liu2022finding}%
  \BibitemOpen
  \bibfield  {author} {\bibinfo {author} {\bibfnamefont {X.}~\bibnamefont {Liu}}, \bibinfo {author} {\bibfnamefont {X.}~\bibnamefont {Wang}}, \bibinfo {author} {\bibfnamefont {S.}~\bibnamefont {Gao}}, \bibinfo {author} {\bibfnamefont {V.}~\bibnamefont {Chang}}, \bibinfo {author} {\bibfnamefont {R.}~\bibnamefont {Tom}}, \bibinfo {author} {\bibfnamefont {M.}~\bibnamefont {Yu}}, \bibinfo {author} {\bibfnamefont {L.~M.}\ \bibnamefont {Ghiringhelli}},\ and\ \bibinfo {author} {\bibfnamefont {N.}~\bibnamefont {Marom}},\ }\bibfield  {title} {\bibinfo {title} {Finding predictive models for singlet fission by machine learning},\ }\href@noop {} {\bibfield  {journal} {\bibinfo  {journal} {npj Comput. Mater.}\ }\textbf {\bibinfo {volume} {8}},\ \bibinfo {pages} {70} (\bibinfo {year} {2022})}\BibitemShut {NoStop}%
\bibitem [{\citenamefont {Singh}\ \emph {et~al.}(2022)\citenamefont {Singh}, \citenamefont {Del~Rose}, \citenamefont {Vazquez}, \citenamefont {Arroyave},\ and\ \citenamefont {Mudryk}}]{singh2022machine}%
  \BibitemOpen
  \bibfield  {author} {\bibinfo {author} {\bibfnamefont {P.}~\bibnamefont {Singh}}, \bibinfo {author} {\bibfnamefont {T.}~\bibnamefont {Del~Rose}}, \bibinfo {author} {\bibfnamefont {G.}~\bibnamefont {Vazquez}}, \bibinfo {author} {\bibfnamefont {R.}~\bibnamefont {Arroyave}},\ and\ \bibinfo {author} {\bibfnamefont {Y.}~\bibnamefont {Mudryk}},\ }\bibfield  {title} {\bibinfo {title} {Machine-learning enabled thermodynamic model for the design of new rare-earth compounds},\ }\href@noop {} {\bibfield  {journal} {\bibinfo  {journal} {Acta Mater.}\ }\textbf {\bibinfo {volume} {229}},\ \bibinfo {pages} {117759} (\bibinfo {year} {2022})}\BibitemShut {NoStop}%
\bibitem [{\citenamefont {Cao}\ \emph {et~al.}(2020)\citenamefont {Cao}, \citenamefont {Ouyang}, \citenamefont {Ghiringhelli}, \citenamefont {Scheffler}, \citenamefont {Liu}, \citenamefont {Carbogno},\ and\ \citenamefont {Zhang}}]{cao2020artificial}%
  \BibitemOpen
  \bibfield  {author} {\bibinfo {author} {\bibfnamefont {G.}~\bibnamefont {Cao}}, \bibinfo {author} {\bibfnamefont {R.}~\bibnamefont {Ouyang}}, \bibinfo {author} {\bibfnamefont {L.~M.}\ \bibnamefont {Ghiringhelli}}, \bibinfo {author} {\bibfnamefont {M.}~\bibnamefont {Scheffler}}, \bibinfo {author} {\bibfnamefont {H.}~\bibnamefont {Liu}}, \bibinfo {author} {\bibfnamefont {C.}~\bibnamefont {Carbogno}},\ and\ \bibinfo {author} {\bibfnamefont {Z.}~\bibnamefont {Zhang}},\ }\bibfield  {title} {\bibinfo {title} {Artificial intelligence for high-throughput discovery of topological insulators: The example of alloyed tetradymites},\ }\href@noop {} {\bibfield  {journal} {\bibinfo  {journal} {Phys. Rev. Mater.}\ }\textbf {\bibinfo {volume} {4}},\ \bibinfo {pages} {034204} (\bibinfo {year} {2020})}\BibitemShut {NoStop}%
\bibitem [{\citenamefont {Foppa}\ \emph {et~al.}(2021)\citenamefont {Foppa}, \citenamefont {Ghiringhelli}, \citenamefont {Girgsdies}, \citenamefont {Hashagen}, \citenamefont {Kube}, \citenamefont {H{\"a}vecker}, \citenamefont {Carey}, \citenamefont {Tarasov}, \citenamefont {Kraus}, \citenamefont {Rosowski} \emph {et~al.}}]{foppa2021materials}%
  \BibitemOpen
  \bibfield  {author} {\bibinfo {author} {\bibfnamefont {L.}~\bibnamefont {Foppa}}, \bibinfo {author} {\bibfnamefont {L.~M.}\ \bibnamefont {Ghiringhelli}}, \bibinfo {author} {\bibfnamefont {F.}~\bibnamefont {Girgsdies}}, \bibinfo {author} {\bibfnamefont {M.}~\bibnamefont {Hashagen}}, \bibinfo {author} {\bibfnamefont {P.}~\bibnamefont {Kube}}, \bibinfo {author} {\bibfnamefont {M.}~\bibnamefont {H{\"a}vecker}}, \bibinfo {author} {\bibfnamefont {S.~J.}\ \bibnamefont {Carey}}, \bibinfo {author} {\bibfnamefont {A.}~\bibnamefont {Tarasov}}, \bibinfo {author} {\bibfnamefont {P.}~\bibnamefont {Kraus}}, \bibinfo {author} {\bibfnamefont {F.}~\bibnamefont {Rosowski}}, \emph {et~al.},\ }\bibfield  {title} {\bibinfo {title} {Materials genes of heterogeneous catalysis from clean experiments and artificial intelligence},\ }\href@noop {} {\bibfield  {journal} {\bibinfo  {journal} {MRS Bull.}\ ,\ \bibinfo {pages} {1}} (\bibinfo {year} {2021})}\BibitemShut {NoStop}%
\bibitem [{\citenamefont {Nordlund}\ \emph {et~al.}(2006)\citenamefont {Nordlund}, \citenamefont {Wallenius},\ and\ \citenamefont {Malerba}}]{nordlund2006molecular}%
  \BibitemOpen
  \bibfield  {author} {\bibinfo {author} {\bibfnamefont {K.}~\bibnamefont {Nordlund}}, \bibinfo {author} {\bibfnamefont {J.}~\bibnamefont {Wallenius}},\ and\ \bibinfo {author} {\bibfnamefont {L.}~\bibnamefont {Malerba}},\ }\bibfield  {title} {\bibinfo {title} {Molecular dynamics simulations of threshold displacement energies in \text{Fe}},\ }\href@noop {} {\bibfield  {journal} {\bibinfo  {journal} {Nucl. Instrum. Methods Phys. Res. B}\ }\textbf {\bibinfo {volume} {246}},\ \bibinfo {pages} {322} (\bibinfo {year} {2006})}\BibitemShut {NoStop}%
\bibitem [{\citenamefont {Steeds}(2011)}]{steeds2011orientation}%
  \BibitemOpen
  \bibfield  {author} {\bibinfo {author} {\bibfnamefont {J.}~\bibnamefont {Steeds}},\ }\bibfield  {title} {\bibinfo {title} {Orientation dependence of near-threshold damage production by electron irradiation of 4h sic and diamond and outward migration of defects},\ }\href@noop {} {\bibfield  {journal} {\bibinfo  {journal} {Nucl. Instrum. Methods Phys. Res., Sect. B}\ }\textbf {\bibinfo {volume} {269}},\ \bibinfo {pages} {1702} (\bibinfo {year} {2011})}\BibitemShut {NoStop}%
\bibitem [{\citenamefont {Steffen}\ \emph {et~al.}(1992)\citenamefont {Steffen}, \citenamefont {Marton},\ and\ \citenamefont {Rabalais}}]{steffen1992displacement}%
  \BibitemOpen
  \bibfield  {author} {\bibinfo {author} {\bibfnamefont {H.}~\bibnamefont {Steffen}}, \bibinfo {author} {\bibfnamefont {D.}~\bibnamefont {Marton}},\ and\ \bibinfo {author} {\bibfnamefont {J.}~\bibnamefont {Rabalais}},\ }\bibfield  {title} {\bibinfo {title} {Displacement energy threshold for \text{Ne}+ irradiation of graphite},\ }\href@noop {} {\bibfield  {journal} {\bibinfo  {journal} {Phys. Rev. Lett.}\ }\textbf {\bibinfo {volume} {68}},\ \bibinfo {pages} {1726} (\bibinfo {year} {1992})}\BibitemShut {NoStop}%
\bibitem [{\citenamefont {Poulin}\ and\ \citenamefont {Bourgoin}(1980)}]{poulin1980threshold}%
  \BibitemOpen
  \bibfield  {author} {\bibinfo {author} {\bibfnamefont {F.}~\bibnamefont {Poulin}}\ and\ \bibinfo {author} {\bibfnamefont {J.}~\bibnamefont {Bourgoin}},\ }\bibfield  {title} {\bibinfo {title} {Threshold energy for atomic displacement in electron irradiated germanium},\ }\href@noop {} {\bibfield  {journal} {\bibinfo  {journal} {Rev. Phys. Appl.}\ }\textbf {\bibinfo {volume} {15}},\ \bibinfo {pages} {15} (\bibinfo {year} {1980})}\BibitemShut {NoStop}%
\bibitem [{\citenamefont {Vajda}\ \emph {et~al.}(1977)\citenamefont {Vajda}, \citenamefont {Biget}, \citenamefont {Lucasson},\ and\ \citenamefont {Lucasson}}]{vajda1977problem}%
  \BibitemOpen
  \bibfield  {author} {\bibinfo {author} {\bibfnamefont {P.}~\bibnamefont {Vajda}}, \bibinfo {author} {\bibfnamefont {M.}~\bibnamefont {Biget}}, \bibinfo {author} {\bibfnamefont {A.}~\bibnamefont {Lucasson}},\ and\ \bibinfo {author} {\bibfnamefont {P.}~\bibnamefont {Lucasson}},\ }\bibfield  {title} {\bibinfo {title} {On the problem of displacement threshold determination in irradiated metals: subthreshold effects and recovery spectrum},\ }\href@noop {} {\bibfield  {journal} {\bibinfo  {journal} {J. Phys. F: Met. Phys.}\ }\textbf {\bibinfo {volume} {7}},\ \bibinfo {pages} {L123} (\bibinfo {year} {1977})}\BibitemShut {NoStop}%
\bibitem [{\citenamefont {DeKock}\ \emph {et~al.}(2012)\citenamefont {DeKock}, \citenamefont {Strikwerda},\ and\ \citenamefont {Eric}}]{dekock2012atomic}%
  \BibitemOpen
  \bibfield  {author} {\bibinfo {author} {\bibfnamefont {R.~L.}\ \bibnamefont {DeKock}}, \bibinfo {author} {\bibfnamefont {J.~R.}\ \bibnamefont {Strikwerda}},\ and\ \bibinfo {author} {\bibfnamefont {X.~Y.}\ \bibnamefont {Eric}},\ }\bibfield  {title} {\bibinfo {title} {Atomic size, ionization energy, polarizability, asymptotic behavior, and the slater--zener model},\ }\href@noop {} {\bibfield  {journal} {\bibinfo  {journal} {Chem. Phys. Lett.}\ }\textbf {\bibinfo {volume} {547}},\ \bibinfo {pages} {120} (\bibinfo {year} {2012})}\BibitemShut {NoStop}%
\bibitem [{\citenamefont {Lindemann}(1910)}]{lindemann1910fa}%
  \BibitemOpen
  \bibfield  {author} {\bibinfo {author} {\bibfnamefont {F.}~\bibnamefont {Lindemann}},\ }\bibfield  {title} {\bibinfo {title} {Fa lindemann, phys. z. 11, 609 (1910)},\ }\href@noop {} {\bibfield  {journal} {\bibinfo  {journal} {Phys. Z}\ }\textbf {\bibinfo {volume} {11}},\ \bibinfo {pages} {609} (\bibinfo {year} {1910})}\BibitemShut {NoStop}%
\bibitem [{\citenamefont {Grimvall}\ and\ \citenamefont {Sj{\"o}din}(1974)}]{grimvall1974correlation}%
  \BibitemOpen
  \bibfield  {author} {\bibinfo {author} {\bibfnamefont {G.}~\bibnamefont {Grimvall}}\ and\ \bibinfo {author} {\bibfnamefont {S.}~\bibnamefont {Sj{\"o}din}},\ }\bibfield  {title} {\bibinfo {title} {Correlation of properties of materials to debye and melting temperatures},\ }\href@noop {} {\bibfield  {journal} {\bibinfo  {journal} {Phys. Scr.}\ }\textbf {\bibinfo {volume} {10}},\ \bibinfo {pages} {340} (\bibinfo {year} {1974})}\BibitemShut {NoStop}%
\bibitem [{\citenamefont {Li}\ and\ \citenamefont {Wu}(2002)}]{li2002empirical}%
  \BibitemOpen
  \bibfield  {author} {\bibinfo {author} {\bibfnamefont {C.}~\bibnamefont {Li}}\ and\ \bibinfo {author} {\bibfnamefont {P.}~\bibnamefont {Wu}},\ }\bibfield  {title} {\bibinfo {title} {Empirical relation of melting temperatures of \text{CsCl}-type intermetallic compounds to their cohesive energies},\ }\href@noop {} {\bibfield  {journal} {\bibinfo  {journal} {Chem. Mater.}\ }\textbf {\bibinfo {volume} {14}},\ \bibinfo {pages} {4833} (\bibinfo {year} {2002})}\BibitemShut {NoStop}%
\bibitem [{\citenamefont {{Ioffe Institute}}(2024{\natexlab{a}})}]{IoffeC-diamond}%
  \BibitemOpen
  \bibfield  {author} {\bibinfo {author} {\bibnamefont {{Ioffe Institute}}},\ }\href@noop {} {\bibinfo {title} {Ioffe institute - semiconductor material - diamond (c)}},\ \bibinfo {howpublished} {Available online: \url{http://www.ioffe.ru/SVA/NSM/Semicond/Diamond/}} (\bibinfo {year} {2024}{\natexlab{a}}),\ \bibinfo {note} {accessed: 2024-05-08}\BibitemShut {NoStop}%
\bibitem [{\citenamefont {{AZoMaterials}}(2024)}]{AZoM}%
  \BibitemOpen
  \bibfield  {author} {\bibinfo {author} {\bibnamefont {{AZoMaterials}}},\ }\href@noop {} {\bibinfo {title} {Azom - materials properties database}},\ \bibinfo {howpublished} {Available online: \url{https://www.azom.com/properties.aspx?ArticleID=516}} (\bibinfo {year} {2024}),\ \bibinfo {note} {accessed: 2024-05-08}\BibitemShut {NoStop}%
\bibitem [{\citenamefont {Crocombette}\ \emph {et~al.}(2016)\citenamefont {Crocombette}, \citenamefont {Van~Brutzel}, \citenamefont {Simeone},\ and\ \citenamefont {Luneville}}]{crocombette2016molecular}%
  \BibitemOpen
  \bibfield  {author} {\bibinfo {author} {\bibfnamefont {J.-P.}\ \bibnamefont {Crocombette}}, \bibinfo {author} {\bibfnamefont {L.}~\bibnamefont {Van~Brutzel}}, \bibinfo {author} {\bibfnamefont {D.}~\bibnamefont {Simeone}},\ and\ \bibinfo {author} {\bibfnamefont {L.}~\bibnamefont {Luneville}},\ }\bibfield  {title} {\bibinfo {title} {Molecular dynamics simulations of high energy cascade in ordered alloys: Defect production and subcascade division},\ }\href@noop {} {\bibfield  {journal} {\bibinfo  {journal} {J. Nucl. Mater.}\ }\textbf {\bibinfo {volume} {474}},\ \bibinfo {pages} {134} (\bibinfo {year} {2016})}\BibitemShut {NoStop}%
\bibitem [{\citenamefont {Agarwal}\ \emph {et~al.}(2021)\citenamefont {Agarwal}, \citenamefont {Lin}, \citenamefont {Li}, \citenamefont {Stoller},\ and\ \citenamefont {Zinkle}}]{agarwal2021use}%
  \BibitemOpen
  \bibfield  {author} {\bibinfo {author} {\bibfnamefont {S.}~\bibnamefont {Agarwal}}, \bibinfo {author} {\bibfnamefont {Y.}~\bibnamefont {Lin}}, \bibinfo {author} {\bibfnamefont {C.}~\bibnamefont {Li}}, \bibinfo {author} {\bibfnamefont {R.}~\bibnamefont {Stoller}},\ and\ \bibinfo {author} {\bibfnamefont {S.}~\bibnamefont {Zinkle}},\ }\bibfield  {title} {\bibinfo {title} {On the use of \text{SRIM} for calculating vacancy production: Quick calculation and full-cascade options},\ }\href@noop {} {\bibfield  {journal} {\bibinfo  {journal} {Nucl. Instrum. Methods Phys. Res. Sect. B Beam Interact. Mater. Atoms}\ }\textbf {\bibinfo {volume} {503}},\ \bibinfo {pages} {11} (\bibinfo {year} {2021})}\BibitemShut {NoStop}%
\bibitem [{\citenamefont {Lin}\ \emph {et~al.}(2023)\citenamefont {Lin}, \citenamefont {Zinkle}, \citenamefont {Ortiz}, \citenamefont {Crocombette}, \citenamefont {Webb},\ and\ \citenamefont {Stoller}}]{lin2023predicting}%
  \BibitemOpen
  \bibfield  {author} {\bibinfo {author} {\bibfnamefont {Y.-R.}\ \bibnamefont {Lin}}, \bibinfo {author} {\bibfnamefont {S.~J.}\ \bibnamefont {Zinkle}}, \bibinfo {author} {\bibfnamefont {C.~J.}\ \bibnamefont {Ortiz}}, \bibinfo {author} {\bibfnamefont {J.-P.}\ \bibnamefont {Crocombette}}, \bibinfo {author} {\bibfnamefont {R.}~\bibnamefont {Webb}},\ and\ \bibinfo {author} {\bibfnamefont {R.~E.}\ \bibnamefont {Stoller}},\ }\bibfield  {title} {\bibinfo {title} {Predicting displacement damage for ion irradiation: Origin of the overestimation of vacancy production in \text{SRIM} full-cascade calculations},\ }\href@noop {} {\bibfield  {journal} {\bibinfo  {journal} {Curr. Opin. Solid State Mater. Sci.}\ }\textbf {\bibinfo {volume} {27}},\ \bibinfo {pages} {101120} (\bibinfo {year} {2023})}\BibitemShut {NoStop}%
\bibitem [{\citenamefont {Vopson}\ \emph {et~al.}(2020)\citenamefont {Vopson}, \citenamefont {Rogers},\ and\ \citenamefont {Hepburn}}]{vopson2020generalized}%
  \BibitemOpen
  \bibfield  {author} {\bibinfo {author} {\bibfnamefont {M.~M.}\ \bibnamefont {Vopson}}, \bibinfo {author} {\bibfnamefont {N.}~\bibnamefont {Rogers}},\ and\ \bibinfo {author} {\bibfnamefont {I.}~\bibnamefont {Hepburn}},\ }\bibfield  {title} {\bibinfo {title} {The generalized lindemann melting coefficient},\ }\href@noop {} {\bibfield  {journal} {\bibinfo  {journal} {Solid State Commun.}\ }\textbf {\bibinfo {volume} {318}},\ \bibinfo {pages} {113977} (\bibinfo {year} {2020})}\BibitemShut {NoStop}%
\bibitem [{\citenamefont {Singh}\ and\ \citenamefont {Phantsi}(2018)}]{singh2018bond}%
  \BibitemOpen
  \bibfield  {author} {\bibinfo {author} {\bibfnamefont {M.}~\bibnamefont {Singh}}\ and\ \bibinfo {author} {\bibfnamefont {T.}~\bibnamefont {Phantsi}},\ }\bibfield  {title} {\bibinfo {title} {Bond theory model to study cohesive energy, thermal expansion coefficient and specific heat of nanosolids},\ }\href@noop {} {\bibfield  {journal} {\bibinfo  {journal} {Chin. J. Phys.}\ }\textbf {\bibinfo {volume} {56}},\ \bibinfo {pages} {2948} (\bibinfo {year} {2018})}\BibitemShut {NoStop}%
\bibitem [{\citenamefont {{Wolfram Research, Inc.}}(2024)}]{Mathematica_ElementData}%
  \BibitemOpen
  \bibfield  {author} {\bibinfo {author} {\bibnamefont {{Wolfram Research, Inc.}}},\ }\href@noop {} {\bibinfo {title} {{Mathematica Version 13.3.1.0}}},\ \bibinfo {howpublished} {{Wolfram Research, Inc.}} (\bibinfo {year} {2024}),\ \bibinfo {note} {data retrieved using the \texttt{ElementData} function in Mathematica.}\BibitemShut {Stop}%
\bibitem [{\citenamefont {{Materials Project}}(2024)}]{MaterialsProject}%
  \BibitemOpen
  \bibfield  {author} {\bibinfo {author} {\bibnamefont {{Materials Project}}},\ }\href@noop {} {\bibinfo {title} {Materials project}},\ \bibinfo {howpublished} {Available online: \url{https://legacy.materialsproject.org/}} (\bibinfo {year} {2024}),\ \bibinfo {note} {accessed: 2024-05-08}\BibitemShut {NoStop}%
\bibitem [{\citenamefont {{Ioffe Institute}}(2024{\natexlab{b}})}]{IoffeGe}%
  \BibitemOpen
  \bibfield  {author} {\bibinfo {author} {\bibnamefont {{Ioffe Institute}}},\ }\href@noop {} {\bibinfo {title} {Ioffe institute - semiconductor material - germanium (ge)}},\ \bibinfo {howpublished} {Available online: \url{http://www.ioffe.ru/SVA/NSM/Semicond/Ge/}} (\bibinfo {year} {2024}{\natexlab{b}}),\ \bibinfo {note} {accessed: 2024-05-08}\BibitemShut {NoStop}%
\bibitem [{\citenamefont {{University of Colorado}}(2001)}]{ColoradoPT}%
  \BibitemOpen
  \bibfield  {author} {\bibinfo {author} {\bibnamefont {{University of Colorado}}},\ }\href@noop {} {\bibinfo {title} {Periodic table of elements}},\ \bibinfo {howpublished} {\url{https://home.cs.colorado.edu/~kena/classes/7818/f01/assignments/pt.html}} (\bibinfo {year} {2001}),\ \bibinfo {note} {accessed on: May 7, 2024}\BibitemShut {NoStop}%
\bibitem [{\citenamefont {{National Center for Biotechnology Information}}(2024)}]{PubChemAtomicRadius}%
  \BibitemOpen
  \bibfield  {author} {\bibinfo {author} {\bibnamefont {{National Center for Biotechnology Information}}},\ }\href@noop {} {\bibinfo {title} {Periodic table - atomic radius}},\ \bibinfo {howpublished} {Available online: \url{https://pubchem.ncbi.nlm.nih.gov/ptable/atomic-radius/}} (\bibinfo {year} {2024}),\ \bibinfo {note} {accessed: 2024-05-08}\BibitemShut {NoStop}%
\bibitem [{\citenamefont {Kittel}(2005)}]{Kittel2005}%
  \BibitemOpen
  \bibfield  {author} {\bibinfo {author} {\bibfnamefont {C.}~\bibnamefont {Kittel}},\ }\href@noop {} {\emph {\bibinfo {title} {Introduction to Solid State Physics}}},\ \bibinfo {edition} {8th}\ ed.\ (\bibinfo  {publisher} {John Wiley \& Sons, Inc},\ \bibinfo {address} {Hoboken, NJ},\ \bibinfo {year} {2005})\ p.~\bibinfo {pages} {50}\BibitemShut {NoStop}%
\bibitem [{\citenamefont {Moelle}\ \emph {et~al.}(1997)\citenamefont {Moelle}, \citenamefont {Klose}, \citenamefont {Sz{\"u}cs}, \citenamefont {Fecht}, \citenamefont {Johnston}, \citenamefont {Chalker},\ and\ \citenamefont {Werner}}]{moelle1997measurement}%
  \BibitemOpen
  \bibfield  {author} {\bibinfo {author} {\bibfnamefont {C.}~\bibnamefont {Moelle}}, \bibinfo {author} {\bibfnamefont {S.}~\bibnamefont {Klose}}, \bibinfo {author} {\bibfnamefont {F.}~\bibnamefont {Sz{\"u}cs}}, \bibinfo {author} {\bibfnamefont {H.}~\bibnamefont {Fecht}}, \bibinfo {author} {\bibfnamefont {C.}~\bibnamefont {Johnston}}, \bibinfo {author} {\bibfnamefont {P.}~\bibnamefont {Chalker}},\ and\ \bibinfo {author} {\bibfnamefont {M.}~\bibnamefont {Werner}},\ }\bibfield  {title} {\bibinfo {title} {Measurement and calculation of the thermal expansion coefficient of diamond},\ }\href@noop {} {\bibfield  {journal} {\bibinfo  {journal} {Diam Relat Mater.}\ }\textbf {\bibinfo {volume} {6}},\ \bibinfo {pages} {839} (\bibinfo {year} {1997})}\BibitemShut {NoStop}%
\bibitem [{\citenamefont {Tsang}\ \emph {et~al.}(2005)\citenamefont {Tsang}, \citenamefont {Marsden}, \citenamefont {Fok},\ and\ \citenamefont {Hall}}]{tsang2005graphite}%
  \BibitemOpen
  \bibfield  {author} {\bibinfo {author} {\bibfnamefont {D.}~\bibnamefont {Tsang}}, \bibinfo {author} {\bibfnamefont {B.}~\bibnamefont {Marsden}}, \bibinfo {author} {\bibfnamefont {S.}~\bibnamefont {Fok}},\ and\ \bibinfo {author} {\bibfnamefont {G.}~\bibnamefont {Hall}},\ }\bibfield  {title} {\bibinfo {title} {Graphite thermal expansion relationship for different temperature ranges},\ }\href@noop {} {\bibfield  {journal} {\bibinfo  {journal} {Carbon}\ }\textbf {\bibinfo {volume} {43}},\ \bibinfo {pages} {2902} (\bibinfo {year} {2005})}\BibitemShut {NoStop}%
\bibitem [{\citenamefont {{{LibreTexts}}}()}]{LibreTextsSolidsDensities}%
  \BibitemOpen
  \bibfield  {author} {\bibinfo {author} {\bibnamefont {{{LibreTexts}}}},\ }\href@noop {} {\bibinfo {title} {{Appendix A - List of Solids Densities}}},\ \bibinfo {howpublished} {\url{https://eng.libretexts.org/Bookshelves/Civil_Engineering/Book\%3A_Slurry_Transport_(Miedema)/11\%3A_Appendices/11.01\%3A_Appendix_A-_List_of_Solids_Densities}},\ \bibinfo {note} {accessed: 2024-05-14}\BibitemShut {NoStop}%
\bibitem [{\citenamefont {{{International Atomic Energy Agency (IAEA)}}}(2024)}]{IAEA_GuideToGraphite}%
  \BibitemOpen
  \bibfield  {author} {\bibinfo {author} {\bibnamefont {{{International Atomic Energy Agency (IAEA)}}}},\ }\href@noop {} {\bibinfo {title} {{What is Graphite?}}},\ \bibinfo {howpublished} {\url{https://nucleus.iaea.org/sites/graphiteknowledgebase/wiki/Guide_to_Graphite/What\%20is\%20Graphite.aspx}} (\bibinfo {year} {2024}),\ \bibinfo {note} {accessed: 2024-05-14}\BibitemShut {NoStop}%
\bibitem [{\citenamefont {Chelikowsky}\ and\ \citenamefont {Louie}(1984)}]{chelikowsky1984first}%
  \BibitemOpen
  \bibfield  {author} {\bibinfo {author} {\bibfnamefont {J.~R.}\ \bibnamefont {Chelikowsky}}\ and\ \bibinfo {author} {\bibfnamefont {S.~G.}\ \bibnamefont {Louie}},\ }\bibfield  {title} {\bibinfo {title} {First-principles linear combination of atomic orbitals method for the cohesive and structural properties of solids: Application to diamond},\ }\href@noop {} {\bibfield  {journal} {\bibinfo  {journal} {Phys. Rev. B}\ }\textbf {\bibinfo {volume} {29}},\ \bibinfo {pages} {3470} (\bibinfo {year} {1984})}\BibitemShut {NoStop}%
\bibitem [{\citenamefont {Weinert}\ \emph {et~al.}(1982)\citenamefont {Weinert}, \citenamefont {Wimmer},\ and\ \citenamefont {Freeman}}]{weinert1982total}%
  \BibitemOpen
  \bibfield  {author} {\bibinfo {author} {\bibfnamefont {M.}~\bibnamefont {Weinert}}, \bibinfo {author} {\bibfnamefont {E.}~\bibnamefont {Wimmer}},\ and\ \bibinfo {author} {\bibfnamefont {A.}~\bibnamefont {Freeman}},\ }\bibfield  {title} {\bibinfo {title} {Total-energy all-electron density functional method for bulk solids and surfaces},\ }\href@noop {} {\bibfield  {journal} {\bibinfo  {journal} {Phys. Rev. B}\ }\textbf {\bibinfo {volume} {26}},\ \bibinfo {pages} {4571} (\bibinfo {year} {1982})}\BibitemShut {NoStop}%
\bibitem [{\citenamefont {Pierson}(2012)}]{pierson2012handbook}%
  \BibitemOpen
  \bibfield  {author} {\bibinfo {author} {\bibfnamefont {H.~O.}\ \bibnamefont {Pierson}},\ }\href@noop {} {\emph {\bibinfo {title} {Handbook of carbon, graphite, diamonds and fullerenes: processing, properties and applications}}}\ (\bibinfo  {publisher} {William Andrew},\ \bibinfo {year} {2012})\BibitemShut {NoStop}%
\end{thebibliography}%

\end{document}